\documentclass[final,3p]{elsarticle}

 \usepackage{graphics}
 \usepackage{graphicx}
\usepackage{caption}
\usepackage{float}
\usepackage{subcaption}
\usepackage{amsmath}
\usepackage{color}
\usepackage{soul}           
\usepackage{ulem}
\usepackage{tikz}
\usepackage{makecell}
\usepackage{wrapfig}
\usepackage{algorithm}
\usepackage{algpseudocode}
\usepackage{makecell}
\usetikzlibrary{shapes.geometric, arrows}

\tikzstyle{normal} = [rectangle, rounded corners, minimum width=3cm, minimum height=1cm,text centered, draw=black, fill=red!30,text width=10em]

\tikzstyle{robust} = [trapezium, trapezium left angle=70, trapezium right angle=110, minimum width=3cm, minimum height=1cm, text centered, draw=black, fill=blue!30,text width=4em]

\usepackage[T1]{fontenc}
\usepackage{hyperref}
\hypersetup{
colorlinks=true
}
\usepackage{epstopdf}
\usepackage{xcolor}

\usepackage{enumitem}  

\biboptions{sort&compress}

\journal{arXiv}

\begin{document}

\begin{frontmatter}
\title{
{A procedure for assessing of machine health index data prediction quality}}

\author[label1]{Daniel Kuzio}
\author[label1]{Rados{\l}aw Zimroz}
\author[label2]{Agnieszka Wy\l oma\'nska}

\address[label1]{Faculty of Geoengineering, Mining and Geology, Wroclaw University of Science and Technology, Na Grobli 15, 50-421 Wroclaw, Poland}

\address[label2]{Faculty of Pure and Applied Mathematics, Hugo Steinhaus Center, Wroclaw University of Science and Technology, Wyspianskiego 27, 50-370 Wroclaw, Poland}


\begin{abstract}

 The paper discusses the challenge of evaluating the prognosis quality of \textcolor{black}{machine} health index (HI) data. Many existing solutions in machine health forecasting involve visual assessing of the quality of predictions to roughly gauge the similarity between predicted and actual samples, lacking precise measures or decisions. In this paper, we introduce a universal procedure with multiple variants and criteria. The overarching concept involves comparing predicted data with true HI time series, but each procedure variant has a specific pattern determined through statistical analysis. Additionally, a statistically established threshold is employed to classify the result as either a reliable or non-reliable prognosis. The criteria include both simple measures (MSE, MAPE) and more advanced ones (Space quantiles-inclusion factor, Kupiec’s POF, and TUFF statistics). Depending on the criterion chosen, the pattern and decision-making process vary. To illustrate effectiveness, we apply the proposed procedure to HI data sourced from the literature, covering both warning (linear degradation) and critical (exponential degradation) stages. While the method yields a binary output, there is potential for extension to a multi-class classification. Furthermore, experienced users can use the quality measure expressed in percentage for more in-depth analysis. 
\end{abstract}


\end{frontmatter}

\section{Introduction}

The modeling of Health Index (HI) data is a crucial and frequently discussed topic in the literature \cite{9234078, app11167175, 9263930, 9311875, Cholette1107, LI2023110213} \textcolor{black}{in machine condition monitoring problem}. 
\textcolor{black}{One of the main objectives of HI data modeling is to enable prediction, ultimately leading to the forecasting of machine condition and the remaining useful life time (RUL).} 
As outlined in \cite{Lei2018799}, a comprehensive machinery prognostic framework typically involves four key steps, including data acquisition \cite{FEMTO_data_acquisition}, the construction of health indicators \cite{9104023, 9896888, 9782410}, evaluation of health stages \cite{JANCZURA2023113399}, and the prediction of RUL \cite{CUI2020103967, CHEN2020107116, YAN2020471}. In this paper, we propose an extension to this procedure by introducing an additional element: the evaluation of the degradation process prediction.


\textcolor{black}{We can categorize degradation modeling into three main types: physics-of-failure-based models, data-driven approaches and hybrid approaches \cite{LEE2014314}. Physics-of-failure-based models capture the failure mechanisms or physical phenomena to formulate a mathematical description of the degradation process. Conversely, data-driven models utilize statistical fitting to degradation data without considering the physical properties of materials \cite{Ameneh299}. Finally, hybrid methods uses advantages of both types of methods and attempts to integrate them.}

\textcolor{black}{Physics-of-failure-based approaches are extensively explored in \cite{Cubillo2016}. The most common, the Paris-Erdogan (PE) model, was introduced in \cite{Paris1963}. Then, Lei et al. modified this model and applied for RUL prediction \cite{Lei20161314}. Additionally, Liao in \cite{6544227} modified PE model into a state space model. A significant challenge associated with this category of methods is that the complexity of the model escalates with the complexity of the considered mechanical system. Precise modeling of a jet engine or a multistage planetary gearbox demands multiple parameters and advanced numerical procedures to simulate the system's response. In many cases, obtaining such data is challenging, as they may not be readily accessible or may require experimental identification.}


\textcolor{black}{The data-driven approach can be categorized into two classes \cite{zio2022prognostics, Lei2018799, LI2023110646, WEN2022110276}: \textcolor{black}{machine learning-based models (ML-based models)}~\cite{szarek2023non, moosavi2022application, XIANG2020343, WU2020241, 9143075, 9089252} and statistical models~\cite{Si20111, zulawinski2023framework, Liu2022360, 9263930, 9311875, Cholette1107}. The application of \textcolor{black}{ML-based models} is increasing due to their effectiveness in modeling complex data \cite{electronics10243126}; however, a drawback is their requirement for a substantial amount of training data. We refer the readers to \cite{soft_computing}, where authors make a compressive summary of the prediction methods (mostly ML-based approaches) indicating their main advantages and disadvantages. On the contrary, statistical models may not be as effective as ML-based approaches for highly complex data, but can handle modeling and prediction with a smaller sample compared to machine learning models \cite{zio2022prognostics}. Additionally, statistical models are easier to interpret than ML-based models.} One drawback of this approach class is the necessity to choose a model that accurately fits the degradation process. Several categories of statistical models can be distinguished in the context of RUL prediction: autoregressive models \cite{Qian20142599}, random coefficient models \cite{Zhou2012793}, Brownian motion-based models \cite{Zhang2021, XU2021107675, YAN2021107378, SI201353, 6135842}, Markov process-based models \cite{7080918, CHIACHIO2020106621, KHAROUFEH2005}, Inverse Gaussian process-based models \cite{Wang2010188}, proportional hazards models \cite{ZHENG2021107964}, L\'evy stable-based models \cite{Liu2022360, LI2023110646, DUAN2021107974}, gamma process-based approaches \cite{Ling201977, 6228785, 7107275, WANG2021107504} etc.

\textcolor{black}{Hybrid approaches commonly use the integration of physics-based models with machine learning (ML) techniques. In \cite{RITTO2021107614}, the authors allowed the use of an interpretable physics-based model to construct a fast digital twin connected to the physical twin, supporting real-time engineering decisions. Shi et al. \cite{SHI2022109347} proposed a physics-informed machine learning method that combines a physics-based degradation model with an LSTM-based approach. The physics-based model takes into account the effects of operating conditions such as cycle time, environmental temperature, and load condition. Yan et al. \cite{YAN2021107378} developed a two-stage physics-based Wiener process that integrates the fatigue crack mechanism and the crack growth law with other minor factors. Furthermore, ML techniques are extensively combined with statistical models. Zemouri and Gouriveau \cite{Zemouri2010} combined artificial neural networks with autoregressive models for the prediction of degradation. In \cite{Hu2011} an ensemble data-driven prognostic approach is proposed that combines multiple member algorithms with a weighted-sum formulation. Maio et al. \cite{Maio2012} combined relevance vector machines to select a low number of significant basis functions and exponential regression to compute and continuously update the residual life estimation.
}

Determining the accuracy of a prognosis and deciding whether to proceed based on the obtained results remains a challenging task. There are only a few methods available in the literature to assess the quality of predictions in the RUL context. In \cite{KAMARIOTIS2024109723}, a novel metric is introduced to evaluate data-driven prognostic algorithms based on their impact on downstream predictive maintenance decisions. The metric is defined in association with a decision setting and a corresponding predictive maintenance policy, with the evaluation of the policy enabling the estimation of the proposed metric. This metric serves as an objective function to optimize heuristic predictive maintenance policies and algorithm hyperparameters. Pater and Mitici \cite{Pater96} proposed novel metrics to evaluate the quality of probabilistic RUL prognostics. The authors estimated the distribution of the RUL of turbofan engines using a Convolutional Neural Network with Monte Carlo dropout. \textcolor{black}{To assess the accuracy and sharpness of the probabilistic prognostics obtained, the authors employed the Continuous Ranked Probability Score (CRPS) and weighted CRPS}. The reliability of the obtained probabilistic prognostics was evaluated using the $\alpha$-Coverage and the Reliability Score. Li et al. \cite{7245050} proposed systemic prediction evaluation parameters to assess the prediction ability of a method for neural network RUL prediction models. Lan et al. \cite{Lan_2023}, utilizing the Sc-LSTM model, employed the relation to conduct a segmental predictive analysis and experimental validation of the overall prediction.

The lack of methods to evaluate predictions is the motivation behind our focus on this topic \textcolor{black}{and our main goal is to verify and assess the prediction quality based on the assumed model}. To illustrate the effectiveness of the proposed methodology, we consider a commonly used degradation model considered in the literature \cite{zulawinski2023framework}, specifically a three-regime degradation model corresponding to the three stages of the machine. These regimes include an almost constant value of the HI data  with no trend and a small linear growth of the variance (healthy stage), a linear trend with a significant linear growth of the scale -- variance equivalent (warning stage), and a nonlinear trend with a nonlinear growth of the scale of the HI values (critical stage). To build a model of long term HI data the procedure proposed by Żuławinski et al. \cite{zulawinski2023framework} is used. It is fully data-driven and take into account mentioned three-stage model. 




Usually, degradation modeling relies on data from the critical stage (third regime). However, such data often displays nonlinear growth of the scale, potential non-Gaussian components (such as impulsive disturbances in the data), and typically a limited number of samples for model building. In such situations, making accurate predictions becomes challenging. Moreover, determining objectively and precisely whether a prediction is deemed good or not becomes even more challenging.


In this article, we introduce a general procedure designed for evaluating predictions based on the assumed model. \textcolor{black}{In our approach we use the statistical-based approach where the uncertainty of the data is modeled by the random part of the adopted model. This approach somehow is related to the fuzzy logic methodology commonly applied to improve the robustness coming from the  uncertainty  \cite{s22114232}.} Although we present the procedure in the specific context of HI data forecasting, its universality allows application to any adopted model, any metric for prediction evaluation, and within the context of any problem or data set. We illustrate this procedure using complex data with a non-homogeneous structure, where both trend and scale change over time. Furthermore, we demonstrate the applicability of this procedure to data corresponding to two regimes (warning and critical stages), which are pivotal in the analysis of HI and RUL data prediction. In our approach, we propose adaptations of several different metrics (mean squared error (MSE), mean absolute percentage error (MAPE), space quantiles-inclusion factor (SQIF), Kupiec's proportion of failures test statistic (Kupiec's - POF), and Kupiec's time until first failure test statistic (Kupiec's - TUFF)). These metrics are known in data modeling or financial analysis \cite{Sikora20191202, RAUF2022111903, electronics11071125, Kamat2021, Moslem2923, Amiri2021, Astuti2021}. \textcolor{black}{According to our knowledge, they were not used in the considered context, i.e. in the problem of the prediction evaluation of long-term data.}
\textcolor{black}{We want to highlight that proposed methodology is not just an application of known metrics but the general procedure is introduced in which one may utilize different measures.}

Our procedure incorporates a decision-making element, providing the user with a binary response (0 or 1, corresponding to good or bad prediction, respectively). In addition, visual evaluation is possible and indirect measures can be employed to evaluate the performance of a given prediction. The decision-making process relies on statistical methods, considering the number of simulations, the patterns of the HI trajectories, and the comparison with the trajectory from actual data. The comparison procedure utilizes the mentioned metrics, and the decision itself is determined by relating the metric to accepted threshold values.

The novelty of this paper lies in the development of a universal procedure for prediction quality evaluation, applicable beyond the field of condition monitoring. In the proposed procedure, any numerical metric and any type of data behavior can be utilized. We specifically present the application of this procedure to complex HI data with a non-homogeneous structure and the efficiency of the procedure is tested in two regimes—linearly changing trend and scale, as well as exponentially changing trend and scale, both crucial in degradation modeling and RUL prediction. Furthermore, we apply our framework to two real data cases with distinct characteristics. \textcolor{black}{As mentioned, even though our procedure is illustrated in the context of machine HI data prognosis, it can be easily adapted to other areas where prognosis precision is crucial and where data with a non-homogeneous structure are analyzed. Potential applications may include forecasting financial data \cite{he2021seasonal}, especially electricity prices \cite{weron2014electricity}, mineral commodity prices \cite{cortez2018alternative}, or stock prices \cite{yu2020stock}; furthermore, prognoses mortality due to disease \cite{otunuga2022stochastic} or predictions of traffic accidents \cite{cai2023different}.}





This paper is organized as follows: In Section \ref{sec_Methodology}, we introduce the methodology and the proposed procedure for evaluating predictions using the metrics specified for this purpose. Section \ref{sec_simul_gauss} provides a comprehensive description of the model used for prediction, along with an assessment of the results obtained for the simulated data from that model. The assessment results for FEMTO {(Franche-Comt\'e Electronics Mechanics Thermal Science and Optics)} data are presented in Section \ref{sec_FEMTO}, and the prediction evaluation for IMS {(Intelligent Maintenance Systems)} data is discussed in Section \ref{sec_IMS}. Section \ref{sec_Discussion_Conclusions} covers the interpretation of the findings and significance of the results with the concluding remarks.

\section{Methodology}\label{sec_Methodology}
In this section, first (Section \ref{main_s}), we provide a detailed description of the proposed methodology to assess prediction quality. The methodology is presented in a general form, considering the availability of a metric $M$ for prediction quality assessment. Various metrics are considered in the literature for use in such contexts. In our subsequent analysis, we utilize five of them, and in Section \ref{metrics}, we present their definitions and provide additional details on the relevant parts of the procedure when a specific metric is applied. It is important to note that the proposed methodology is universal and allows for the use of any other metric in a similar manner.
\subsection{Description of the proposed procedure}\label{main_s}
\begin{figure}[H]
        \centering
        \includegraphics[width=0.96\textwidth]{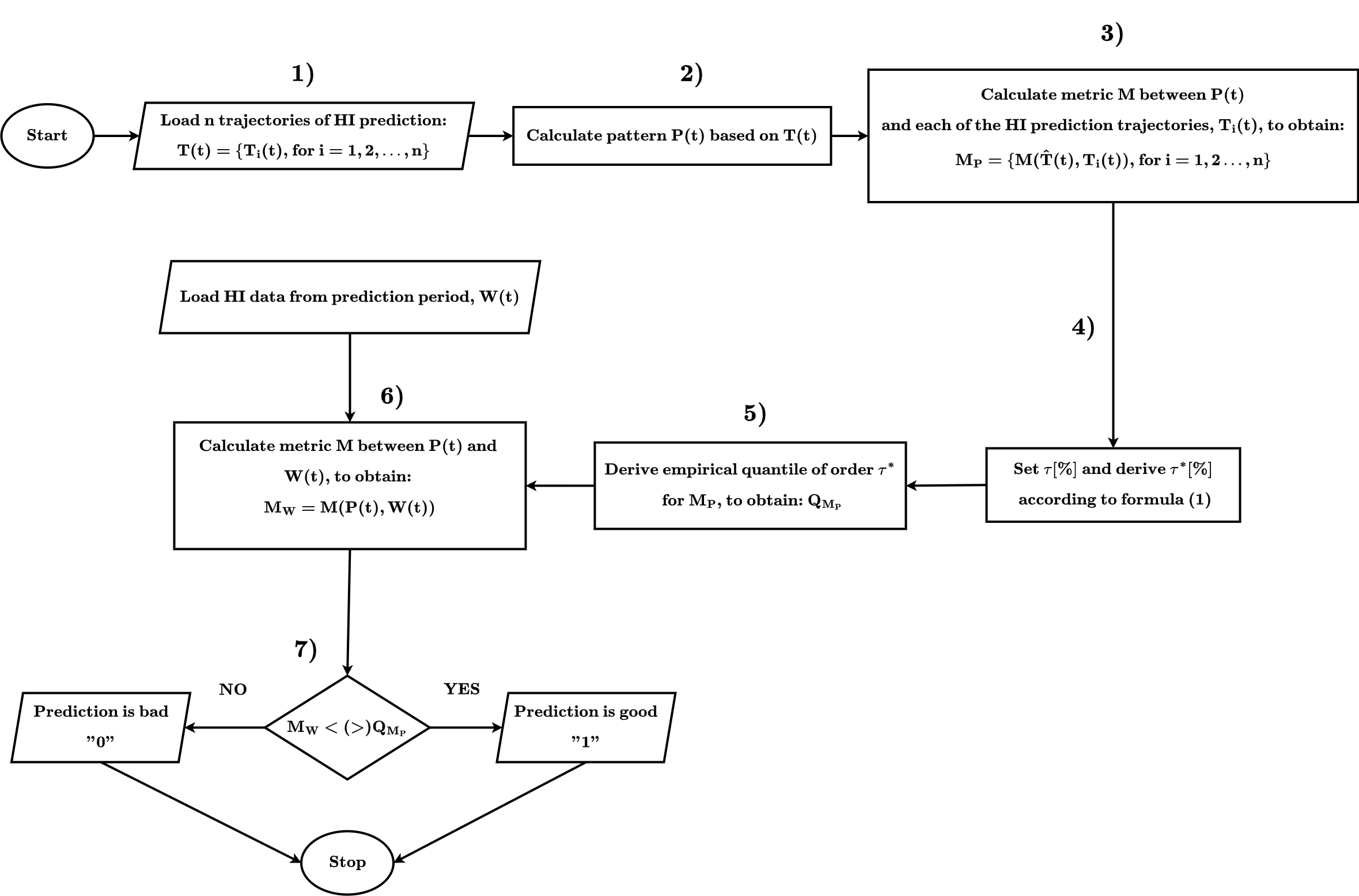}
        \caption{{Flowchart of the proposed procedure for prediction quality assessment. *According to point 7. we check whether $M_W<Q_{M_P}$ if a lower value of a metric $M$ means better prediction or we check whether $M_W>Q_{M_P}$ if a higher value of a metric $M$ means better prediction}}
    \label{fig1}
\end{figure}

The flowchart of the proposed procedure is presented in Fig. \ref{fig1}. Here, we assume that there are $n$ available trajectories of HI prediction for given period. In practice, the trajectories are obtained on the basis of the assumed model fitted to the data from the training period. The estimation of the assumed model is a crucial step in the analysis; however, here we omit it and refer the readers to the bibliography positions where the estimation of the model to HI data is discussed, see e.g. \cite{zulawinski2023framework}. The proposed procedure consists of the following steps (please note that the numbers of the consecutive steps correspond to the numbers on the flowchart):
 \begin{enumerate}
    \item Load $n$ trajectories of HI prediction. We denote them by {$T(t)=\{T_i(t),~~i=1,2,\ldots,n,~t\in V\}$, where V} is a prediction period.
    \item Based on $n$ available trajectories, derive the pattern $P(t)$, where \{$t\in V$.\} \textcolor{black}{Depending on the used metric for prediction quality assessment, the pattern may have very different form}. In Section \ref{metrics} we present the definition of five metrics used and describe the corresponding patterns.  
    \item Based on metric $M$ calculate the distance between the derived pattern $P(t)$ and each of the HI prediction trajectories $T(t)$. In consequence, we obtain a set of $n$ numbers $M_P=\{M_{P_1},M_{P_2},\ldots,M_{P_n}\}$. The details related to such values calculation for each of the metric used we describe in Section \ref{metrics}.
    \item Set a threshold $\tau$, which corresponds to the percentage value of the prediction assessment, above which the prediction is considered as acceptable. Based on $\tau$, derive $\tau^*$ in the following way:
    \begin{equation}
        \tau^*=
        \begin{cases}
            \tau[\%] &\text{if higher value of metric $M$ means better
            prediction}\\
            100-\tau[\%] &\text{otherwise.}
        \end{cases}
    \end{equation}
    \item Derive an empirical quantile of order $\tau^*$ based on $M_P$ set and denote it by $Q_{M_P}$.
    \item Load the true HI data from the prediction period on the basis of which we will assess the prediction quality. We denote these data as $W(t)$, $t\in {V}$. Then calculate the metric $M$ between the pattern, $P(t)$ (obtained in point 2) and $W(t)$ data. This value we denote as $M_W$. 
    \item Compare $M_W$ and $Q_{M_P}$. If a lower value of the metric $M$ means better prediction, then we verify whether $M_W<Q_{M_P}$. Otherwise, we verify whether $M_W>Q_{M_P}$.  Based on that, we can assess whether the prediction is good (the outcome of the procedure is "1") or bad (the outcome of the procedure is "0").
\end{enumerate}

\subsection{Metrics for prediction quality assessment}\label{metrics}

In this subsection, we present five metrics used for assessing prediction quality. In addition, we show how to calculate the pattern $P(t)$ (see step 2 of the procedure mentioned above). Finally, for exemplary trajectories of the assumed model, we show the patterns for the considered metrics. For consistency, in this part, we use the same notation as in the description of the procedure presented in Section \ref{main_s}.  
\begin{itemize}
        \item \textbf{Mean squared error (MSE)} for trajectory $T_i(t)$ and pattern $P(t)$ is calculated as follows:
\begin{eqnarray}
    M_{P_i}=\dfrac{1}{m}\sum\limits_{j=1}^{m} (P(t_j)-T_i(t_j))^2,
\end{eqnarray}
where $m$ is a length of $P(t)$ vector, i.e. number of data points in the prediction period {V}. Let us note that the lower value of MSE means better prediction. 
\item \textbf{Mean absolute percentage error (MAPE)}  for trajectory $T_i(t)$ and pattern $P(t)$ given by 
\begin{eqnarray}
M_{P_i}=\dfrac{1}{m}\sum\limits_{j=1}^{m} \dfrac{|P(t_j)-T_i(t_j)|}{|P(t_j)|},
\end{eqnarray} where, similar as in the previous case, $m$ is a length of $P(t)$. {Similar to MSE}, the lower MAPE value means better prediction.
\item \textbf{Space quantiles-inclusion factor (SQIF)} is proposed in  \cite{Sikora20191202} and in our case it is constructed as follows. First, based on available trajectories $T_i(t)$, $i=1,2,\ldots,n$ we calculate the empirical quantile lines on levels $q+(100-q)/2$ and $q+(100-q)/2$ for $q=0,10,\ldots,100[\%]$. By empirical quantile lines, we mean empirical quantiles for each $t\in {V}$. Next, for each $i=1,2,\ldots,n $ we calculate the fraction of $T_i(t)$ values that lies between the estimated quantile lines. This fraction we denote as $\phi_i(q)$. Next, we calculate the average square distance between $\phi_i(q)$ and the identity line (e.g. function $f(q)=q$). Thus, the metric used for prediction assessment is defined as follows:
\begin{equation}\label{factor}
  M_{P_i} = \frac{1}{\#Q} \sum_{q \in Q} (\phi_i(q)-q)^2 ,
\end{equation}
where $Q$ is a set containing all possible quantile orders (in our case $Q=\{0,10,20,\ldots,100\}$), while $\#Q$ is the number of elements of this set. The lower value of $M_{P_i}$ means better prediction. More details about the interpretation and construction of the procedure described above can be found in \cite{Sikora20191202}.
\item \textbf{Kupiec's POF statistic}, described in \cite{Kupiec1995} in our case is calculated based on the increments of the HI trajectories. We denote them by $S_i(t)$, $i=1,2,\ldots,n$. The corresponding metric is defined as follows:
\begin{equation}\label{POF_test}
  M_{P_i} = 
  \begin{cases}
            -2\left[(N-x)\text{log}\left(\dfrac{N(1-p^*)}{N-x}\right)+x\text{log}\left(\dfrac{Np^*}{x}\right)\right]&x\in(0,N)\\
            -2N\text{log}(1-p^*) &x=0\\
            -2N\text{log}(p^*) &x=N,
        \end{cases}
\end{equation}
where $N=m-1$ is the length of the vector $S_i(t)$ while $m$ is the length of $T_i(t)$. Furthermore, $x$ is the number of observations in $S_i(t)$, which exceeded $q_p$, where $q_p(\cdot)$ is a quantile of order $p[\%]$ and $p^* = 1 - \dfrac{p [\%]}{100}$. In this paper, we assume $p={51}$. This order of quantile is chosen because of two aspects. First, the value of $q$ around 50\% gives a wide range of received percentage values of quantile exceedance. Second, we choose $q_{51}$ instead of $q_{50}$ to get the asymmetry in statistic test values \footnote{E.g. if we consider $N=200$ and $p^*=0.5$, thus the expected value of $x$ is 100, then for $x=101$ and for $x=99$, the statistic value is the same in opposite to case, when we consider $N=200$ and $p^*=0.51$, thus the expected value of $x$ is 102, then for $x=103$ and for $x=101$, the statistic values are different.}. 
The lower value of $M_{P_i}$ means better prediction.
\item \textbf{Kupiec's TUFF statistic} defined in  \cite{Kupiec1995}, similar as in the previous case, is calculated for the increments of HI data. The corresponding metric is defined as follows:
\begin{equation}\label{TUFF_test}
  M_{P_i} = 
  \begin{cases}
            -2\left[\text{log}(p^*)+(x-1)\text{log}(1-p^*)+x\text{log}(x)-(x-1)\text{log}(x-1)\right]&x\in(1,N)\\
            -2N\text{log}(p^*) &x=1\\
            -2N\text{log}(1-p^*) &x\in \emptyset,
        \end{cases}
\end{equation}
where $x$ 
is the observation number in $S_i(t)$ of the  
first exceedance $q_p$, where $q_p(\cdot)$ -- quantile of order $p[\%]$ and $p^* = 1 - \dfrac{p [\%]}{100}$. In this paper, for Kupiec's TUFF statistic, we propose to consider order of $q_p$ in such a way, 
to get equal probability of no exceedances and probability of first exceedance in the first observation. Thus, we derive $p$ by solving the equation
\begin{equation}\label{p_TUFF}
    (1-p)^N=p,
\end{equation}
where $N=m-1$ is the length of the vector $S_i(t)$ while $m$ is the length of $T_i(t)$. The lower value of $M_{P_i}$ means better prediction.
\end{itemize}
Recall in the procedure presented in Section \ref{main_s}, in step 2 we need to derive the pattern based on $n$ available trajectories. As mentioned, the pattern may be different depending on the metric used for prediction quality assessment. Here, we consider three groups of patterns:
    \begin{itemize}
        \item For \textbf{MSE} and \textbf{MAPE} the pattern is an average predicted trajectory of HI, namely
        \begin{eqnarray}
        \textcolor{black}{P(t_j)=}\bar{T}(t_j)=\dfrac{1}{n}\sum\limits_{i=1}^{n} T_i(t_j).
        \end{eqnarray}
        In Fig. \ref{block_diagram_avg} we present the exemplary trajectories of HI prediction for model described in Section \ref{sec_simul_gauss} (second regime) together with the pattern corresponding to MSE and MAPE
        \item For \textbf{SQIF} the pattern is constructed as quantile lines calculated based on $T_i(t)$ trajectories at levels\\ $q \in \{0,5,10,\ldots,100\}\}[\%]$ \textcolor{black}{($P(t)=\{P_q(t)\}$)}. In  Fig. \ref{block_diagram_quantiles} we present the exemplary trajectories of HI prediction for model described in Section \ref{sec_simul_gauss} (second regime) 
        and the pattern corresponding to SQIF.
        \textcolor{black}{Denote $T_k^{sort}(t), k=1,2,\ldots,n$ -- sorted ascending values of \{$T_i(t),i=1,2,\ldots,n$\}. Then }
         \begin{equation}\label{quantile_lines}
          \textcolor{black}{P_q(t)=
          \begin{cases}
            T_{k^*}^{sort}(t) + \dfrac{\dfrac{q}{100}-\dfrac{k^*-0.5}{n}}{\dfrac{k^*+1-0.5}{n} - \dfrac{k^*-0.5}{n}}\cdot \left(T_{k^*+1}^{sort}(t)-T_{k^*}^{sort}(t)\right) & \text{if } q\in(0,100)[\%]\\
            \text{min}\{T_k^{sort}(t)\} & \text{if } q=0[\%]\\
            \text{max}\{T_k^{sort}(t)\} & \text{if } q=100[\%],
            \end{cases}}
            \end{equation}
        \textcolor{black}{where $k^*=\text{max}\left\{k,\dfrac{k-0.5}{n}<\dfrac{q}{100}\right\}$.}
        \item For the \textbf{Kupiec's POF} and \textbf{Kupiec's TUFF} statistics the pattern is constructed as quantile line of increments $S_i(t)$ of the trajectories $T_i(t)$, $i=1,2,\ldots,n$ \textcolor{black}{analogically as in Eq. (\ref{quantile_lines}) for a single value of $q$}.
        In Fig. \ref{block_diagram_kupiec_quantile} we present the exemplary trajectories of HI prediction for second regime, according to model described in Section \ref{sec_simul_gauss} 
        and the  exemplary pattern corresponding to such metrics (quantile line of the level $p=80[\%]$).
Due to discrete values in considered variants of Kupiec's statistics, in this paper we propose a modification of the classical metrics. First, we calculate the percentage of values of $M_{P}$ higher than $M_W$ (denote $\gamma_1$) and the percentage of values of $M_{P}$ higher than or equal to $M_W$ (denote $\gamma_2$), where $M_P$ are the metric values derived for $S_i(t)$ and $M_W$ is the metric value between the pattern and the HI data from the prediction period. Then, we calculate $\gamma^*$ -- the average of these two values $\gamma_1$ and $\gamma_2$. Finally, if the quantile $M_P$ of order $\gamma^*$ is less than $Q_{M_P}$, we recognize the prediction as good ("1"); otherwise, the prediction is recognized as bad ("0"). 
    \end{itemize}

\section{Analysis of simulated data from the degradation model}\label{sec_simul_gauss}

\subsection{Model description}\label{model_description}

In this section, we demonstrate the efficiency of the proposed methodology for simulated data from the adopted degradation model. This model was proposed in \cite{zulawinski2023framework} on the basis of various real HI data sets. The complete description and motivation for using this model are given in the mentioned bibliography position. Thus, in this paper, we present only its brief description. 

In the theoretical degradation model, we assume three regimes, corresponding to three stages of the machine, namely: healthy, warning, and critical. Each of the regimes consists of two components: deterministic and random. In general, the model can be formulated as follows
\begin{equation}\label{sdef}
    S(t) = D(t) + R(t), 
\end{equation}
where $D(t)$ represents the deterministic component and $R(t)$ is a random component of the model. We assume that the random component is the noise with a time-varying scale. Thus, it is given by
\begin{equation}
    R(t) = R2(t)\cdot SC(t),
\end{equation}
where $t\in(1,m)$, and $m$ is the length of trajectory.
In the above,  $R2(t)$ is a sequence of independent identically distributed (i.i.d.) random variables while $SC(t)$ is the time-varying scale component defined as follows
\begin{eqnarray}\label{SC_t}
    SC(t) = 
    \begin{cases}
        a_1 t + b_1 & t \in (1,t^*_1),\\
        a_2 t + b_2 & t \in (t^*_1+1,t^*_2),\\
        a_3e^{b_3 t} & t \in (t^*_2+1,m),
    \end{cases}
\end{eqnarray}
where $t^*_1$ is the first-second regime changing point, $t^*_2$ is the second-third regime changing point and constants $a_1$, $b_1$, $a_2$, $b_2$, $a_3$, $b_3$ are derived in such a way: $SC(1) = \sigma_1$, $SC(t^*_1)=\sigma_2$, $SC(t^*_2)=\sigma_3$, $SC(m)=\sigma_4$ (to calculate $a_2$, $b_2$, $a_3$, $b_3$, we include $t^*_1$ and $t^*_2$ in the second and third regimes, respectively) and $t^*_1$ -- first-second regime changing point, $t^*_2$ -- second-third regime changing point, $m$ -- length of trajectory.
Let us mention that in \cite{zulawinski2023framework} the degradation model was introduced in a general form in such a way as to be adopted for Gaussian and non-Gaussian data. In this paper, we take its simplified version and assume that the random component $R2(t)$ is a sequence of i.i.d. random variables from the Gaussian distribution, $R2(t) \sim \mathcal{N}(0,1)$.  The deterministic component $D(t)$ in Eq. (\ref{sdef}) is as follows:
\begin{eqnarray}
    D(t) = 
    \begin{cases}
        c_1 & t \in (1,t^*_1),\\
        a_2 t + c_2 & t \in (t^*_1 + 1,t^*_2),\\
        a_3e^{b_3 t} + c_3 & t \in (t^*_2 + 1,m),
    \end{cases}
\end{eqnarray}
where $t^*_1$, $t^*_2$, $m$, $a_2$, $a_3$, $b_3$ are the same as in Eq. (\ref{SC_t}), $c_1=\text{const}$ and $c_2$, $c_3$ are derived to obtain the continuity in $D(t)$ for $t=t^*_1$ and $t=t^*_2$, respectively.

\begin{figure}[H]
    \centering
        \includegraphics[width = 10cm]{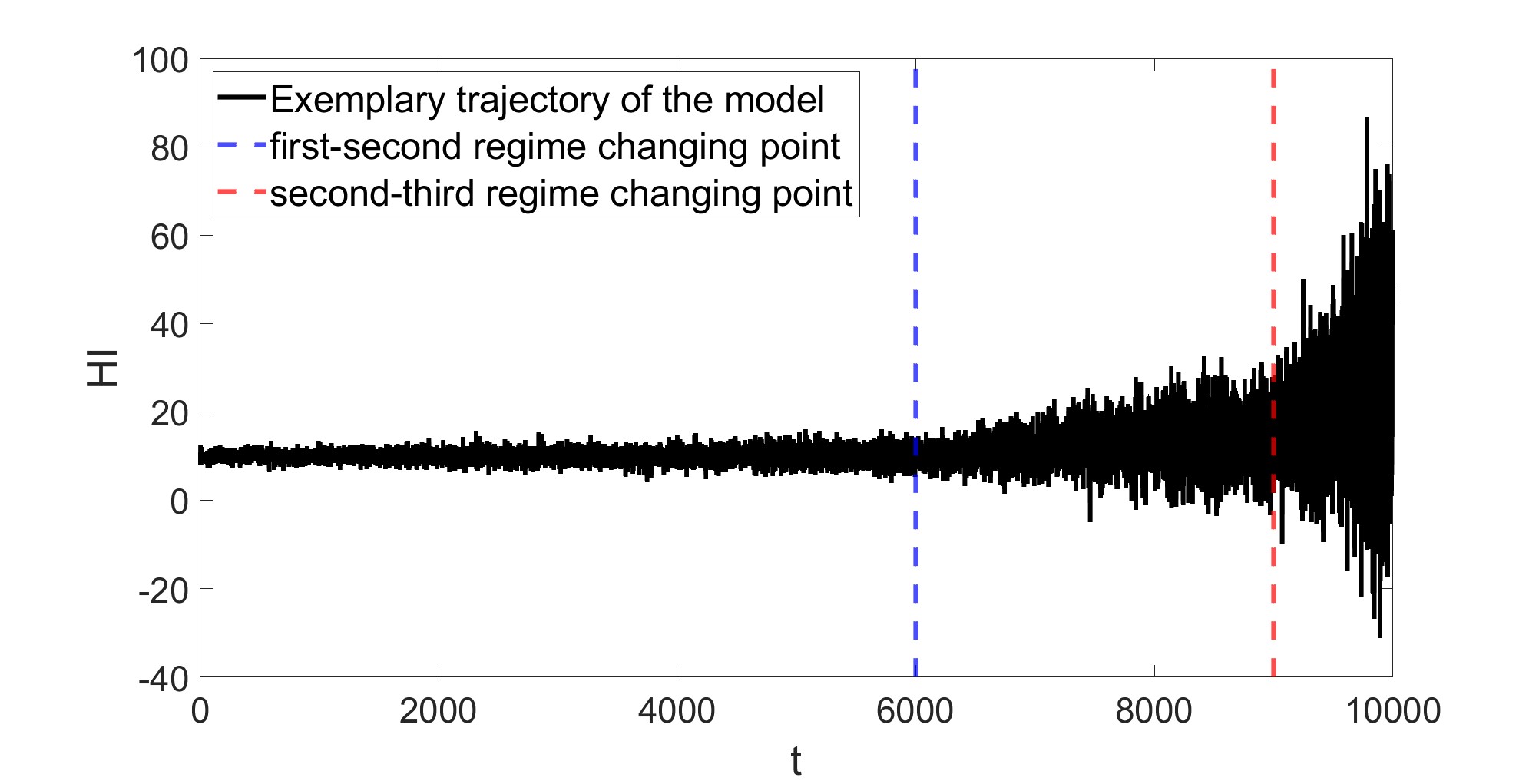}
        \caption{Exemplary simulated data from the adopted degradation model.}
    \label{Simulator_all_regimes}
\end{figure}

In Fig. \ref{Simulator_all_regimes} we present the exemplary trajectory of the model defined in Eq. (\ref{sdef}) with marked regime changing points. To simulate the data, we assumed the following parameters: $t^*_1=6000$, $t^*_2=9000$, $m=10000$, $\sigma_1=1$, $\sigma_2=2$, $\sigma_3=7$, $\sigma_4=25$, $a_1\approx 1.6669\cdot 10^{-4}$, $b_1\approx 0.9998$, $c_1=10$, $a_2\approx 1.6667\cdot 10^{-3}$, $b_2=-8$, $c_2=0$, $a_3\approx 7.4049\cdot 10^{-5}$, $b_3\approx 1.273\cdot 10^{-3}$, $c_3=8$. The first regime includes observations from 1 to $t^*_1=6000$. We can see here that the scale changes only slightly over time, and the trend is almost constant.  The second regime ranges from $t^*_1 + 1 = 6001$ to $t^*_2=9000$. Here, the scale is increasing significantly. Furthermore, the values on average also increase with time. In the third regime, from $t^*_2+1=9001$ to $m=10000$, we observe a huge difference between the average values at the beginning compared to the end of the regime. The variance of the values is also the largest compared to both previous regimes. Since the behavior of the simulated data in the three mentioned regimes is significantly different, the verification of the proposed methodology is performed separately for the second and third regimes.

\subsection{Analysis of the data from the second regime}\label{sec_simul_second}

\begin{figure}[H]
    \centering
        \includegraphics[width = 10cm]{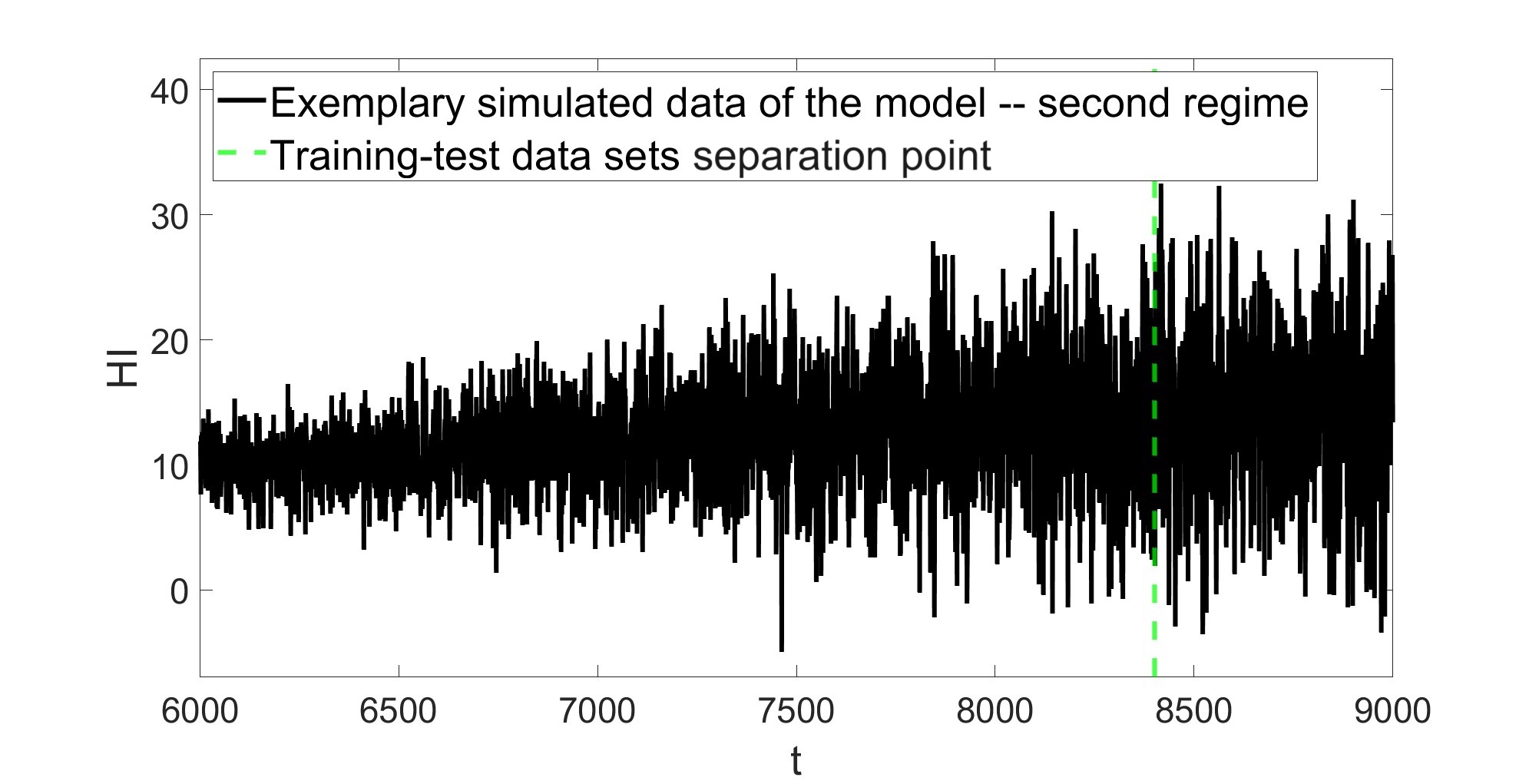}
        \caption{Simulated data from the second regime and the separation point between training and test data sets}
    \label{simul_second_regime}
\end{figure}

In practice, the first regime is rarely considered for degradation modeling; thus, in this paper, only the second and third regimes are analyzed. The simulated data corresponding to the second regime are presented in Fig. \ref{simul_second_regime}. To verify the methodology, we split the data into two sets, namely training and test data sets. The separation point between these sets is marked by a green dashed line. The training data set contains $80\%$ data, while the test data set contains $20\%$ data points. In real-life applications, where the parameters of the adopted model are not known, before the evaluation of the prediction quality, the model parameters should first be estimated and then a number of trajectories corresponding to the prediction period can be simulated. In case of simulated data analysis, when the parameters of the model are known, the estimation is not needed, and for the simulation, we take theoretical values of the parameters. However, in Section \ref{sec_FEMTO} and Section \ref{sec_IMS}, when we analyze the real data, estimation is also performed. 

In Fig. \ref{simul_second_regime} we can see a significant positive linear trend and a linearly increasing scale of the data. It is clearly visible that, at the end of the second regime, on average the values are larger than for the first observations. Later in this paper, for the second regime, we analyze only the test data set that includes observation numbers in the range [8401,9000].

\begin{figure}[H]
    \centering
    \begin{minipage}[t]{0.48\textwidth}
        \centering
        \includegraphics[width=1\textwidth]{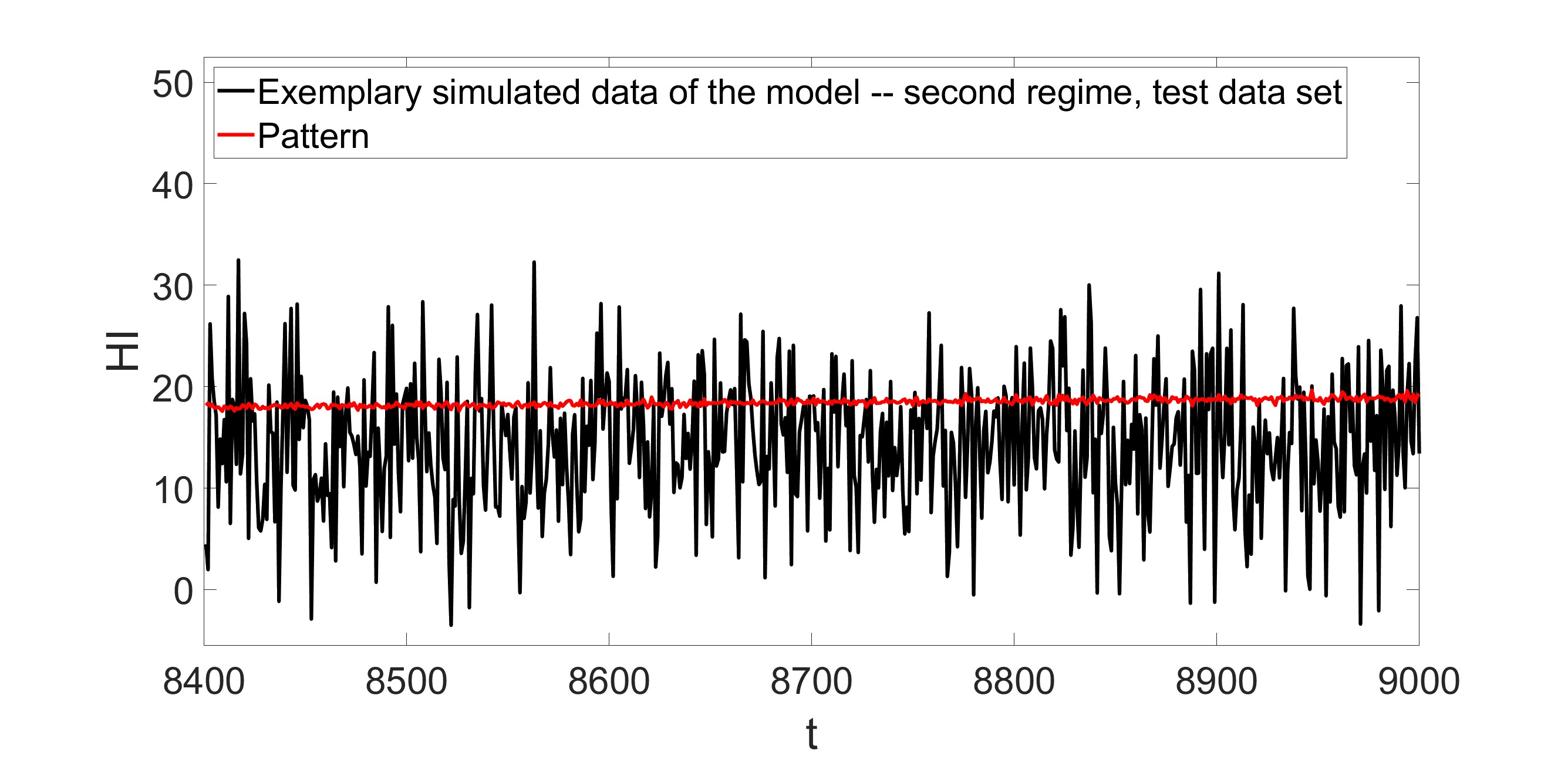}
        \caption{Comparison of the test data from the model corresponding to the second regime with the pattern for MSE and MAPE}
    \label{simul_second_regime_average}
    \end{minipage}
    \hfill
    \begin{minipage}[t]{0.48\textwidth}
        \centering
        \includegraphics[width=1\textwidth]{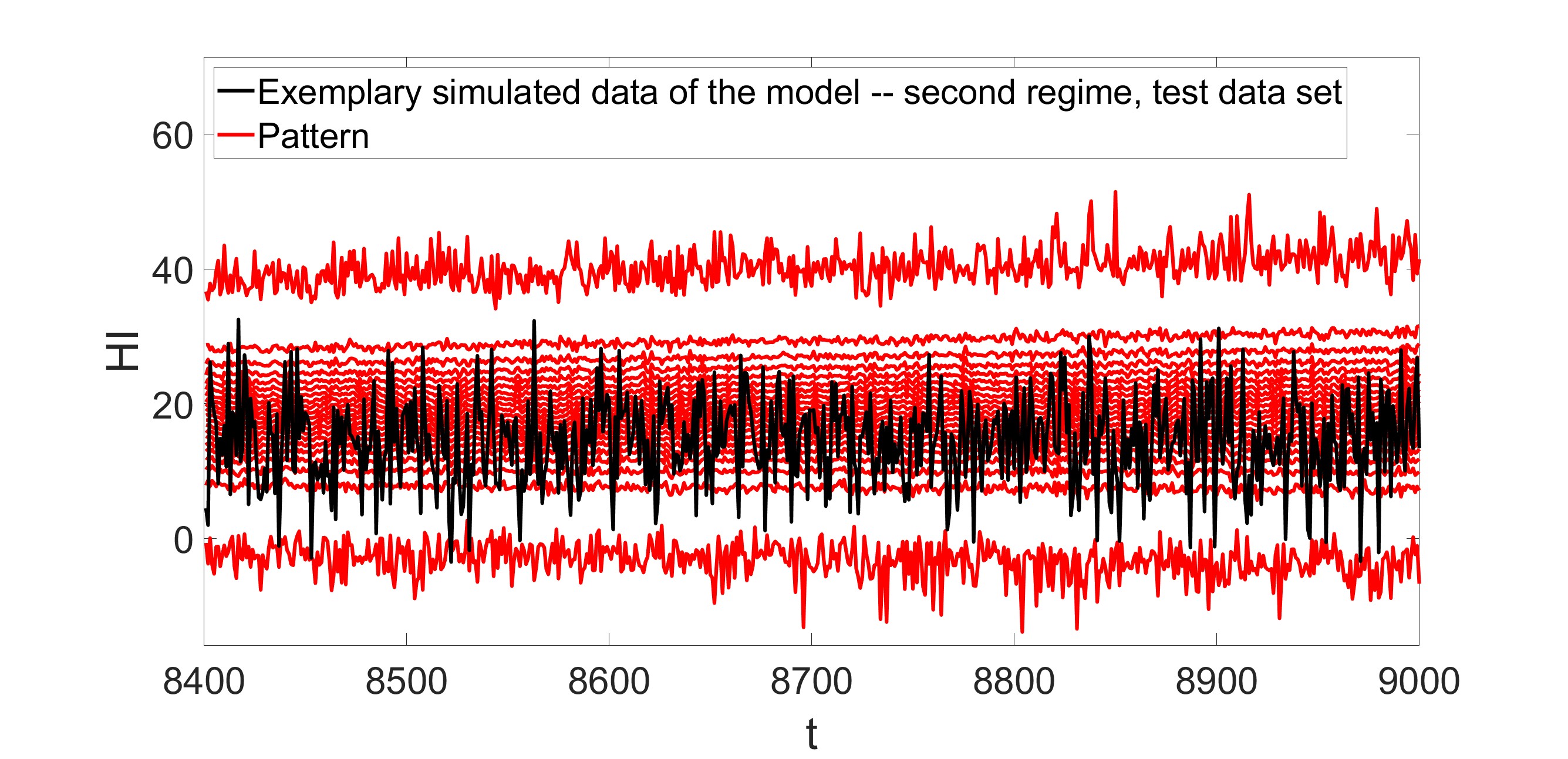}
        \caption{Comparison of the test data from the model corresponding to the second regime with the pattern for SQIF}
    \label{simul_second_regime_space}
    \end{minipage}
\end{figure}

In Fig. \ref{simul_second_regime_average} we can see the predicted HI time series and its comparison with the analyzed test data set of the second regime. The average trajectory is considered here as a pattern for the MSE and MAPE metrics. In ideal conditions, this pattern should be in the middle of the trajectory. However, we can see here that the pattern is slightly higher than most of the observations in the trajectory. This is caused by a random component and is the reason why we generated 1000 test trajectories to minimize the effect of random component.

Similarly, as in the pattern for the MSE and MAPE metrics, the pattern for SQIF \textcolor{black}{(multiple quantile lines calculated as in Eq. (\ref{quantile_lines}))} appears to be shifted relative to the exemplary trajectory of the model (see Fig. \ref{simul_second_regime_space}). There are only a few values of the trajectory that exceed the 95\% quantile lines, while there are many values of the trajectory that exceed the 5\% quantile line (values are smaller than the 5\% quantile line). There is no observation in the trajectory that achieves the 100\% quantile line, but there are still points that achieve the 0\% quantile line.

\begin{figure}[H]
    \centering
    \begin{minipage}[t]{0.48\textwidth}
        \centering
        \includegraphics[width=1\textwidth]{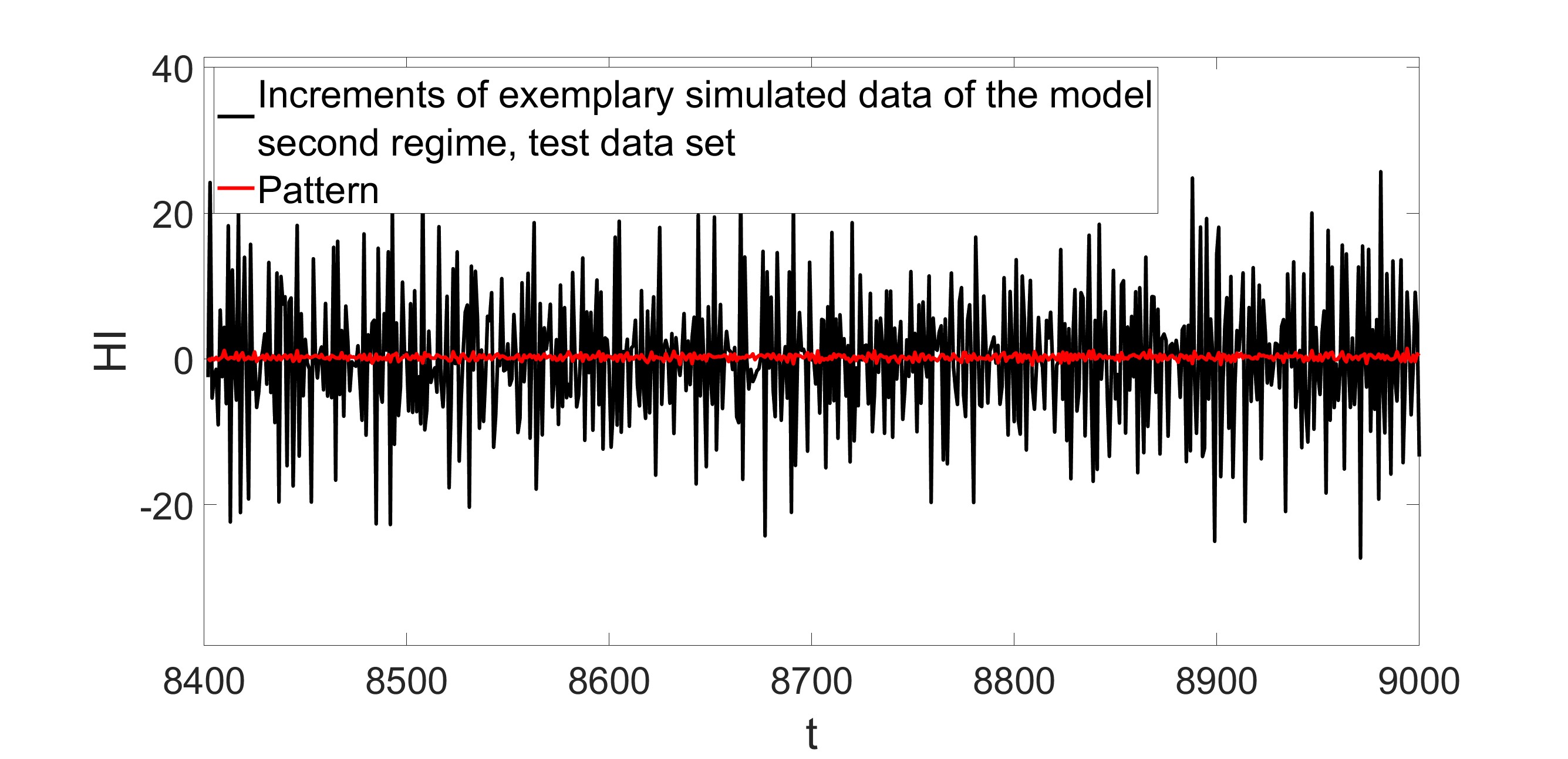}
        \caption{Comparison of the increments of the test data from the model corresponding to the second regime with the pattern for Kupiec's POF}
    \label{simul_second_regime_kupiec_pof}
    \end{minipage}
    \hfill
    \begin{minipage}[t]{0.48\textwidth}
        \centering
        \includegraphics[width=1\textwidth]{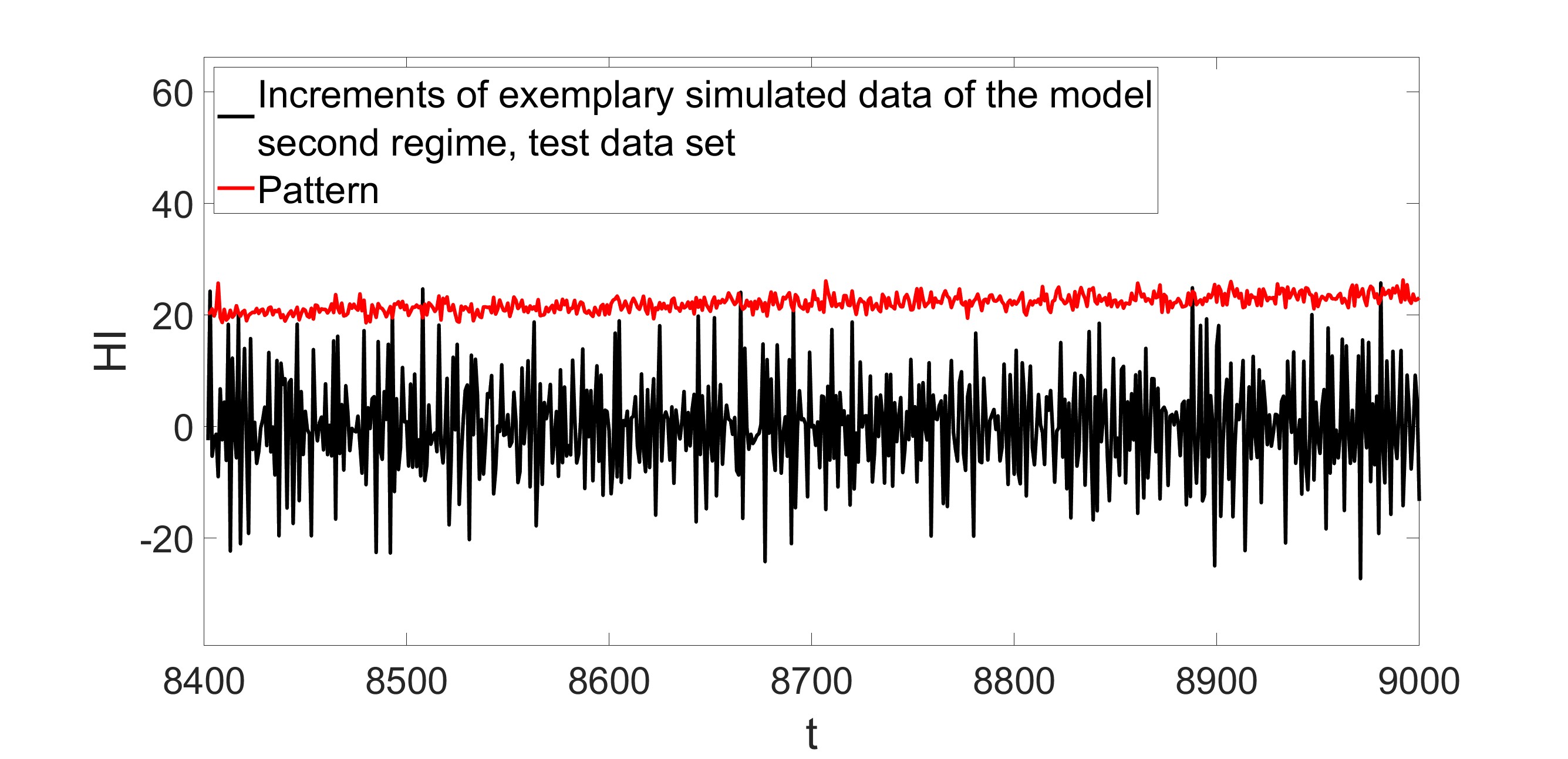}
        \caption{Comparison of the increments of the test data from the model corresponding to the second regime with the pattern for Kupiec's TUFF}
    \label{simul_second_regime_kupiec_tuff}
    \end{minipage}
\end{figure}

If we analyze the increments calculated for the test data of the second regime of the model, the quantile line of order 51\% is almost in the middle of the exemplary trajectory values (see Fig. \ref{simul_second_regime_kupiec_pof}). It can show us that even if the behavior of the trajectory differs from the pattern, there is still the possibility to recognize appropriately the pattern for the increments of the trajectory. This is also one of the reasons why it is important to evaluate the prognosis with different metrics.

In Fig. \ref{simul_second_regime_kupiec_tuff}, we can see that the pattern for the second regime follows a linear trend. Still, only a few observations exceed the presented quantile line. It is important to note that one of the first increments of the exemplary trajectory of the model exceeds this quantile line. This means that for this particular trajectory, the prediction assessed via Kupiec's TUFF-based metric will be recognized as bad.

In Tab. \ref{table_simul_second}, results for 1000 independent trajectories are compared with the patterns derived from 1000 independent trajectories of the same model. Finally, we present here a percentage of good predictions through considered metrics for different $\tau^*$ levels. The majority of the received results differ from the true one by at most 3 percentage points. There are only a few cases where the error is larger; for example, for the MSE metric for $\tau^*=60\%$, the error is 4.9 percentage points. However, for the second regime for the SQIF metric, there is no case with a difference from the proper value being greater than 4 percentage points. This means that most test trajectories follow the behavior of training trajectories.

\begin{table}[H]
    \centering
    \scalebox{0.75}{
    \begin{tabular}{|c|c|c|c|c|c|}
    \hline
    & \multicolumn{5}{c|}{Metric}  \\  \hline
        $\tau^*$ & MSE & MAPE & SQIF & Kupiec's POF & Kupiec's TUFF \\ \hline
        10 & 87,7 & 88,5 & 89,5 & 87,3 & 88,7 \\ \hline
        20 & 77,7 & 76,9 & 77,6 & 77,3 & 78,0 \\ \hline
        30 & 67,0 & 64,8 & 66,2 & 66,2 & 68,1 \\ \hline
        40 & 57,9 & 57,3 & 57,4 & 58,3 & 58,0 \\ \hline
        50 & 47,6 & 47,9 & 50,3 & 45,1 & 48,4 \\ \hline
        60 & 35,1 & 35,8 & 38,2 & 35,7 & 39,9 \\ \hline
        70 & 27,5 & 27,3 & 30,1 & 25,3 & 30,8 \\ \hline
        80 & 19,5 & 18,6 & 21,8 & 19,4 & 19,6 \\ \hline
        90 & 11,2 & 9,3 & 11,7 & 10,0 & 12,0 \\ \hline
    \end{tabular}}
    \caption{Percentage of good predictions for $\tau^*\in\{10,20,\ldots,90\}\%$}
    \label{table_simul_second}
\end{table}

\subsection{Analysis of the data from the third regime}\label{sec_simul_third}

\begin{figure}[H]
    \centering
        \includegraphics[width = 10cm]{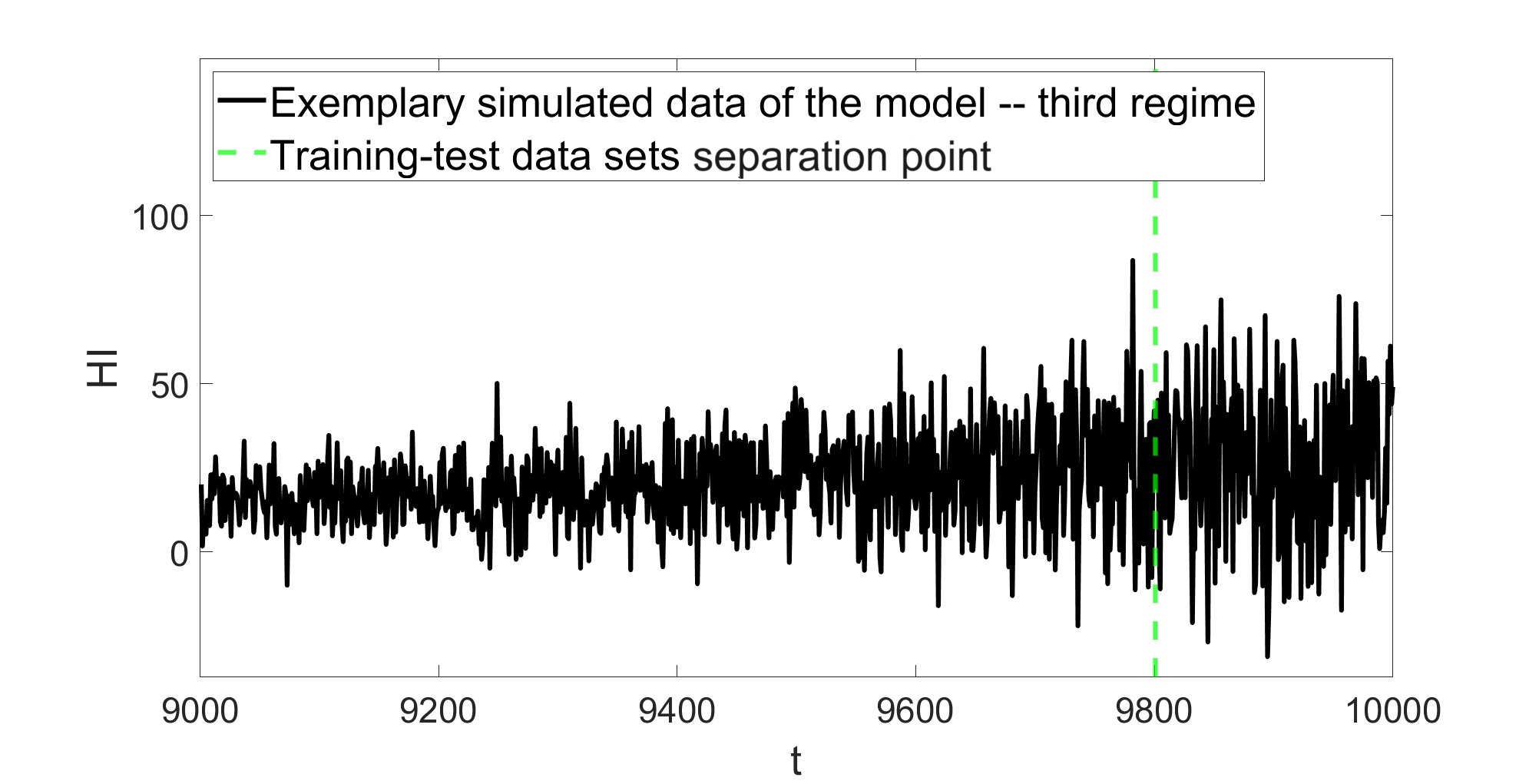}
        \caption{Simulated data from the third regime and the separation point between training and test data sets}
    \label{simul_third_regime}
\end{figure}

Lastly, we repeat the procedure for the third regime. It is characterized by an exponential trend and exponential growth of the scale. The same as for the second regime, we proceed as follows: we divide the third regime of the test trajectories $W_i(t)$ into training data -- first 80\% observations of the third regime (observations within the range [9001,9800]) and test data -- last 20\% observations of the third regime (observations within the range [9801,10000]), we compare the patterns calculated for considered metrics based on $T_i(t)$ with the test data of the test trajectories, $W_i(t)$, and finally we calculate the percentage of good predictions for 1000 test trajectories $W_i(t)$. In Fig. \ref{simul_third_regime} we can see the division of the third regime of the simulated data of the model, $W_1(t)$ into training and test parts. As we can see, there is a small exponential trend, but the most visible is the rapidly increasing scale of the random part.

\begin{figure}[H]
    \centering
    \begin{minipage}[t]{0.48\textwidth}
        \centering
        \includegraphics[width=1\textwidth]{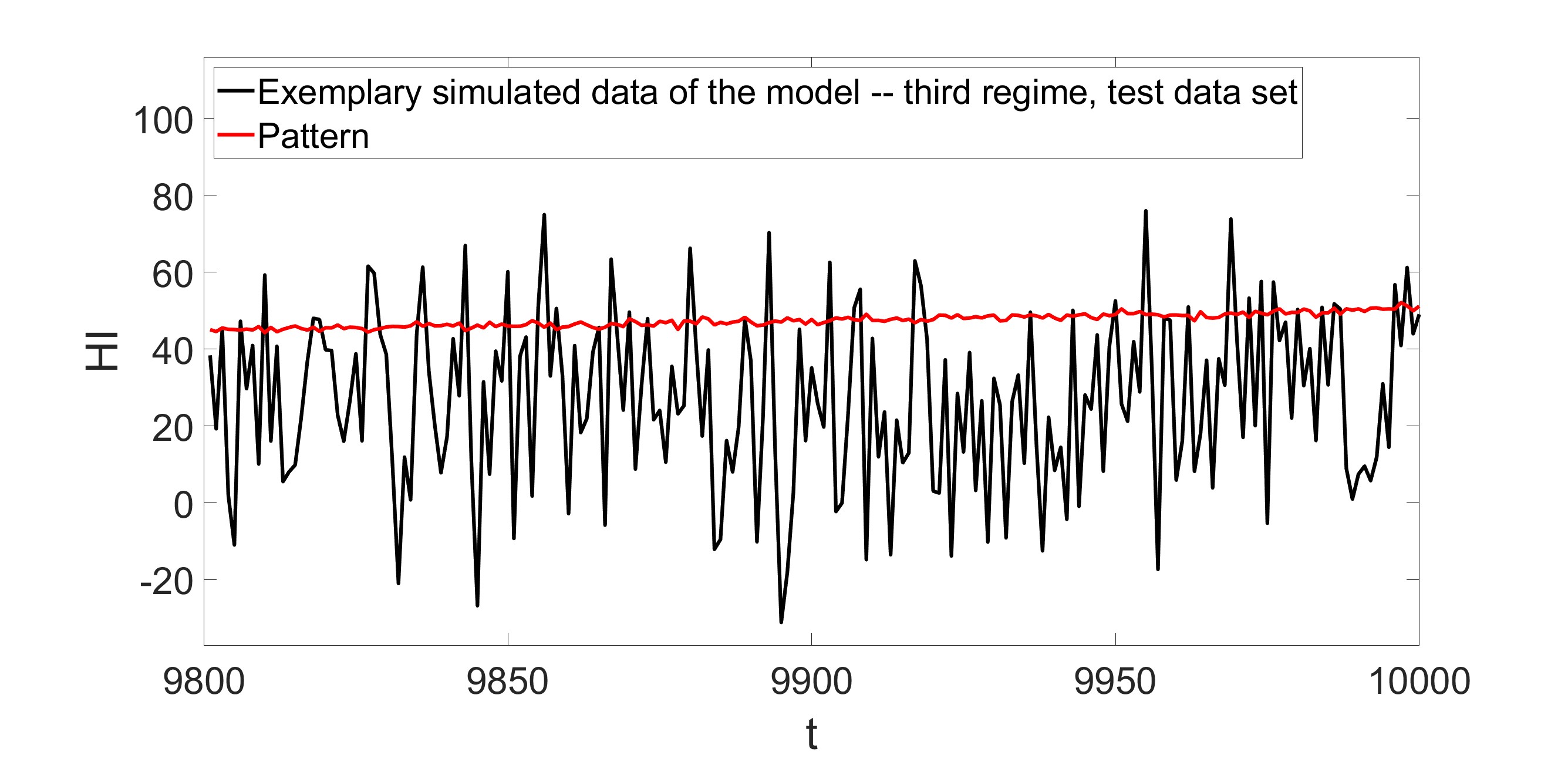}
        \caption{Comparison of the test data from the model corresponding to the third regime with the pattern for MSE and MAPE}
    \label{simul_third_regime_average}
    \end{minipage}
    \hfill
    \begin{minipage}[t]{0.48\textwidth}
        \centering
        \includegraphics[width=1\textwidth]{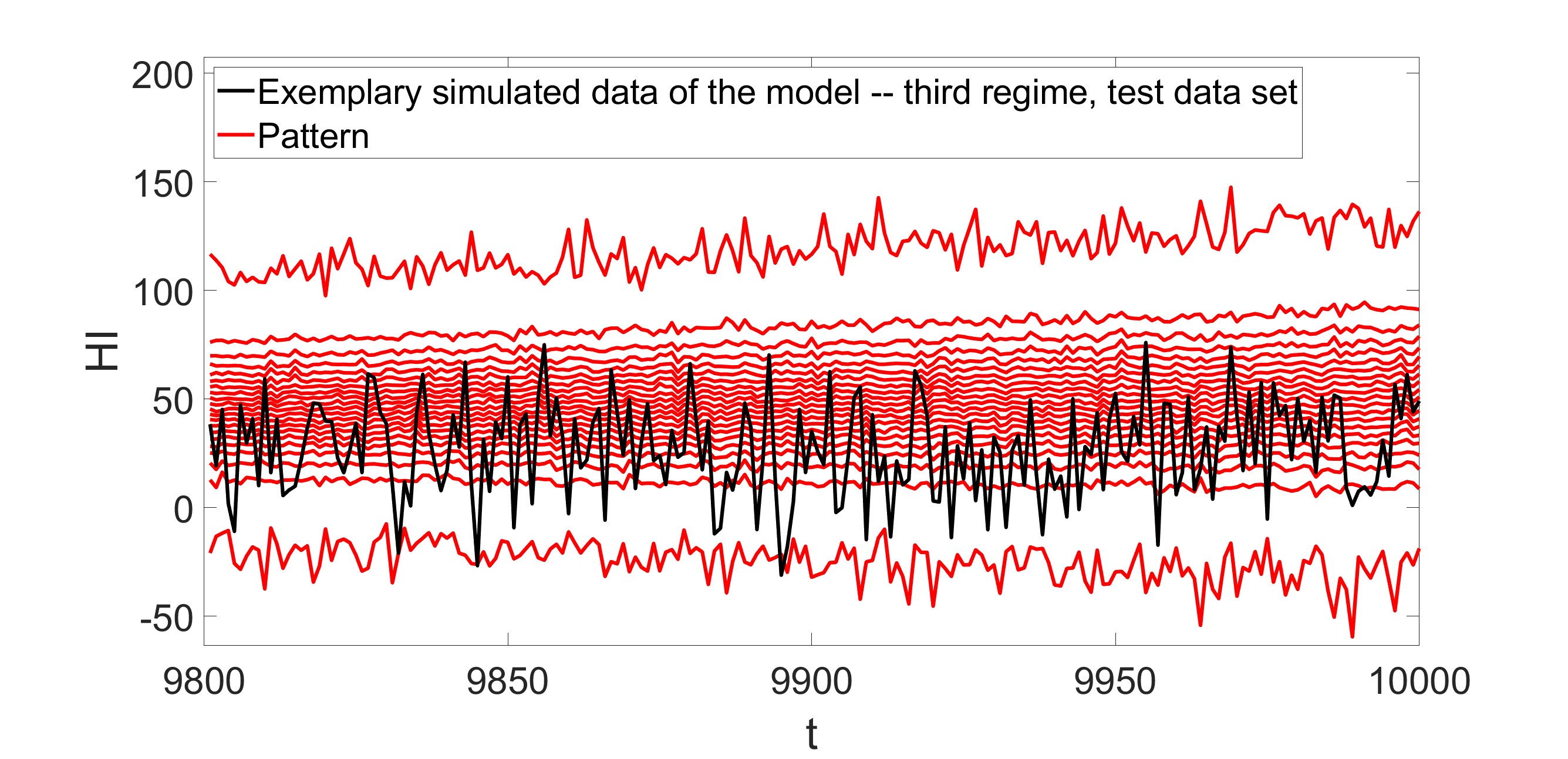}
        \caption{Comparison of the test data from the model corresponding to the third regime with the pattern for SQIF}
    \label{simul_third_regime_space}
    \end{minipage}
\end{figure}

In Fig. \ref{simul_third_regime_average} we can see that even though the pattern for MSE and MAPE metrics is the average of predicted trajectories, thus about half of the observations should be above the quantile line, there are significantly more observations smaller than the pattern than observations larger than the pattern. We can also see that the scale of the increments is high in comparison to the previous regimes.

The impact of the random component is also visible in Fig. \ref{simul_third_regime_space}. Here, we can see the asymmetry. It manifests itself in such a way that, for example, some of the observations exceed the quantile line of order 0\%, but there is no observation that exceeds even the quantile of order 95\%. It looks like the whole trajectory is shifted in relation to the pattern, which is in this case quantile lines of the orders in set $\{0,5,10,\ldots,100\}\%$. Once again, we have to mention the importance of analyzing a large number of independent trajectories to properly identify behavior by minimizing the effect of random component.

\begin{figure}[H]
    \centering
    \begin{minipage}[t]{0.48\textwidth}
        \centering
        \includegraphics[width=1\textwidth]{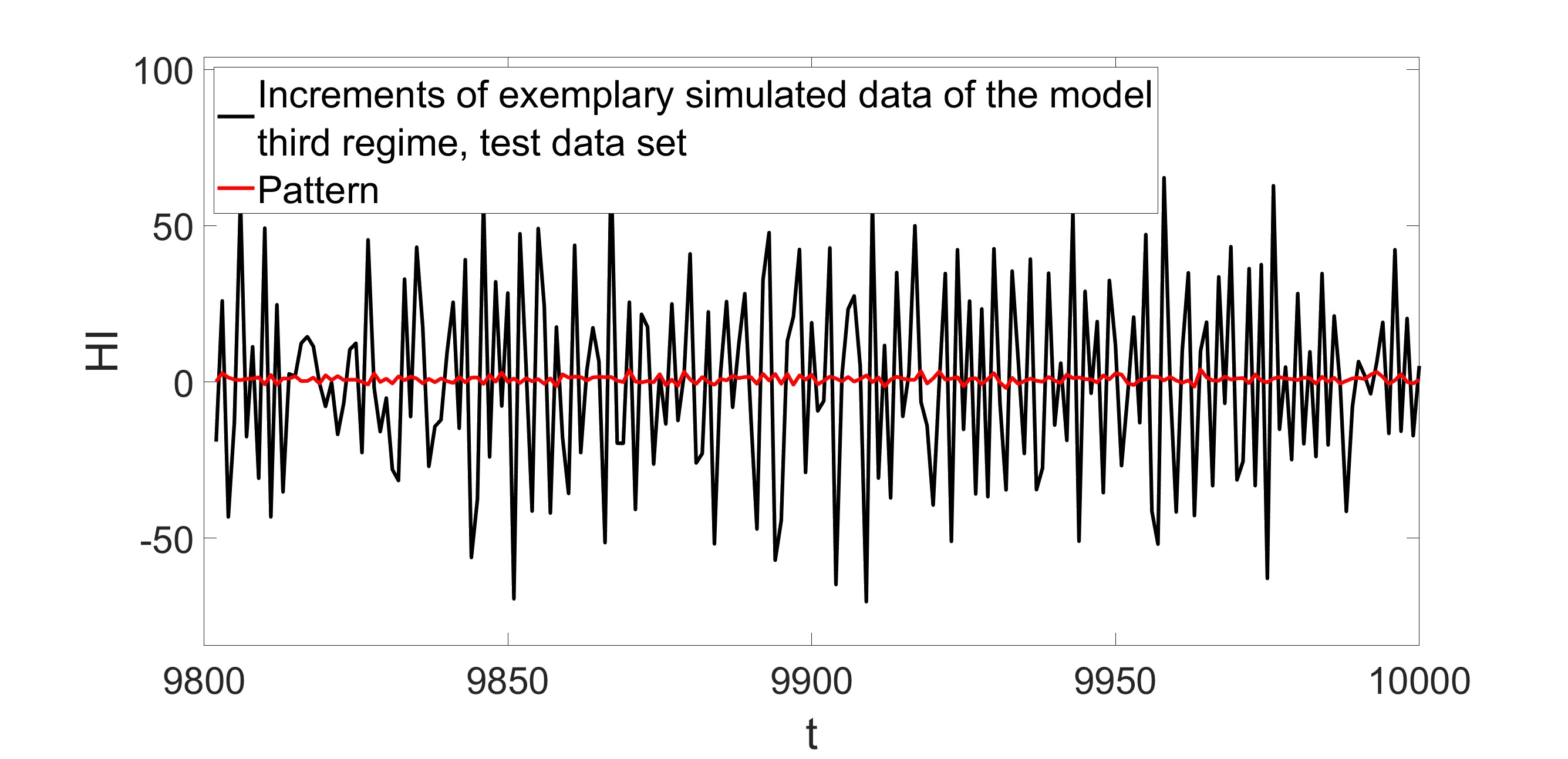}
        \caption{Comparison of the increments of the test data from the model corresponding to the third regime with the pattern for Kupiec's POF}
    \label{simul_third_regime_kupiec_pof}
    \end{minipage}
    \hfill
    \begin{minipage}[t]{0.48\textwidth}
        \centering
        \includegraphics[width=1\textwidth]{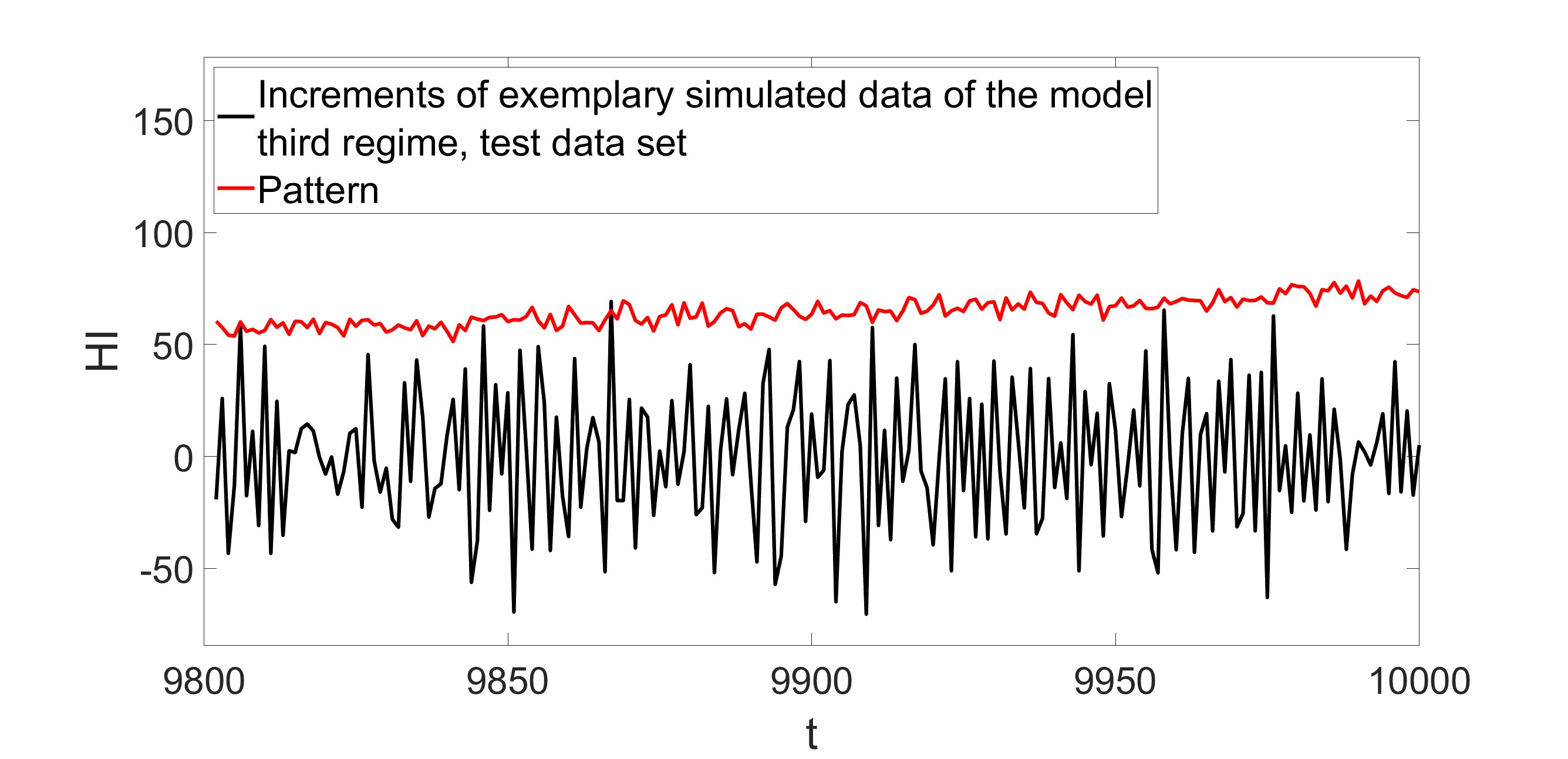}
        \caption{Comparison of the increments of the test data from the model corresponding to the third regime with the pattern for Kupiec's TUFF}
    \label{simul_third_regime_kupiec_tuff}
    \end{minipage}
\end{figure}

Fig. \ref{simul_third_regime_kupiec_pof} shows the increments derived for the test data of the third regime of the exemplary test trajectory, $W_1(t)$ and its comparison with the pattern of the Kupiec POF statistic -- quantile line of order 51\% for the increments derived for the predicted trajectories, $T_i(t)$. We can see that about 50\% of the observations are below the pattern. The scale is high, and, in the presented data, we cannot see the exponential trend or exponential scale growth of the increments. The values vary about 0.

In the second variant considered of the Kupiec TUFF statistic (see Fig. \ref{simul_third_regime_kupiec_tuff}), there are only two values that exceed the pattern. However, in this metric, the number of exceedances is not important as we here analyze the first exceedance time. For this particular example, the fourth increment exceeds the derived quantile line. Taking into account the total number of increments (here, 199) and knowing the method by which we derive the order of that quantile line, we can assume that in this case we are dealing with an extreme trajectory in some respects.

The final assessment for the third regime is presented in the Tab. \ref{table_simul_third}. The results are based on 1000 independent trajectories ($W_i(t)$) and patterns derived through 1000 independent trajectories ($T_i(t)$) of the same model. We can see that every considered metric gives approximately proper results. Although in the test data of the third regime we have the least number of observations (200), this is the most challenging case to properly model and predict the trend and also the time-changing scale. However, we can still reasonably decide whether the prediction is good or not. For all the considered $\tau^*$ values, the percentage values received are close to the expected values. We can also see an ideal result here, for the MSE metric for $\tau^*=90\%$ we get exactly 10\% good predictions, which is consistent with the theory.

\begin{table}[H]
    \centering
    \scalebox{0.75}{
    \begin{tabular}{|c|c|c|c|c|c|}
    \hline
    & \multicolumn{5}{c|}{Metric}  \\  \hline
        $\tau^*$ & MSE & MAPE & SQIF & Kupiec's POF & Kupiec's TUFF \\ \hline
        10 & 89,6 & 90,1 & 91,0 & 88,1 & 90,2 \\ \hline
        20 & 82,0 & 80,5 & 76,5 & 82,2 & 80,1 \\ \hline
        30 & 68,9 & 67,6 & 68,6 & 72,5 & 69,6 \\ \hline
        40 & 57,7 & 57,1 & 59,0 & 61,6 & 61,4 \\ \hline
        50 & 46,4 & 49,0 & 50,5 & 45,9 & 51,5 \\ \hline
        60 & 36,8 & 36,8 & 39,1 & 37,9 & 42,1 \\ \hline
        70 & 27,9 & 26,7 & 30,5 & 26,6 & 32,8 \\ \hline
        80 & 17,9 & 17,3 & 20,5 & 18,4 & 22,9 \\ \hline
        90 & 10,0 & 9,5 & 9,8 & 9,2 & 7,5 \\ \hline
    \end{tabular}}
    \caption{Percentage of good predictions for $\tau^*\in\{10,20,\ldots,90\}\%$}
    \label{table_simul_third}
\end{table}

\section{FEMTO data set}\label{sec_FEMTO}

\subsection{Data set description}

\begin{figure}[H]
    \centering
    \begin{minipage}[t]{0.48\textwidth}
        \centering
      \includegraphics[width=0.9\textwidth, height=3.8cm]{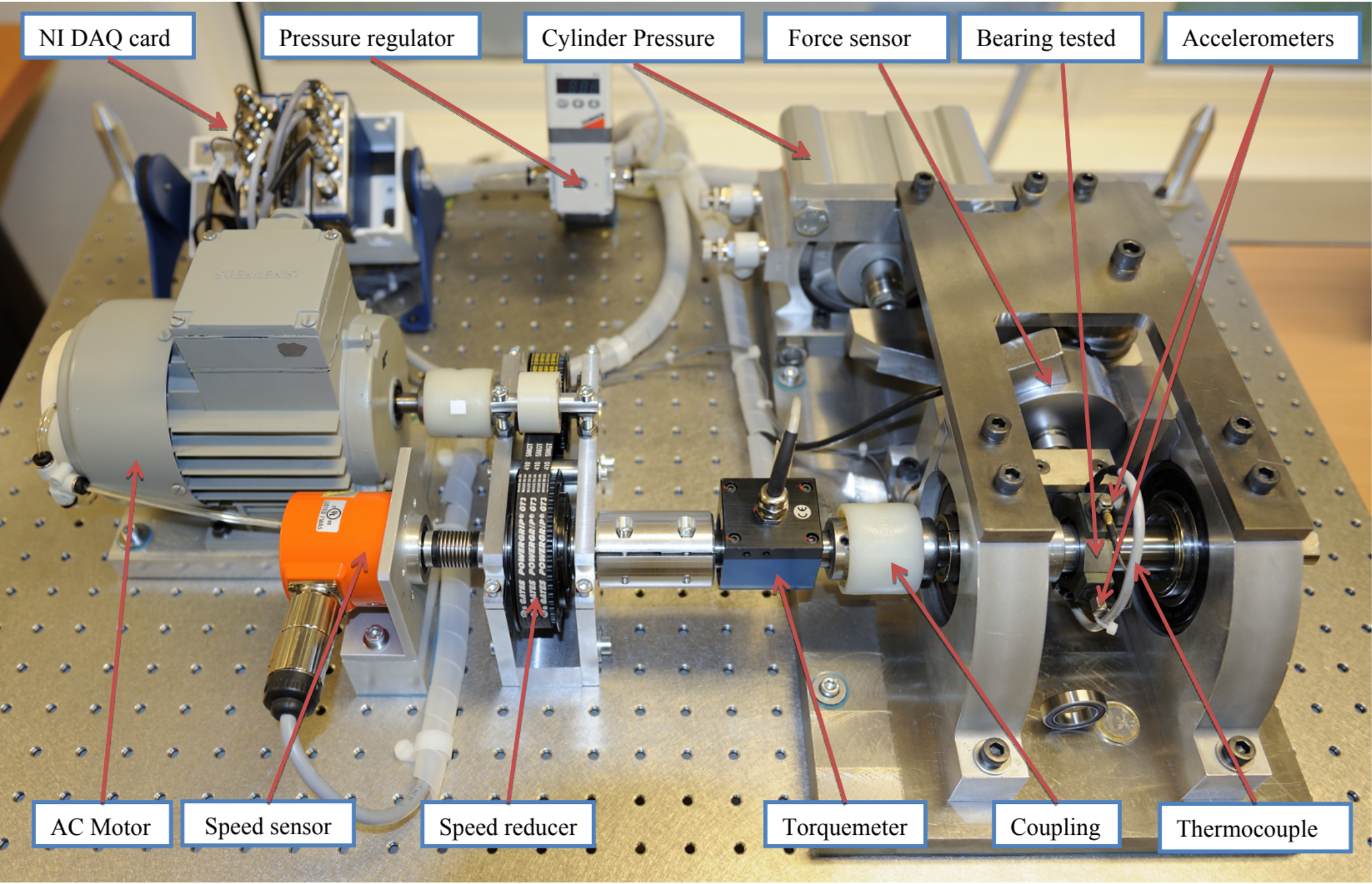}
      \caption{FEMTO test rig \cite{nectoux2012pronostia}}
      \label{fig:FEMTO_testrig}
    \end{minipage}
    \hfill
    \begin{minipage}[t]{0.48\textwidth}
        \centering
        \includegraphics[width=1\textwidth]{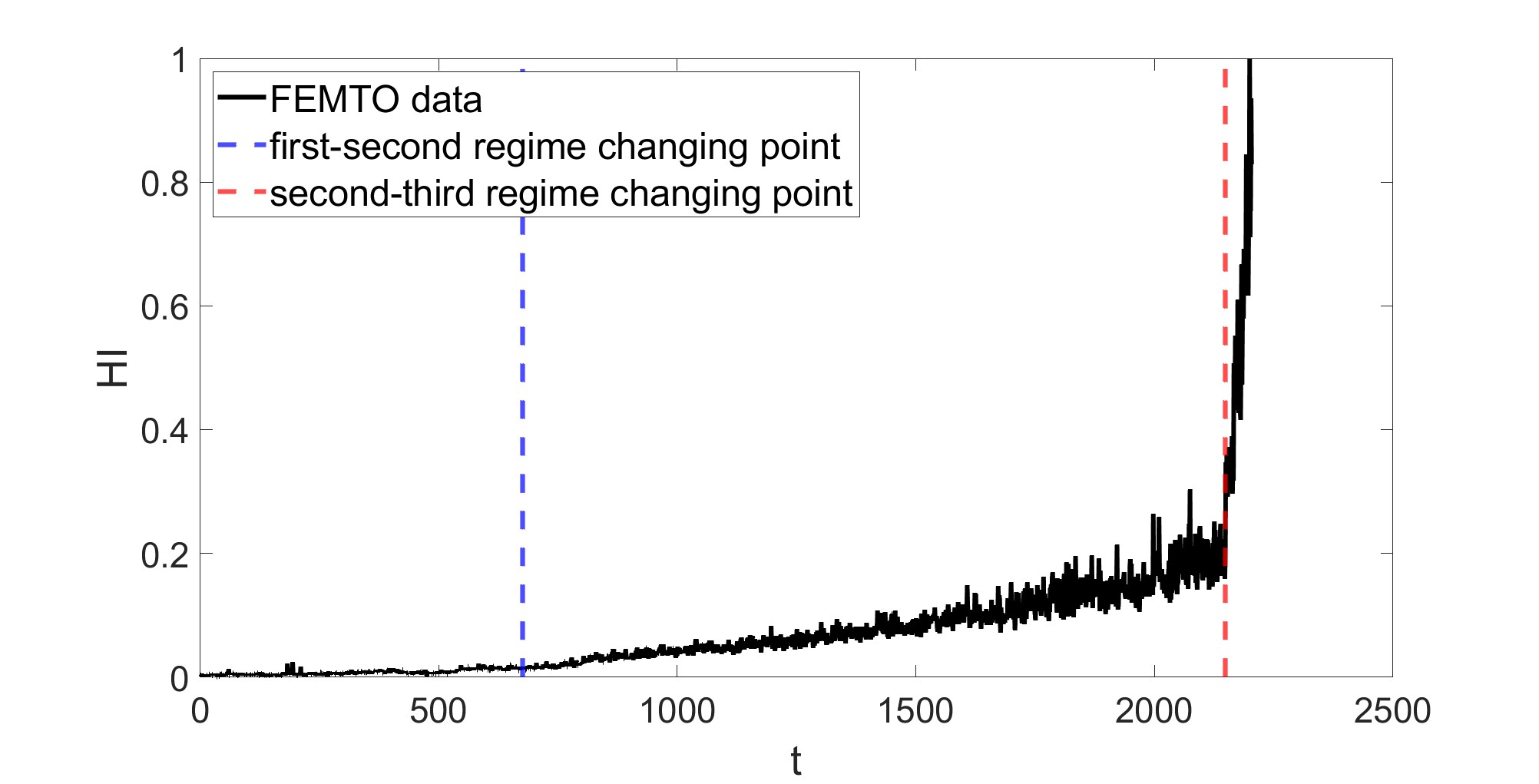}
        \caption{FEMTO HI data}
    \label{FEMTO_data}
    \end{minipage}
\end{figure}

The FEMTO data set was obtained by the Franche-Comté Electronics Mechanics Thermal Science and Optics–Sciences and Technologies Institute from the PRONOSTIA platform; see Fig. \ref{fig:FEMTO_testrig}. This data set comprises 17 instances of historical bearing degradation. The rotational speed of the shaft was maintained at a constant level during the tests. 

In numerous publications, this data set was analyzed for various purposes, such as constructing health indices (HI)~\cite{mosallam2014time,loutas2013remaining,javed2014enabling,singleton2014extended,zhang2016degradation,hong2014adaptive,lei2016model,nie2015estimation}, segmenting the degradation process~\cite{liu2016remaining,kimotho2013machinery,zurita2014distributed,guo2016multifeatures,jin2016anomaly}, and predicting the RUL ~\cite{li2013rolling,huang2017remaining,wang2016two,pan2014machine,wang2015reliability,xiao2017novel}.

In Fig. \ref{FEMTO_data}, we present the HI FEMTO data set. Here, we can distinguish three regimes. First regime, including observations from 1 to 676, second regime -- observations from 677 to 2149, and third regime -- observations from 2150 to 2204. Regime changing points are derived according to \textcolor{black}{methodology presented in } \cite{Maraj2023113495} \textcolor{black}{which is based on the Isolate-Detect (ID) algorithm. However, let us mention that there are known other approaches that have been used for such long-term data segmentation, see, e.g., an approach based on the Hidden Markov Model (HMM)  \cite{JANCZURA2023113399} or Kalman filer-based approach \cite{SHIRI2023110472}.}

In the first regime, there is no significant trend. Observations vary around the constant value. The scale of the noise is very small. In the second regime, we can see that the values increase over time and we can model them by a linear trend. If we compare the first values of the second regime with the last values of the second regime, it is clearly visible that the scale of noise for the first observations is much larger than for the last observations. The third regime for the FEMTO data is very short and there are only 55 observations. There is a significant positive trend and, therefore, the values grow rapidly. In this paper, we model the third regime using an exponential trend. Moreover, the scale of the noise is also significantly larger than for both the first and second regimes. The scale for the third regime is also modeled here by an exponential function.

\subsection{Analysis of the data from the second regime}

\begin{figure}[H]
    \centering
        \includegraphics[width = 10cm]{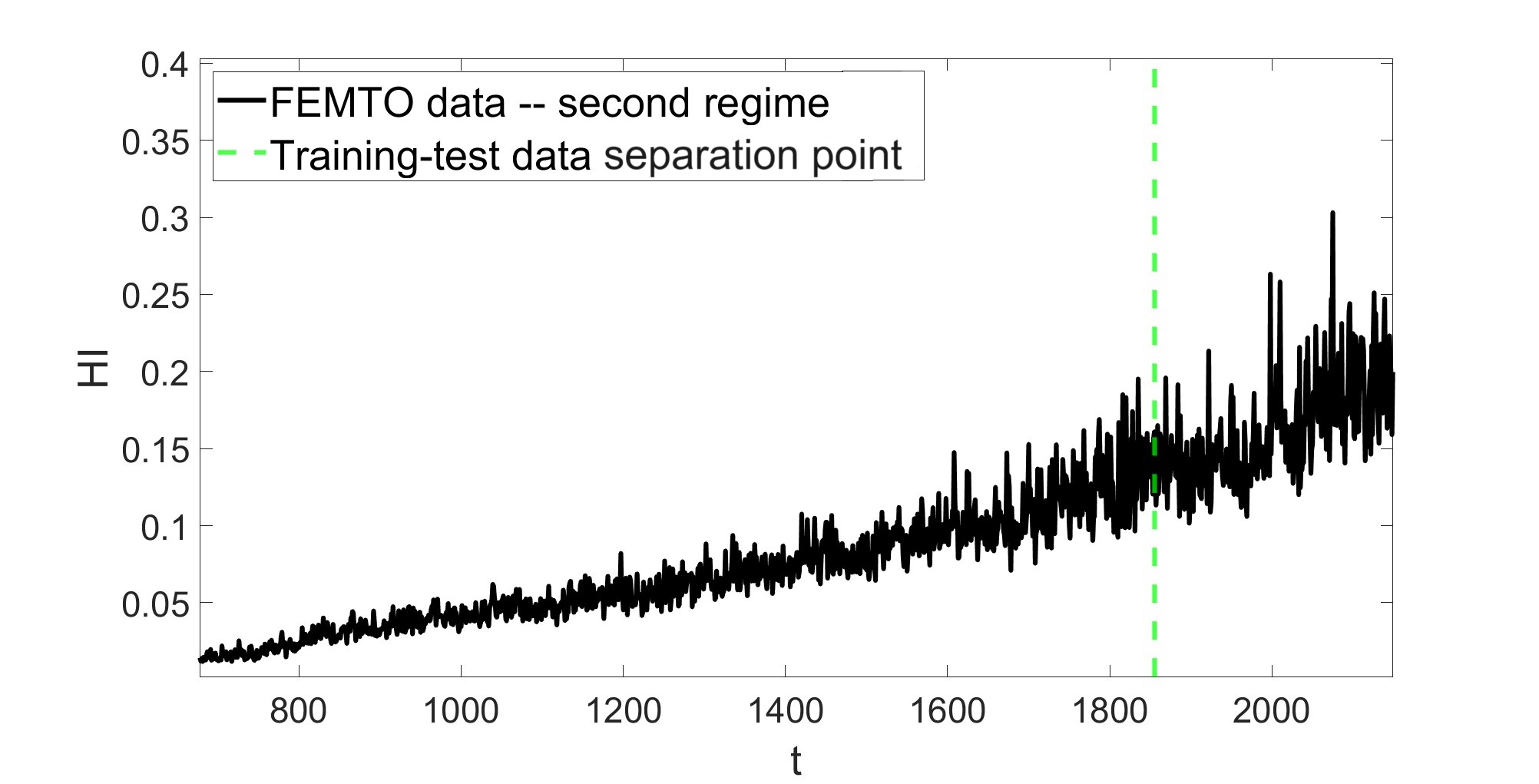}
        \caption{The second regime of the FEMTO data and the separation point between training and test data sets}
    \label{FEMTO_second_regime}
\end{figure}

In this section, we present the results for the second regime of the FEMTO data, where the linear trend and the linearly changing scale occur. We divide the second regime into training and test data sets in proportions 80\% to 20\%, respectively. The trajectory with training-test data separation point we can see in Fig. \ref{FEMTO_second_regime}. We fit a model based on the entire second regime and check the prognosis only for the test data set. We can see that there is a clearly visible linear trend. Furthermore, the scale of the noise also increases significantly over time. In the test data set, we can see some extreme observations. They can have an influence on the metrics values, especially for the MSE, SQIF, and Kupiec's TUFF metrics, which are very sensitive to outliers. Other metrics are more resistant to outliers, so this is one of the reasons to analyze different metrics to obtain a more objective prediction evaluation.


\begin{figure}[H]
    \centering
    \begin{minipage}[t]{0.48\textwidth}
        \centering
        \includegraphics[width=1\textwidth]{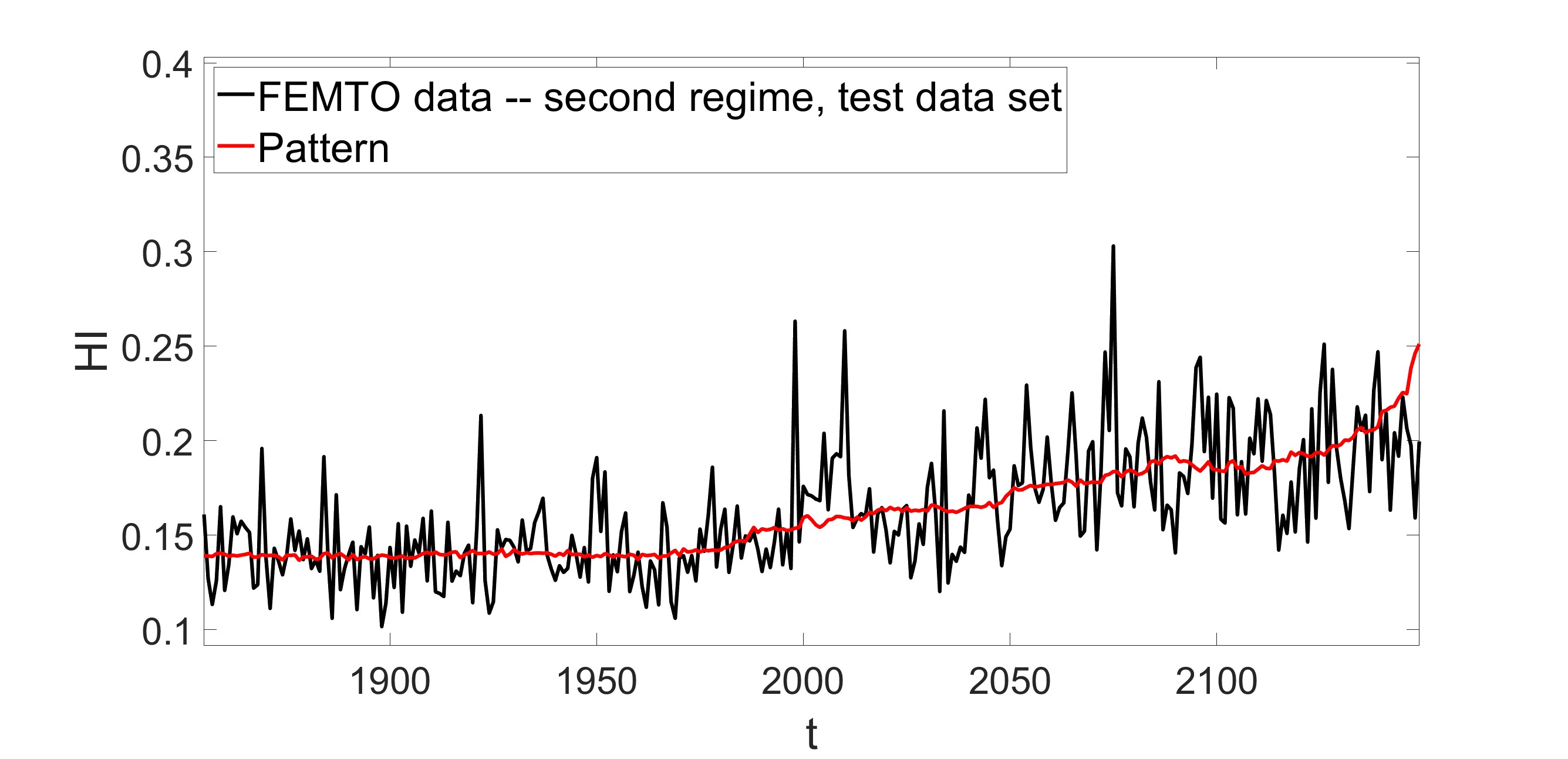}
        \caption{Comparison of the test data from the second regime of the FEMTO data with the pattern for MSE and MAPE}
    \label{FEMTO_second_regime_average}
    \end{minipage}
    \hfill
    \begin{minipage}[t]{0.48\textwidth}
        \centering
        \includegraphics[width=1\textwidth]{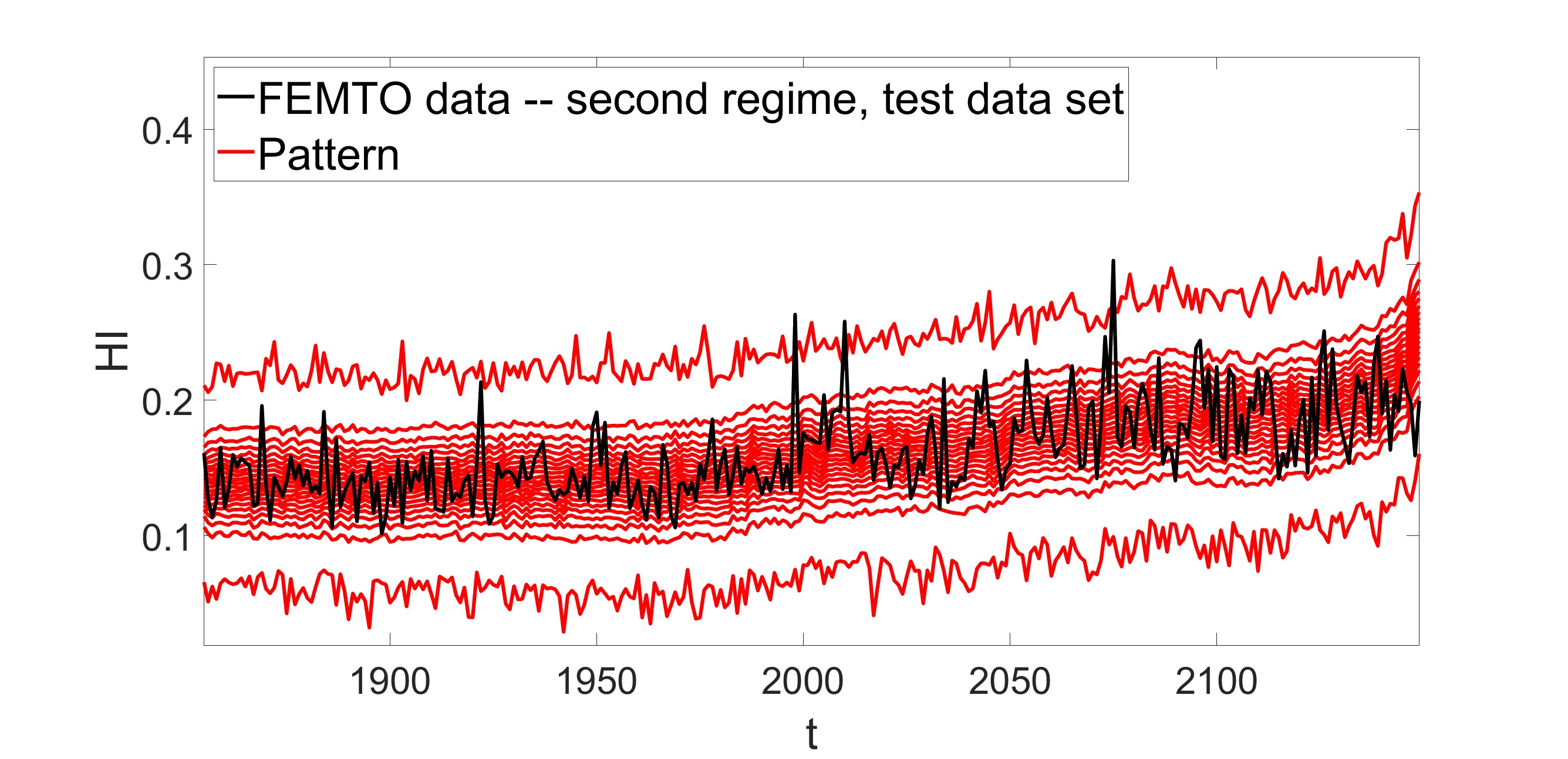}
        \caption{Comparison of the test data from the second regime of the FEMTO data with the pattern for SQIF}
    \label{FEMTO_second_regime_space}
    \end{minipage}
\end{figure}

In Fig. \ref{FEMTO_second_regime_average} we present the pattern for the MSE and MAPE metrics. As we can see, the red line properly follows the trajectory. We can also see the extreme values that are far above the pattern and can have a huge impact on the MSE metric, because in this metric we square the differences between observations and the pattern. The pattern also shows a positive linear trend, especially after $t=1980$.

Extreme values also have a significant impact on the SQIF metric (see Fig. \ref{FEMTO_second_regime_space}). We can see that the mentioned extreme observations exceed the quantile lines of order 100\%. This means that there is no trajectory $T_i(t)$ that achieves these values in the corresponding time. Thus, we can expect here the poor prediction quality assessment. Other values vary between inner quantile lines, however, still a lot of observations are far from the middle quantile lines of the order 45\% and 55\%.

\begin{figure}[H]
    \centering
    \begin{minipage}[t]{0.48\textwidth}
        \centering
        \includegraphics[width=1\textwidth]{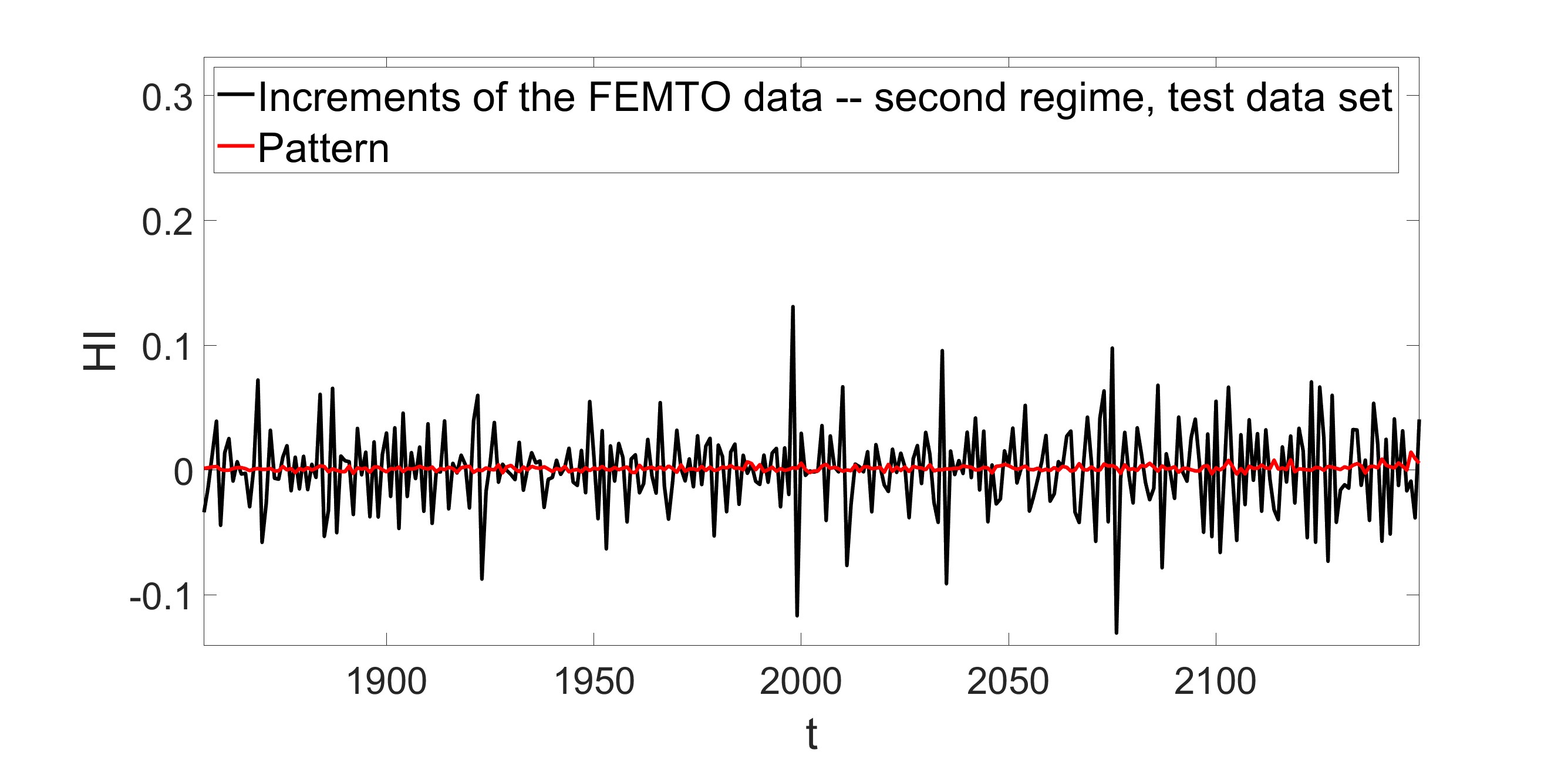}
        \caption{Comparison of the increments of the test data from the second regime of the FEMTO data with the pattern for Kupiec's POF}
    \label{FEMTO_second_regime_kupiec_pof}
    \end{minipage}
    \hfill
    \begin{minipage}[t]{0.48\textwidth}
        \centering
        \includegraphics[width=1\textwidth]{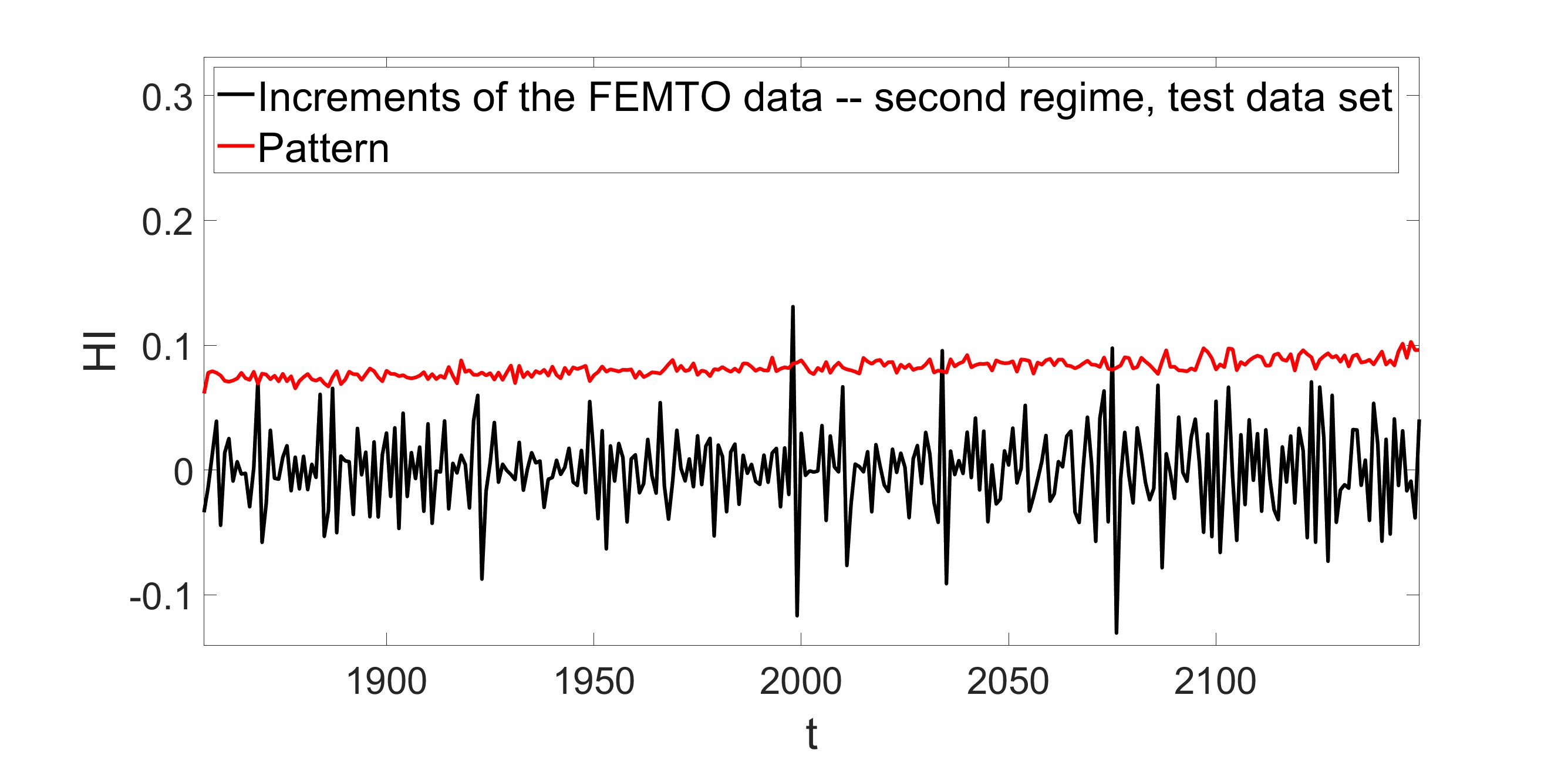}
        \caption{Comparison of the increments of the test data from the second regime of the FEMTO data with the pattern for Kupiec's TUFF}
    \label{FEMTO_second_regime_kupiec_tuff}
    \end{minipage}
\end{figure}

The comparison of pattern for Kupiec's POF metric with FEMTO data is presented in Fig. \ref{FEMTO_second_regime_kupiec_pof}. The number of observations above the pattern is similar to the number of observations below the pattern, which indicates a proper prediction. Here, the extreme values have no such impact on the metric results since we are not calculating here the distance from the pattern, but the number of observations above the pattern. Thus, the large values do not have such a large influence on the metric results.

In comparison to the metric mentioned above, the outliers play a crucial role for Kupiec's TUFF metric (see Fig. \ref{FEMTO_second_regime_kupiec_tuff}). Here, these extreme values exceed the quantile line (pattern for Kupiec's TUFF metric), and thus the first exceedance time is strongly connected with the time where the outliers occur. We can see that the pattern is slightly exceeded at the beginning of the data, which can result in a low prediction quality assessment.

\begin{figure}[H]
    \centering
        \includegraphics[width = 16cm]{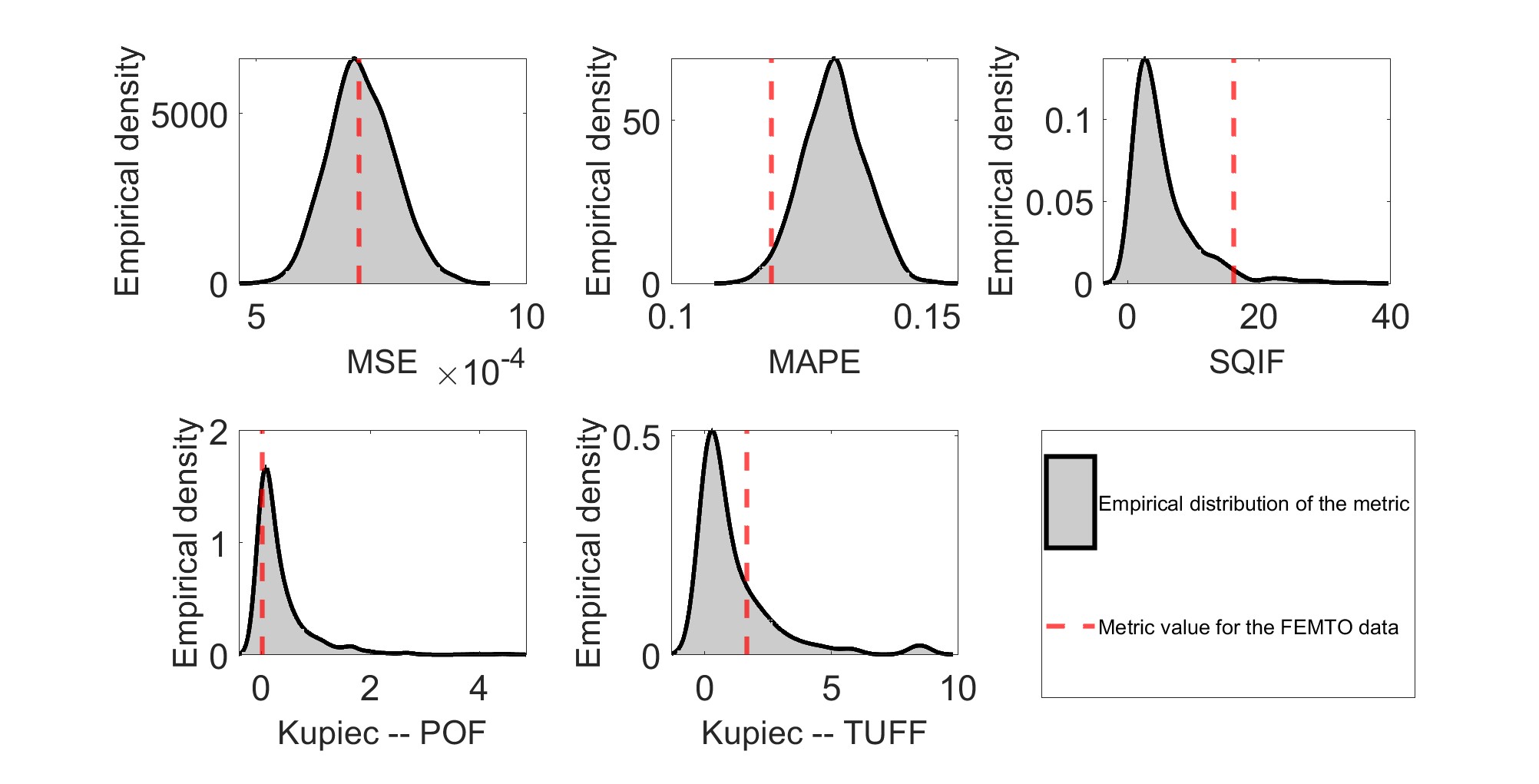}
        \caption{Comparison of metric distribution for 1000 training trajectories and for the test data from the second regime of the FEMTO data}
    \label{FEMTO_second_regime_densities}
\end{figure}

The impact of the extreme values is visible in Fig. \ref{FEMTO_second_regime_densities}. Here, we compare the metrics distributions (empirical probability density function) for prognoses, $T_i(t)$, $i=1,2,\ldots,1000$, and the metric values for the test data of the second regime of FEMTO time series. As mentioned, the highest impact of the outliers we have for MSE, SQIF and Kupiec's TUFF metrics. So, if we compare the left plot in the upper panel (MSE) with the middle plot in the upper panel (MAPE), we can see that for MAPE we get a significantly better prediction assessment. The MAPE value is smaller than for most prognoses, and this is the reason why the mass of the distribution of MAPE values for $T_i(t)$, $i=1,2,\ldots,1000$, is on the right side of the red dashed line. For MSE the metric value for FEMTO data is in the middle of the MSE distribution for prognoses. Similarly, due to outliers, we get very poor prediction quality assessment based on SQIF metric and Kupiec's TUFF metric (see the right plot in the upper panel and the middle plot in the lower panel, respectively). For Kupiec's POF metric, the majority of prognosed HI time series $T_i(t)$, $i=1,2,\ldots,1000$, gives larger metric values, so the prediction evaluation based on Kupiec's POF is greater than 50\% and in Tab. \ref{FEMTO_second_regime_table} we can see that it is evaluated as 80\% which is the second highest result from the considered metrics (the highest result is for MAPE, 90\% and the worst for SQIF, 3\%). There is no metric that gives a prediction quality result less than 3\%. For every metric except SQIF, the prediction assessment is higher than or equal to 20\%.

\subsection{Analysis of the data from the third regime}

\begin{figure}[H]
    \centering
        \includegraphics[width = 10cm]{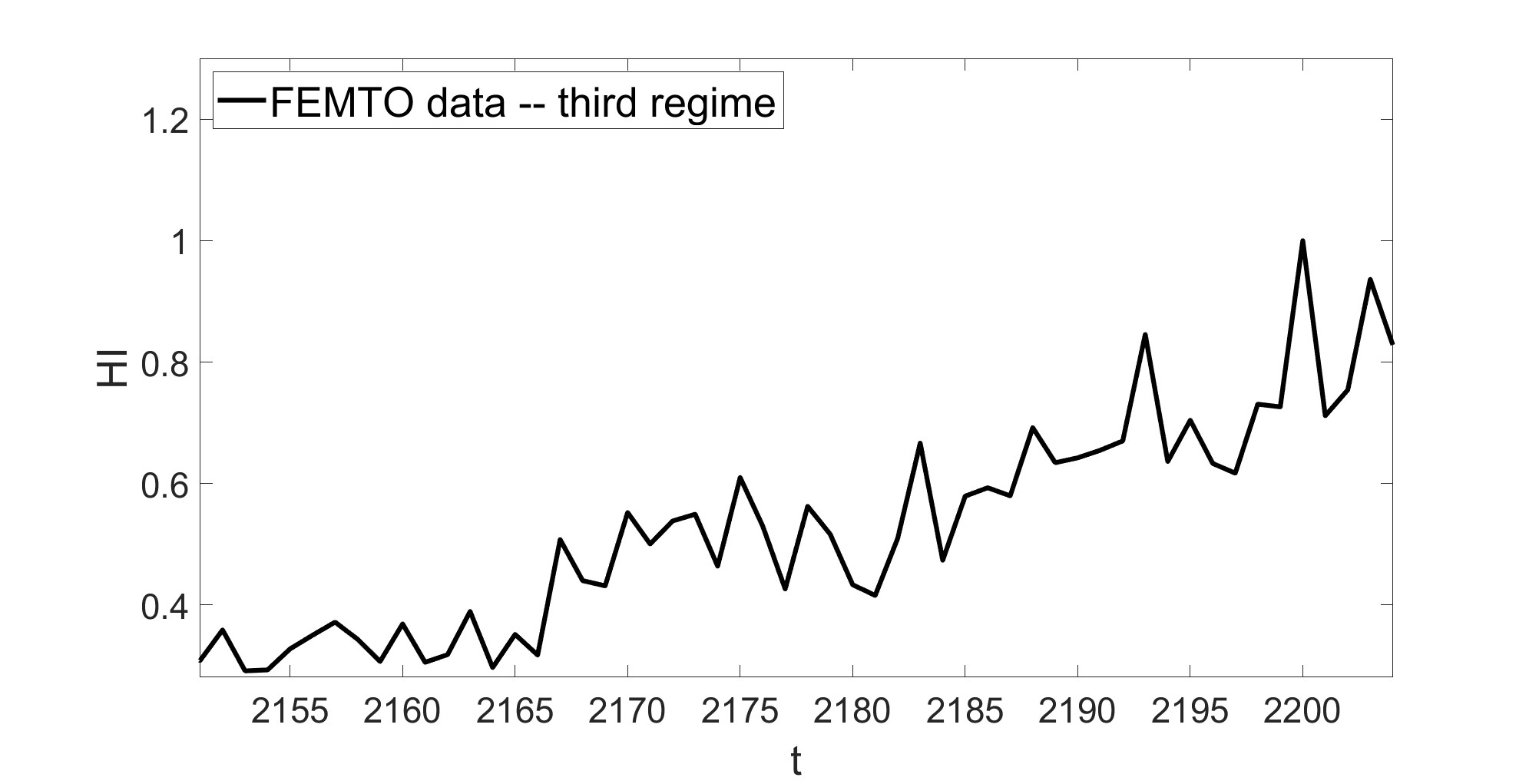}
        \caption{The third regime of the FEMTO data}
    \label{FEMTO_third_regime}
\end{figure}

Due to the very small number of observations in the third regime of the FEMTO data, we do not divide them into training and test data sets. Instead of this, here we present results of the prediction quality assessment for the entire third regime. We fit a model to the entire third regime and check the prognosis for the same time series. In Fig. \ref{FEMTO_third_regime} we can see that values increase significantly over time.  Moreover, the amplitude of the random part also increases. 

\begin{figure}[H]
    \centering
    \begin{minipage}[t]{0.48\textwidth}
        \centering
        \includegraphics[width=1\textwidth]{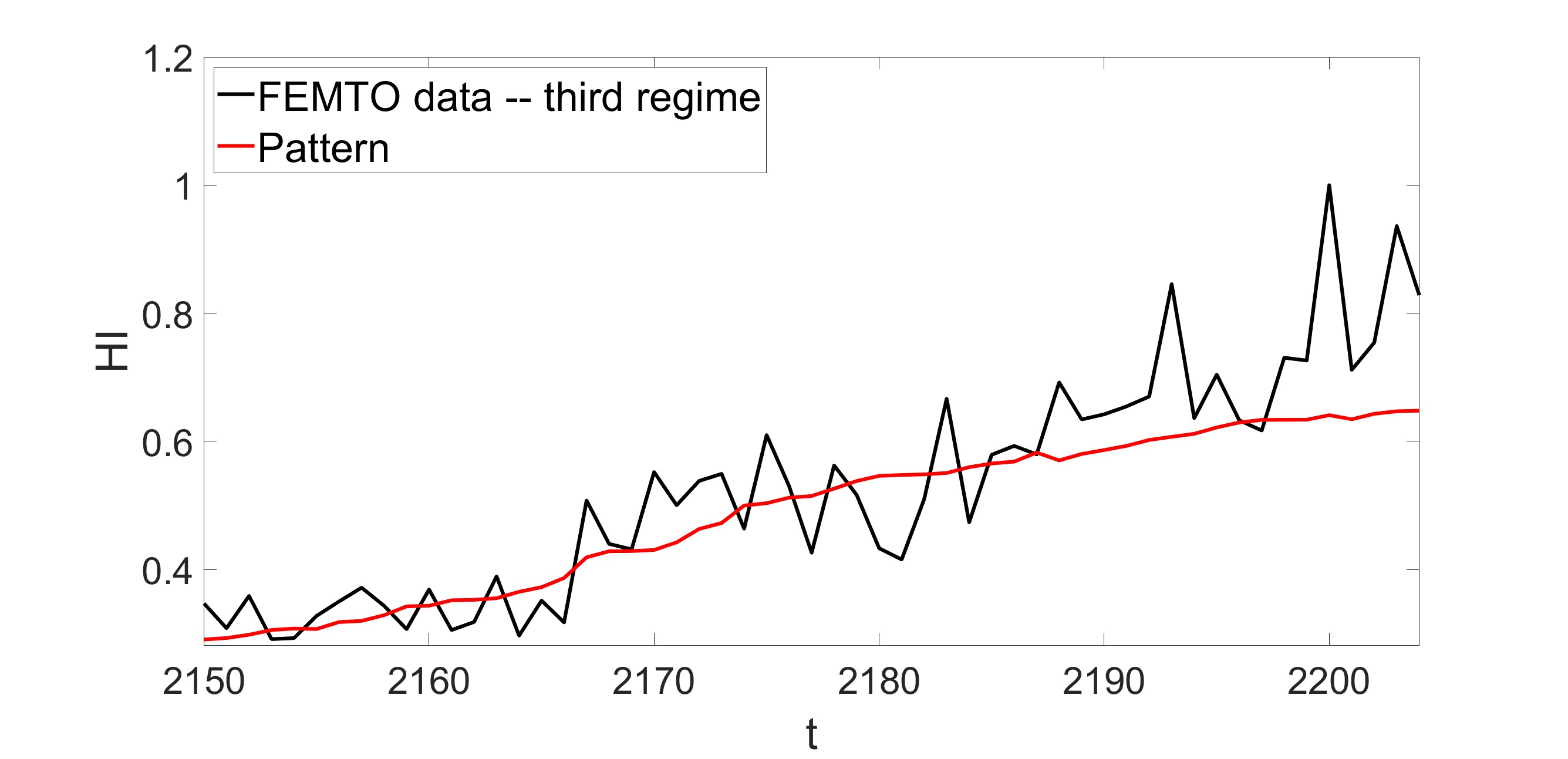}
        \caption{Comparison of the entire third regime of the FEMTO data with the pattern for MSE and MAPE}
    \label{FEMTO_third_regime_average}
    \end{minipage}
    \hfill
    \begin{minipage}[t]{0.48\textwidth}
        \centering
        \includegraphics[width=1\textwidth]{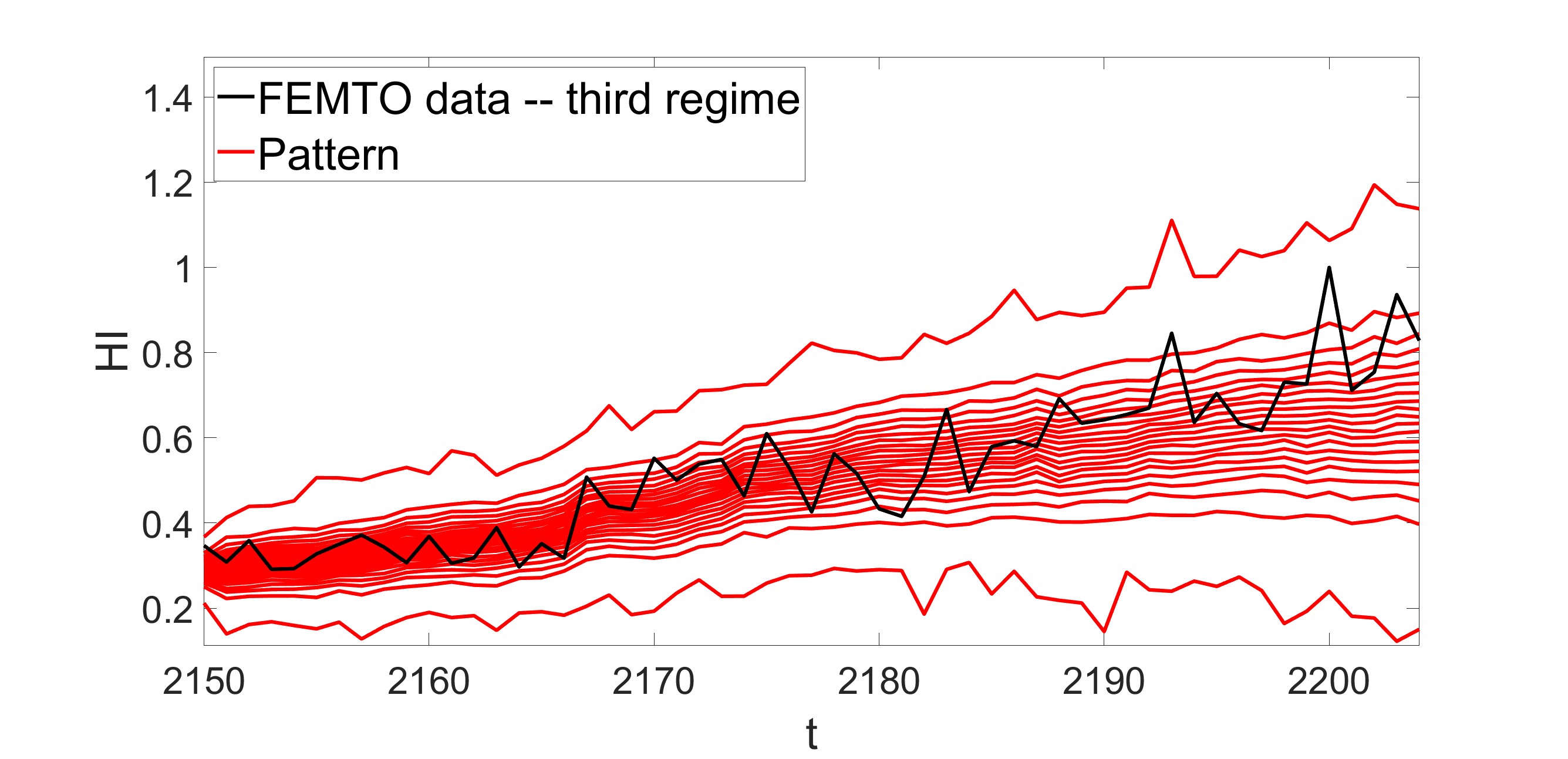}
        \caption{Comparison of the entire third regime of the FEMTO data with the pattern for SQIF}
    \label{FEMTO_third_regime_space}
    \end{minipage}
\end{figure}

In Fig. \ref{FEMTO_third_regime_average} we present the third regime of the FEMTO data compared with the pattern for the MSE and MAPE metrics. We can see that until $t=2185$, the pattern properly follows the trajectory, only at the end of the third regime, observations of the FEMTO data are slightly above the pattern.

The pattern for the SQIF metric is presented in Fig. \ref{FEMTO_third_regime_space}. There are no observations that exceed the outer quantile lines of orders 0\% and 100\%. Most observations lie between quantile lines of order 5\% and 95\%. There are only 3 values which exceed the 95\% quantile line, and they occur at the end of the third regime.

\begin{figure}[H]
    \centering
    \begin{minipage}[t]{0.48\textwidth}
        \centering
        \includegraphics[width=1\textwidth]{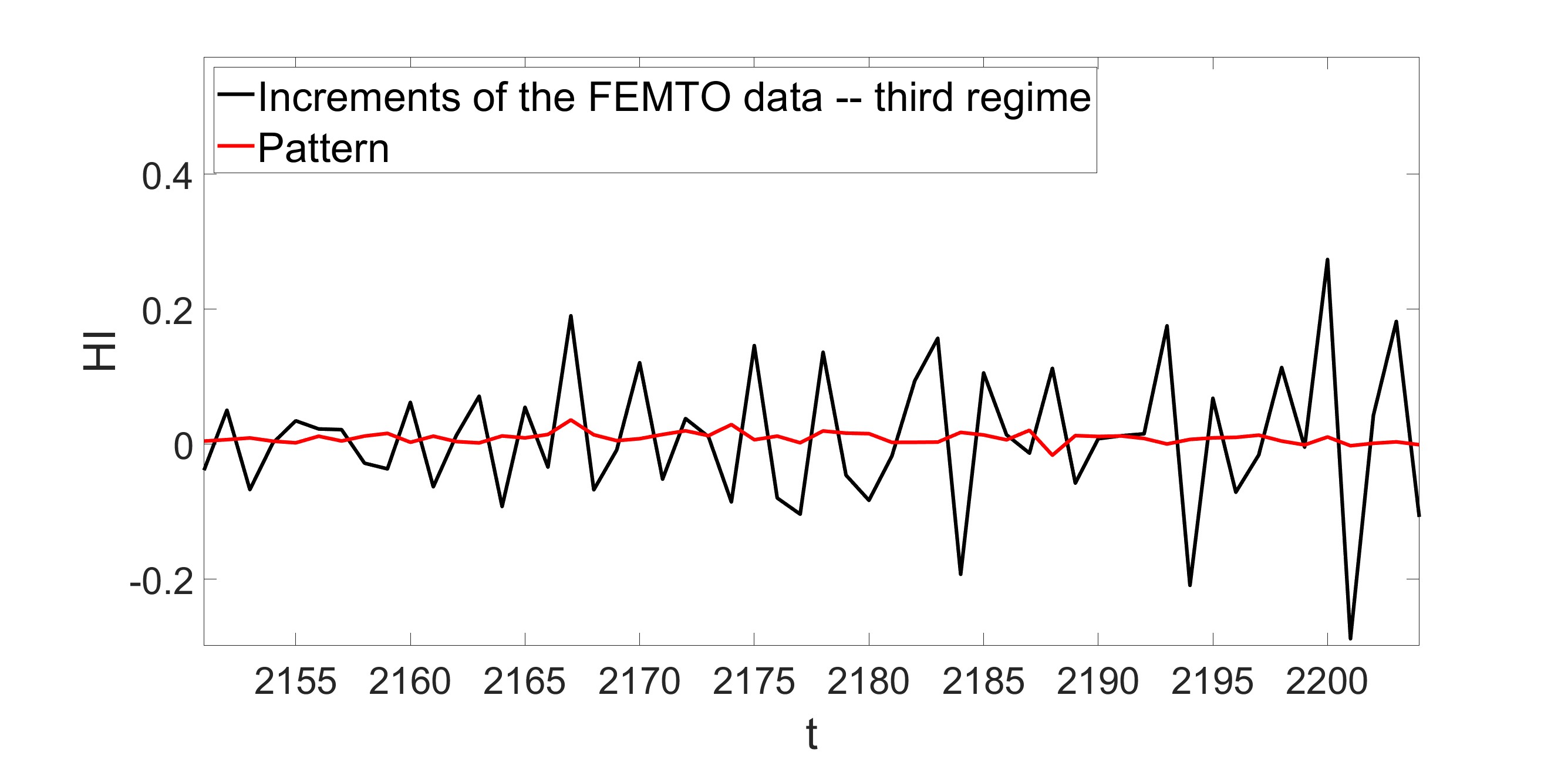}
        \caption{Comparison of the increments of the entire third regime of the FEMTO data with the pattern for Kupiec's POF}
    \label{FEMTO_third_regime_kupiec_pof}
    \end{minipage}
    \hfill
    \begin{minipage}[t]{0.48\textwidth}
        \centering
        \includegraphics[width=1\textwidth]{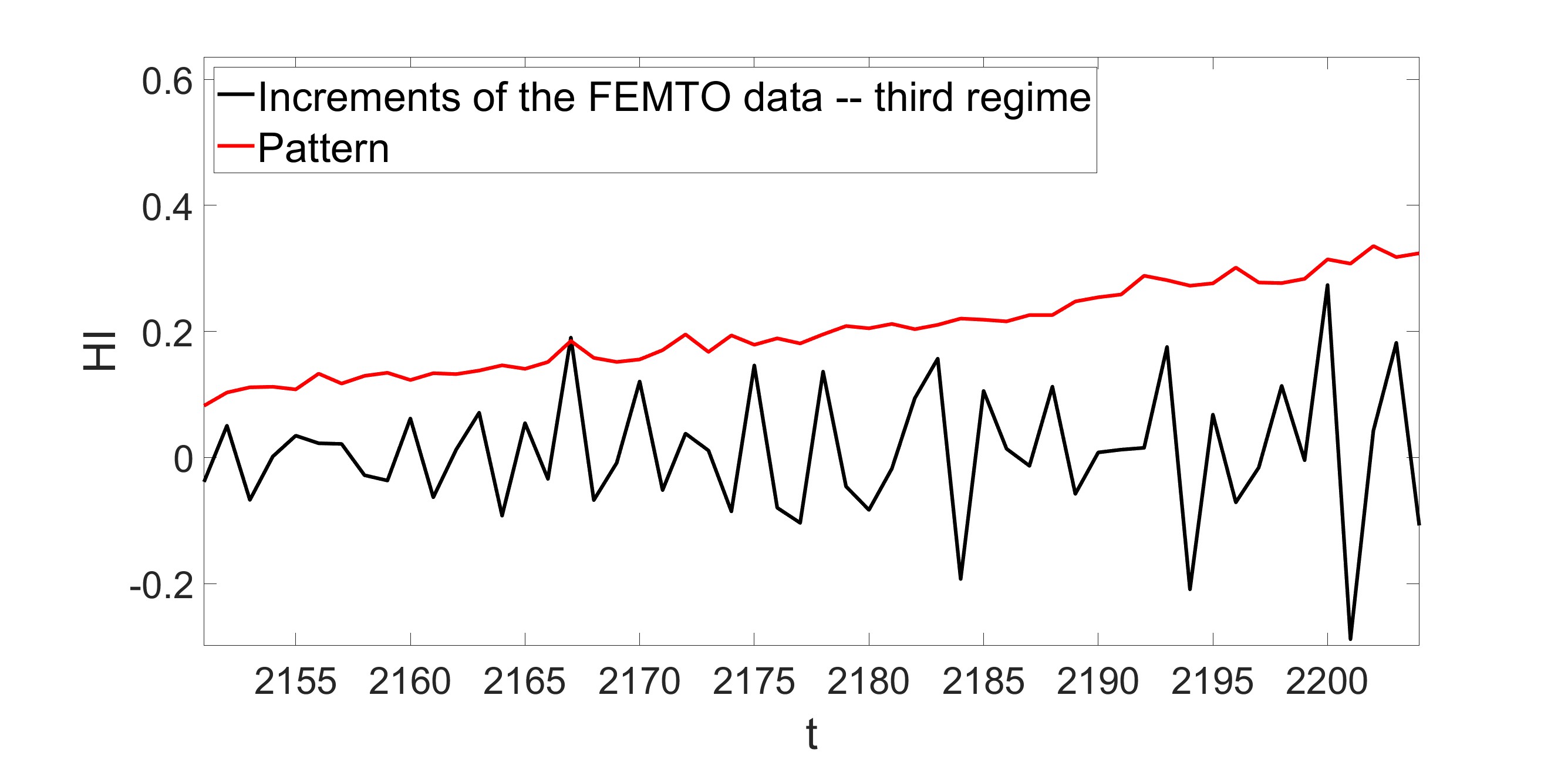}
        \caption{Comparison of the increments of the entire third regime of the FEMTO data with the pattern for Kupiec's TUFF}
    \label{FEMTO_third_regime_kupiec_tuff}
    \end{minipage}
\end{figure}

Increasing amplitudes of the random part are clearly visible if we analyze increments of the FEMTO data in the third regime (see Fig. \ref{FEMTO_third_regime_kupiec_pof}). As we can see here, the values vary around the pattern; however, the distances between observation and the pattern significantly increase over time. For the increments, we cannot see any important positive trend.

Similarly, we analyze increments to derive a pattern for Kupiec's TUFF statistic (see Fig. \ref{FEMTO_third_regime_kupiec_tuff}). Here, only one observation exceeds the quantile line for $t=2167$. There are some other observations which are close to the quantile lines, especially in the middle and at the end of the third regime; however, in Kupiec's TUFF statistic, we proceed with only the first exceedance time. 

\begin{figure}[H]
    \centering
        \includegraphics[width = 16cm]{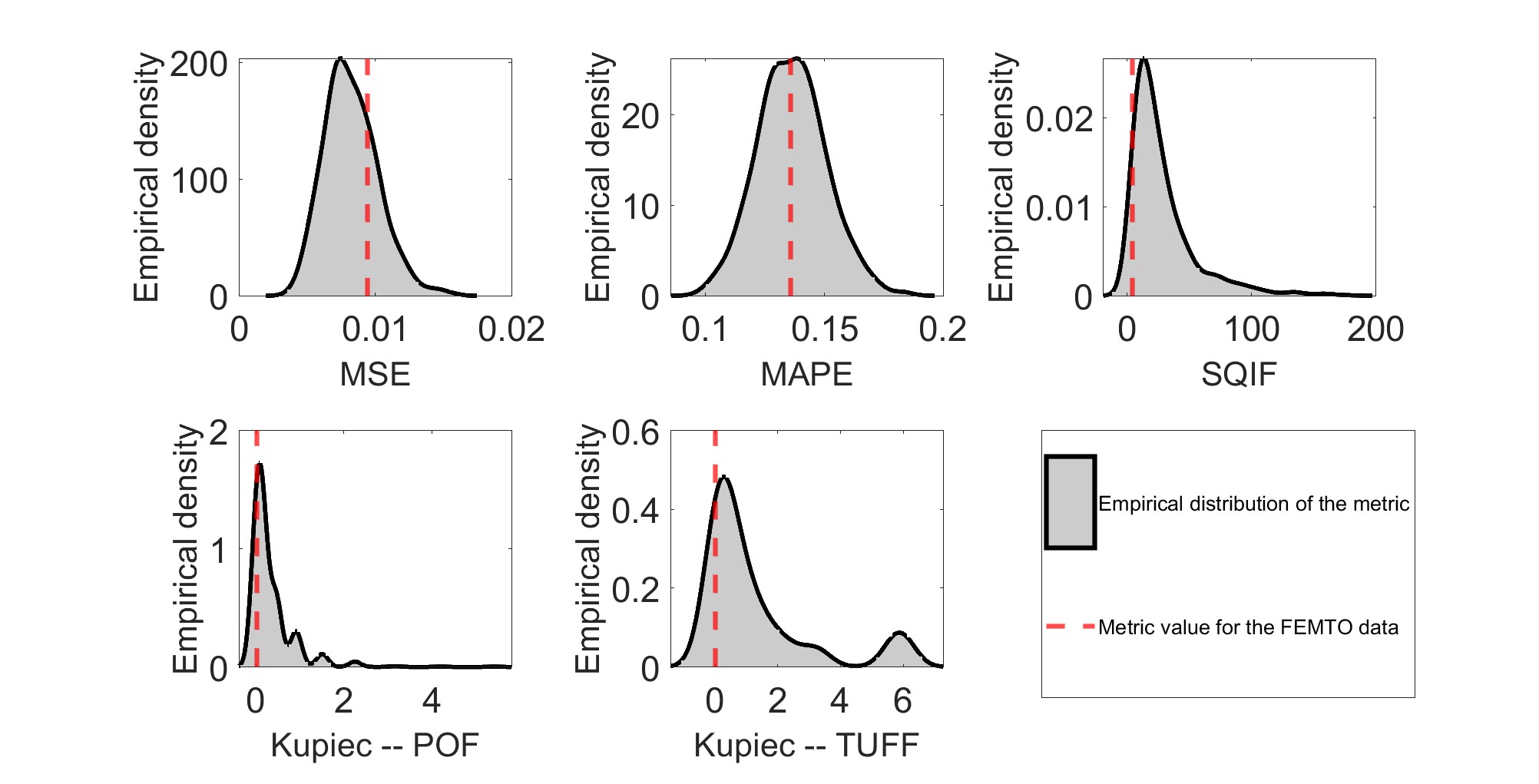}
        \caption{Comparison of metric distribution for 1000 training trajectories and for the entire third regime of the FEMTO data}
    \label{FEMTO_third_regime_densities}
\end{figure}

Finally, we present the summary for the third regime prediction evaluation based on considered metrics (see Fig. \ref{FEMTO_third_regime_densities}). The worst prediction assessment we get for the MSE metric (left plot in the upper panel). We can see here that the derived MSE metric value for the FEMTO data is larger than for most prognosed HI time series. Furthermore, the prediction quality for MSE is evaluated as 20\% (see Tab. \ref{FEMTO_third_regime_table}). For MAPE (middle plot in the upper panel), the derived metric value is in the middle of the distribution, which indicates that the prediction is evaluated approximately as 50\%. Precisely in Tab. \ref{FEMTO_third_regime_table} we can see that this is exactly 50\%. For SQIF, Kupiec's POF and Kupiec's TUFF metrics, we get a very high quality assessment of the prediction, which manifests itself in the fact that the mass of the metric distributions is in the right side of the red dashed lines, which corresponds to metric value for FEMTO data. In Tab. \ref{FEMTO_third_regime_table} it is shown that for both SQIF, Kupiec's POF and Kupiec's TUFF metrics, the prediction quality is evaluated as 90\%. As a result, we obtain at least 20\% quality assessment for every of the considered metrics.

\section{IMS data set}\label{sec_IMS}

\subsection{Data set description}
\begin{figure}[H]
    \centering
    \begin{minipage}[t]{0.48\textwidth}
        \centering
      \includegraphics[width=0.9\textwidth, height=3.8cm]{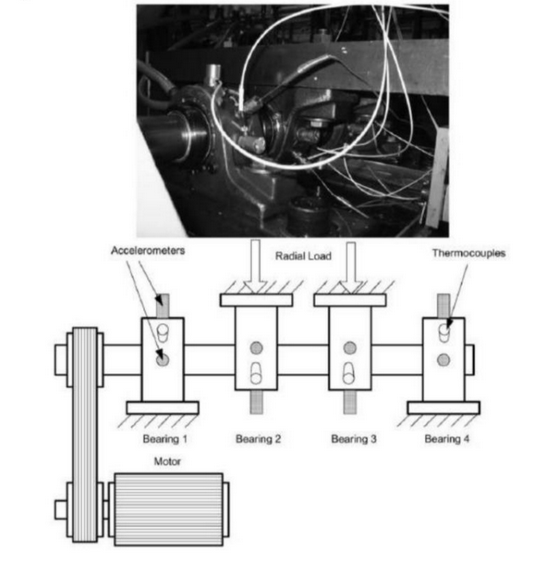}
      \caption{IMS test rig}
      \label{fig:IMS_testrig}
    \end{minipage}
    \hfill
    \begin{minipage}[t]{0.48\textwidth}
        \centering
        \includegraphics[width=1\textwidth]{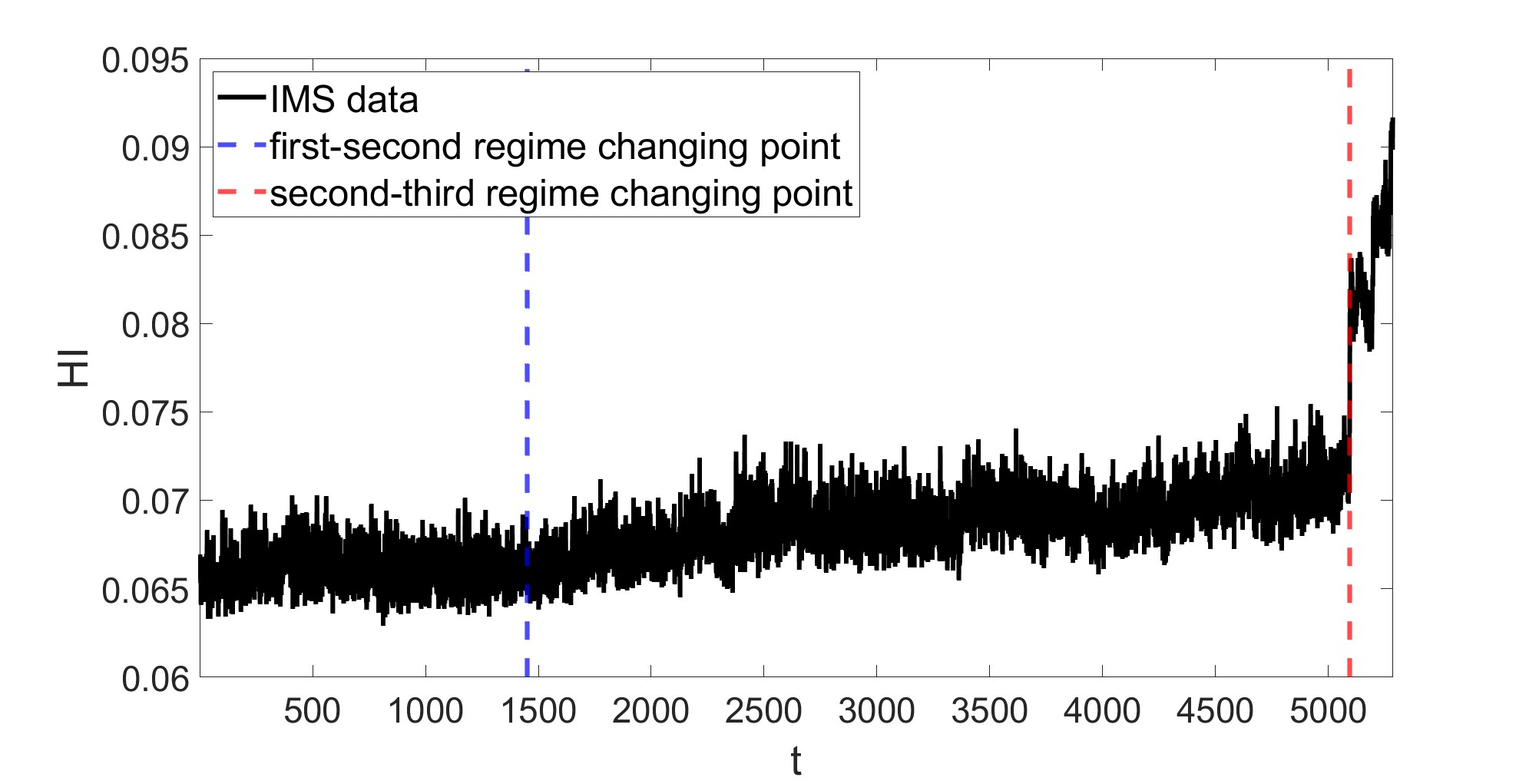}
        \caption{IMS HI data}
    \label{IMS_data}
    \end{minipage}
\end{figure}

The IMS data set was collected by the Intelligent Maintenance System (IMS) laboratory at the University of Cincinnati, which covers three subsets of bearing degradation tests. \textcolor{black}{In this paper data from subset 3 and bearing 3 are analyzed (Subset3\_Bearing3\_1)}. Throughout the degradation experiments, four Rexnord ZA-2115 double-row bearings were employed on a shaft, with accelerometers affixed to the bearing housings. Following the tests, the degradation patterns were meticulously recorded by inspecting the bearings. Fig. \ref{fig:IMS_testrig} illustrates the IMS test rigs. The bearings in the IMS dataset have a longer and more complicated degradation trend than other benchmark datasets, which increase the similarity of the dataset with actual application in industries, also the difficulty of RUL prediction. 
\textcolor{black}{The amplitude of vibration increased at the beginning of the degradation process. This is caused by the initial surface defect, such as cracks or spalling. However, later the amplitude decreases due to smoothed the initial surface defect by continous rolling contact. The amplitude of vibrations increases again, when the damage spreads to a larger area.} The nonlinear degradation trends with fluctuations bring significant challenges to the prediction of RUL.

In Fig. \ref{IMS_data} the entire degradation process is presented for the IMS data set. Here, we can distinguish three main regimes: first regime, including observations from 1 to 1452, second regime -- observations from 1453 to 5095 and third regimes with observations from 5096 to 5286. Regime changing points are marked by blue and red dashed lines for the first-second and second-third regime changing points, respectively, derived as in \cite{Maraj2023113495}. We can see that in the first regime, there is no significant positive trend. Values on average slightly increase for observation numbers from 1 to about 300 but similarly decrease for observation numbers from about 300 until the end of the first regime. The variation of the random part is constant throughout the first regime. As we can see, the behavior of the data is changing in the second regime. Although the variance of the noise is still at the constant level, similar to the first regime, there is a positive linear trend. The most extreme values are achieved in the third regime. The values are increasing drastically from about 0.07 to about 0.09. Thus, the change of HI during the third regime, which is the shortest one, is almost four times larger than the change during the first and second regimes together.

\subsection{Analysis of the data from the second regime}

\begin{figure}[H]
    \centering
        \includegraphics[width = 10cm]{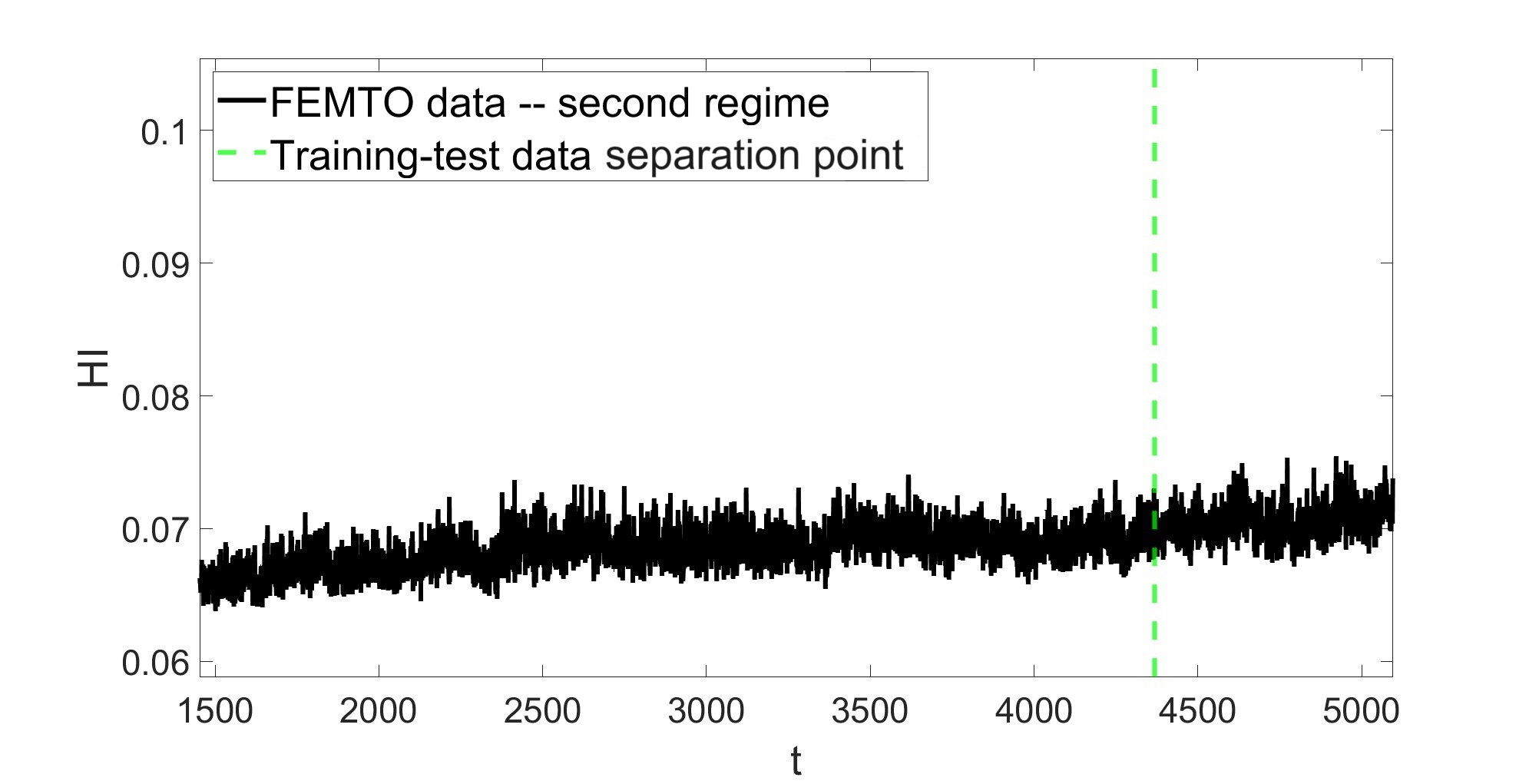}
        \caption{The second regime of the IMS data and the separation point between training and test data sets}
    \label{IMS_second_regime}
\end{figure}

Similarly, as for the FEMTO data, we analyze here the second and third regimes of the IMS data, separately. In Fig. \ref{IMS_second_regime} we can see the second regime of IMS data with training-test data separation point marked by a green dashed line. The same as for the FEMTO data, we fit the model described in Section \ref{model_description} to the entire second regime and calculate the metrics only for the test part, which constitutes the last 20\% observations of the second regime. We can see here that the behavior in the test part of the data does not differ much from the training part. There is a similar variance of the noise as well as a slope of trend.

\begin{figure}[H]
    \centering
    \begin{minipage}[t]{0.48\textwidth}
        \centering
        \includegraphics[width=1\textwidth]{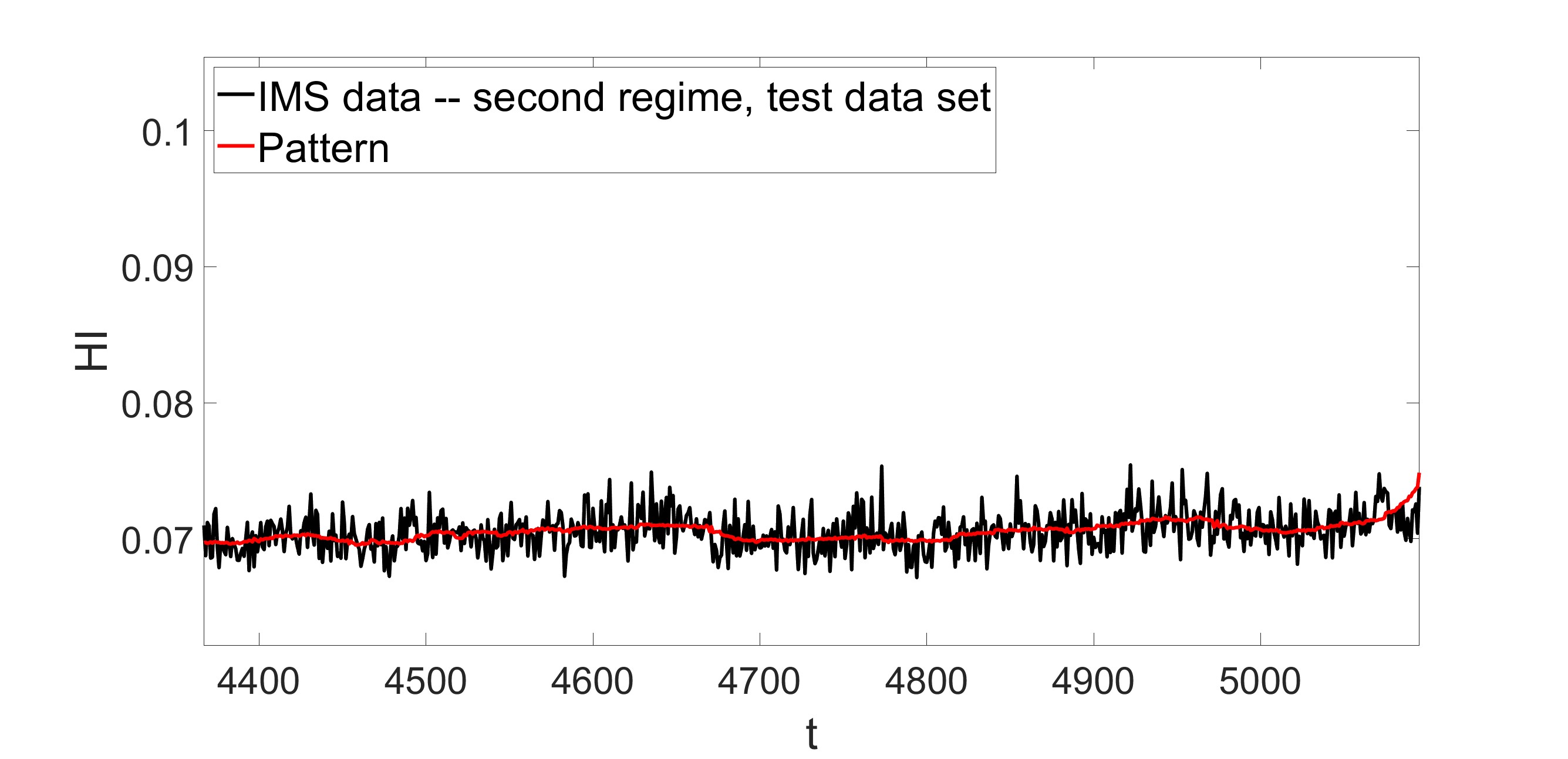}
        \caption{Comparison of the test data from the second regime of the IMS data with the pattern for MSE and MAPE}
    \label{IMS_second_regime_average}
    \end{minipage}
    \hfill
    \begin{minipage}[t]{0.48\textwidth}
        \centering
        \includegraphics[width=1\textwidth]{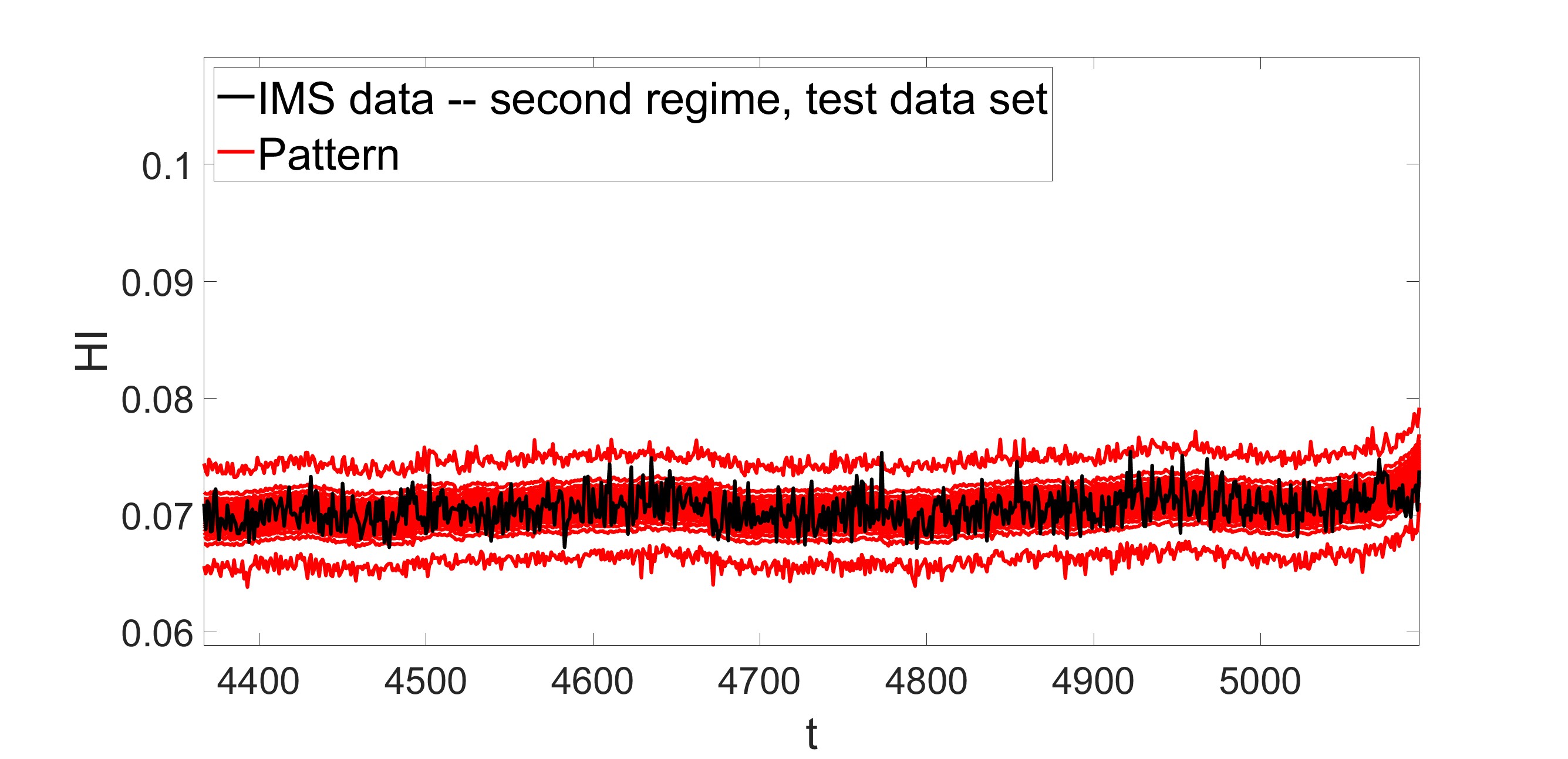}
        \caption{Comparison of the test data from the second regime of the IMS data with the pattern for SQIF}
    \label{IMS_second_regime_space}
    \end{minipage}
\end{figure}

To assess the prediction based on MSE and MAPE metrics, we have to calculate the corresponding pattern, which in these cases is the average predicted trajectory. In Fig. \ref{IMS_second_regime_average} we can see its comparison with the test part of the second regimes of IMS data. It is clearly visible that the pattern follows the IMS data in a proper way. The significant difference between pattern and IMS data is only at the end of the second regime. Pattern values are increasing, while IMS data are decreasing. For other observations, we can see that the derived pattern is in the middle of the IMS degradation data, indicating a proper prediction.

Similarly, we proceed with the SQIF metric. In this case, the pattern is described by quantile lines (see Fig. \ref{IMS_second_regime_space}). There are only a few observations that exceed the 100\% quantile line, while there are no observations that are below the 0\% quantile line. The IMS data values vary mainly between the quantile lines 5\% and 95\%.

\begin{figure}[H]
    \centering
    \begin{minipage}[t]{0.48\textwidth}
        \centering
        \includegraphics[width=1\textwidth]{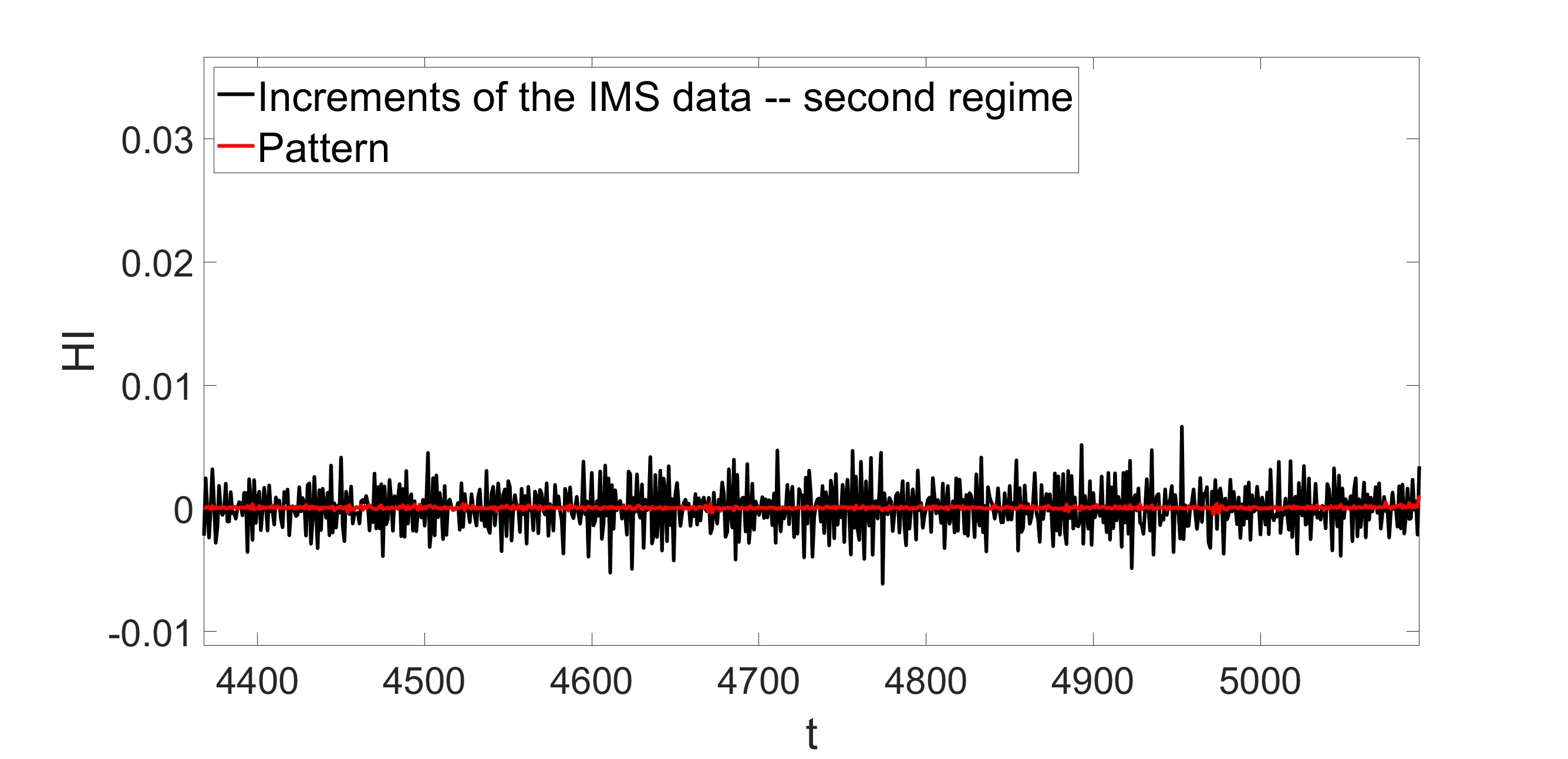}
        \caption{Comparison of the increments of the test data from the second regime of the IMS data with the pattern for Kupiec's POF}
    \label{IMS_second_regime_kupiec_pof}
    \end{minipage}
    \hfill
    \begin{minipage}[t]{0.48\textwidth}
        \centering
        \includegraphics[width=1\textwidth]{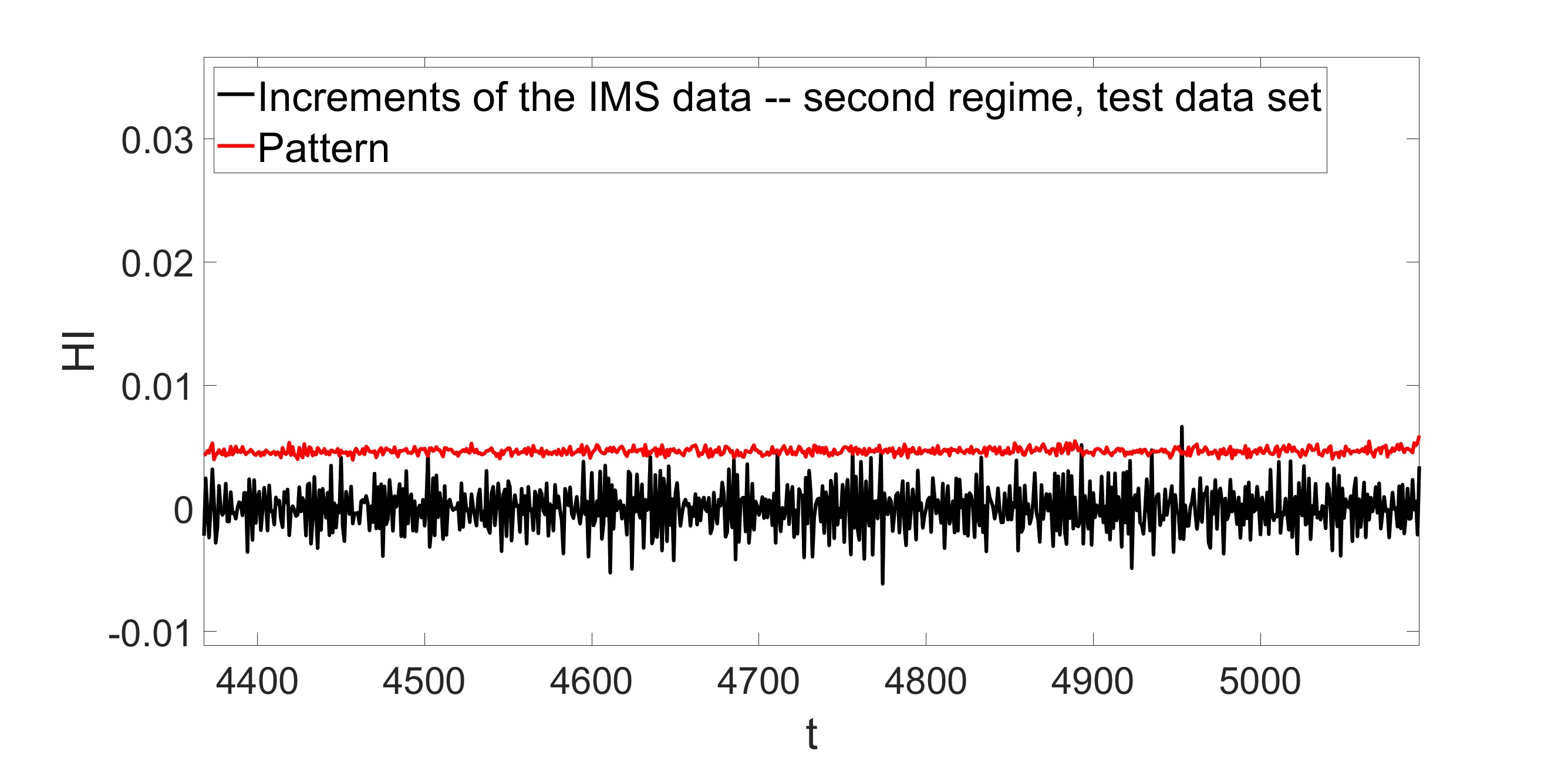}
        \caption{Comparison of the increments of the test data from the second regime of the IMS data with the pattern for Kupiec's TUFF}
    \label{IMS_second_regime_kupiec_tuff}
    \end{minipage}
\end{figure}

For Kupiec's POF and Kupiec's TUFF metrics, we derive a pattern based on increments of simulated data from the fitted model. In Fig. \ref{IMS_second_regime_kupiec_pof} there is a comparison of increments of the test part of the second regime of IMS data with the quantile line of order 51\%. We can see that the pattern follows increments of IMS data in a proper way.

In Fig. \ref{IMS_second_regime_kupiec_tuff} we present the pattern for Kupiec's TUFF. As we can see, we compare here increments of IMS data with a quantile line of order calculated according to the formula described in Section \ref{metrics}. There is only one observation about $t=4950$ that significantly exceeds the quantile level; however, there are a few observations that take values similar to the pattern.

\begin{figure}[H]
    \centering
        \includegraphics[width = 16cm]{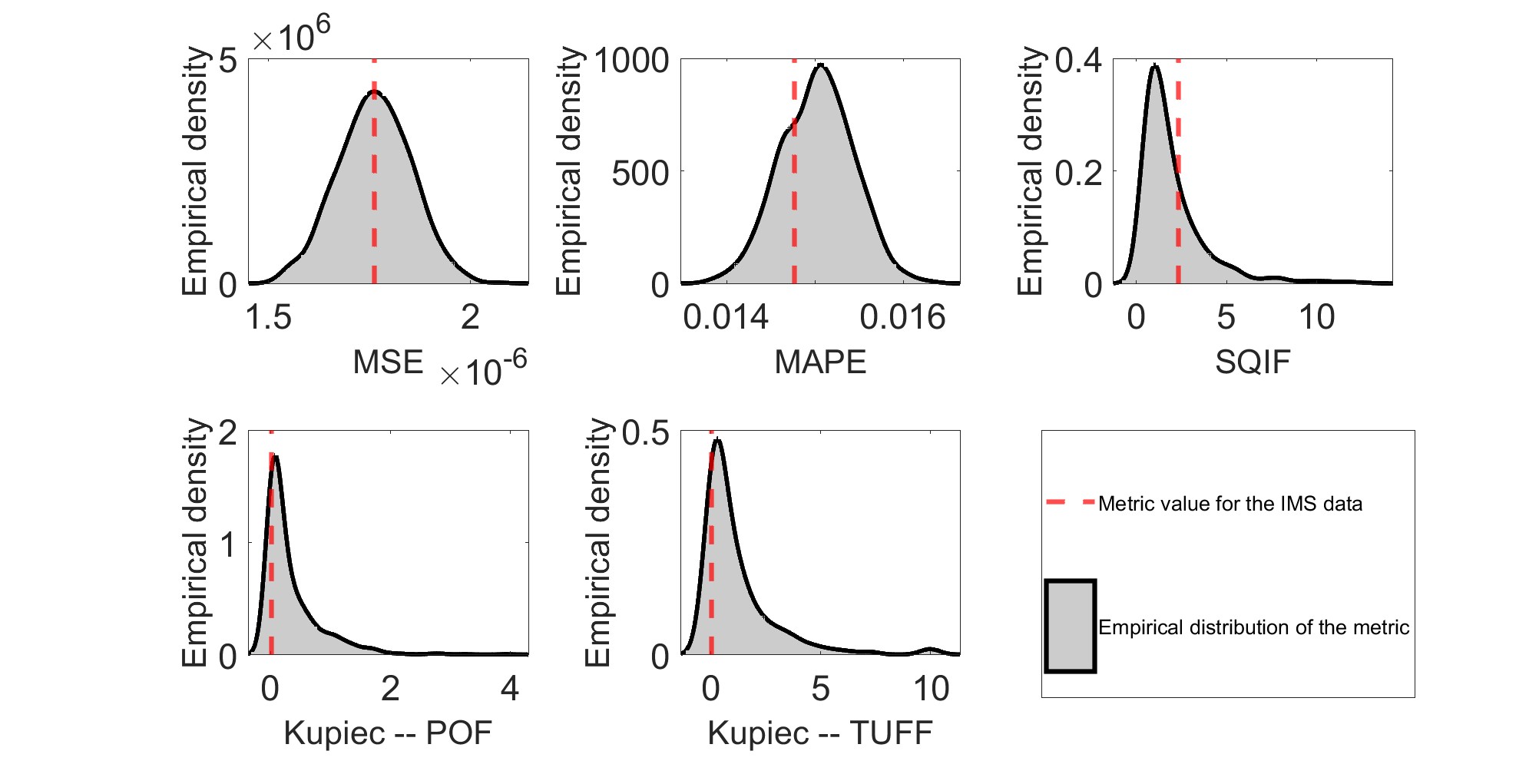}
        \caption{Comparison of metric distribution for 1000 training trajectories and for the test data from the second regime of the IMS data}
    \label{IMS_second_regime_densities}
\end{figure}

The final evaluation of the second regime prediction for the IMS data is presented in Fig. \ref{IMS_second_regime_densities}. For the MSE metric, the calculated value for IMS data is about in the middle of the empirical density derived for simulated data from the fitted model, in the center of the mass of distribution. This means that there are 50\% simulated trajectories with higher MSE, as well as 50\% simulated trajectories with lower MSE metric values. The MAPE metric calculated for the IMS data is smaller than most of the simulated trajectories of the model; therefore, we can expect that the prediction evaluation is higher than 50\% -- for at least 50\% of the prognosed HI time series, the obtained metric was larger than for the true HI time series (in Tab. \ref{IMS_second_regime_table} we can see that the prediction assessment is at the level 70\%). The worst prediction evaluation, we obtain for the SQIF metric (right plot in the upper panel). As we can see, the mass of the distribution is to the left of the SQIF metric for the IMS data. The opposite situation is for both Kupiec's POF and Kupiec's TUFF metrics, where the majority of derived metric values for simulated data were greater than those obtained for IMS data. In Tab. \ref{IMS_second_regime_table} we can see more detailed analysis. It is shown there that for every metric we get prediction evaluation at least as 20\%, where 20\%-evaluation is obtained for SQIF metric. For the other metrics, the assessment is 50\% or higher and for the Kupiec's TUFF statistic the assessment value achieves 90\%.

\subsection{Analysis of the data from the third regime}

\begin{figure}[H]
    \centering
        \includegraphics[width = 10cm]{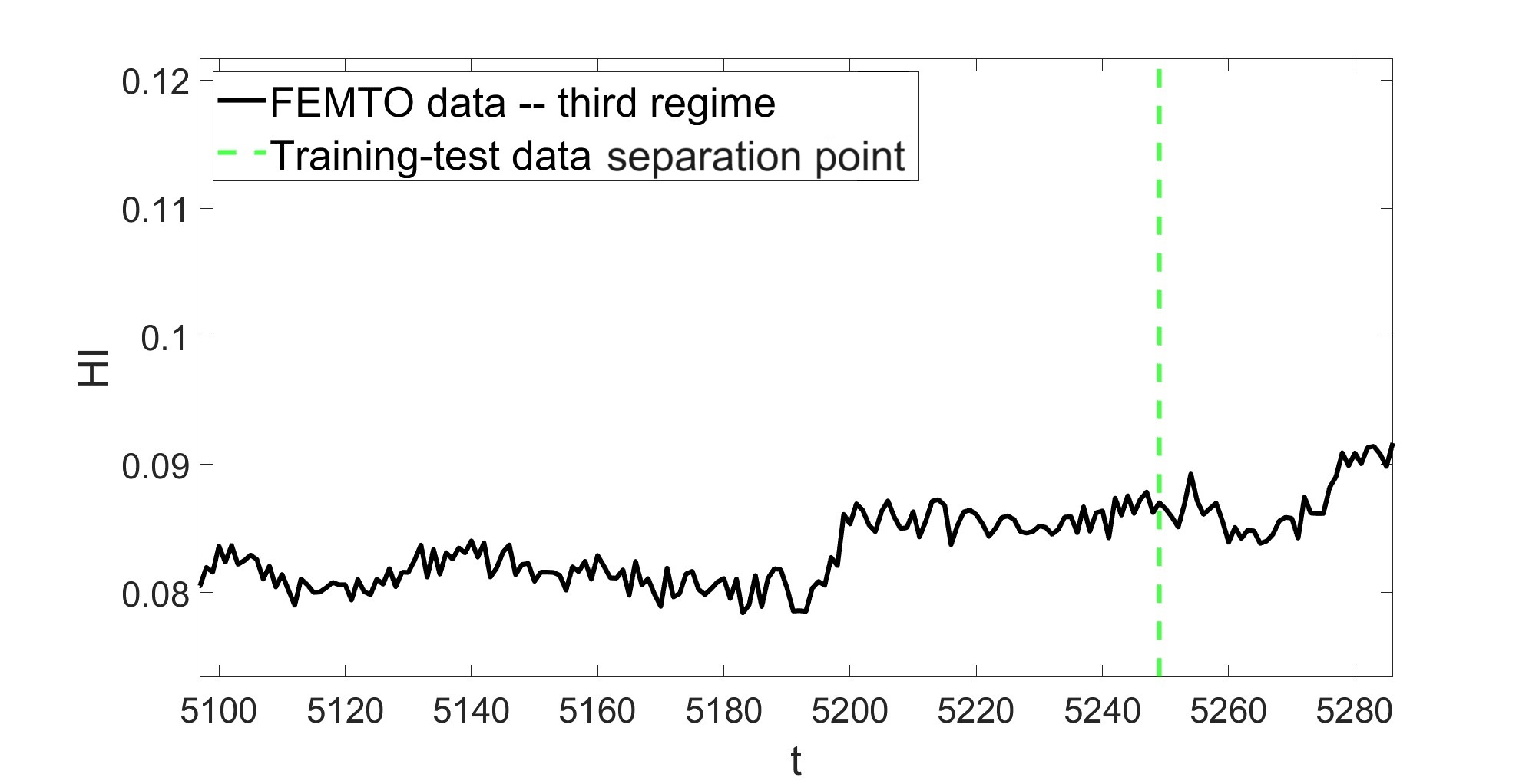}
        \caption{The third regime of the IMS data and the separation point between training and test data sets}
    \label{IMS_third_regime}
\end{figure}

In Fig. \ref{IMS_third_regime} degradation data is shown that come from the third regime of IMS time series. The green dashed line denotes the training-test data separation point. We can see that the third regime consists of significantly less number of observations than the second regime. This makes both the prediction and prediction assessment harder; however, typically damage of a machine occurs in the third regime, thus, it is required to be able to assess the prediction despite having a small number of observations. We can see that the values increase over time; however, in $t\approx 5190$ and $t\approx 5270$, they are increasing faster than in the rest of the data. Furthermore, one can say that in the rest time periods, the observations rather vary around a constant value.

\begin{figure}[H]
    \centering
    \begin{minipage}[t]{0.48\textwidth}
        \centering
        \includegraphics[width=1\textwidth]{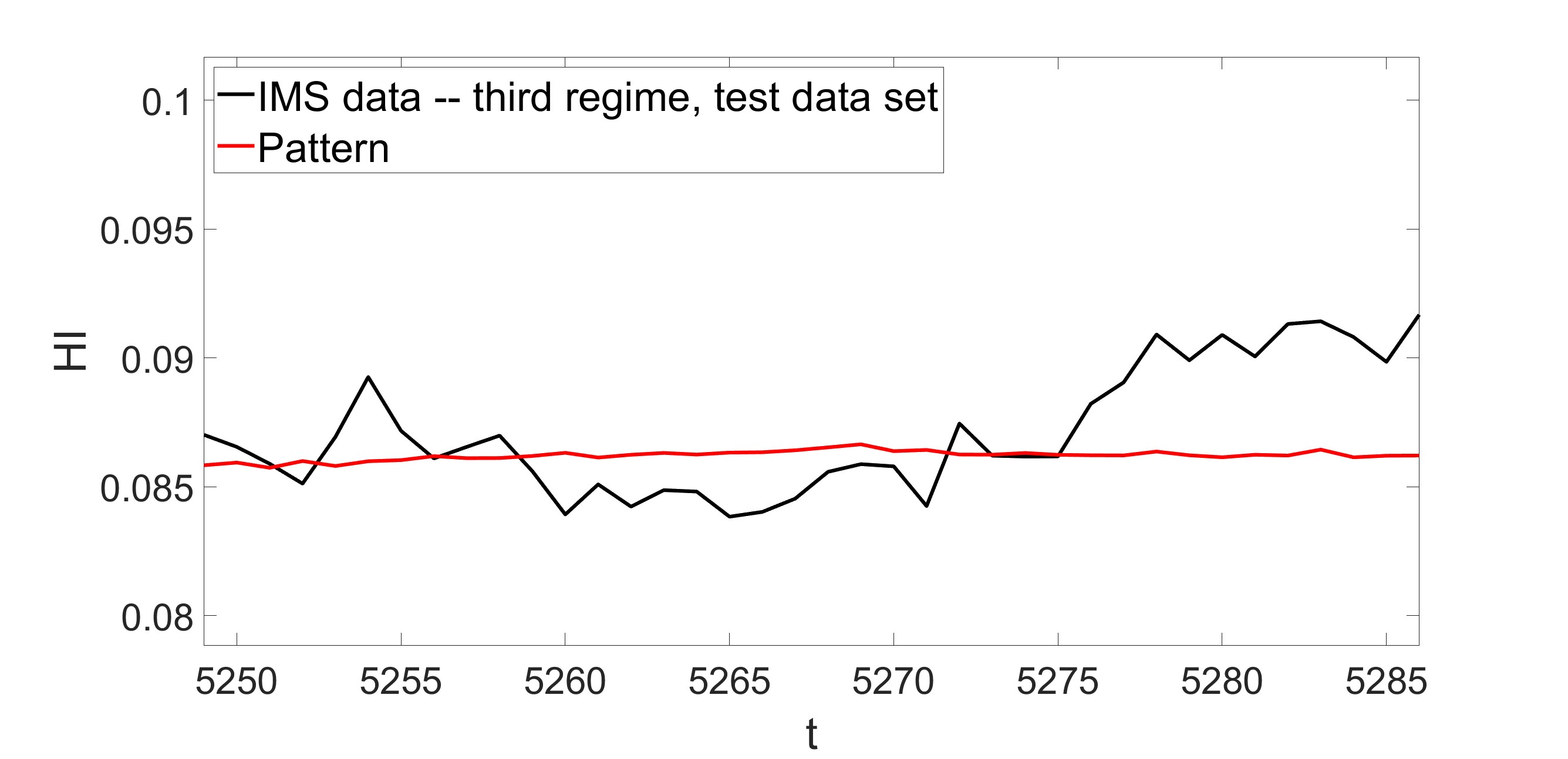}
        \caption{Comparison of the test data from the third regime of the IMS data with the pattern for MSE and MAPE}
    \label{IMS_third_regime_average}
    \end{minipage}
    \hfill
    \begin{minipage}[t]{0.48\textwidth}
        \centering
        \includegraphics[width=1\textwidth]{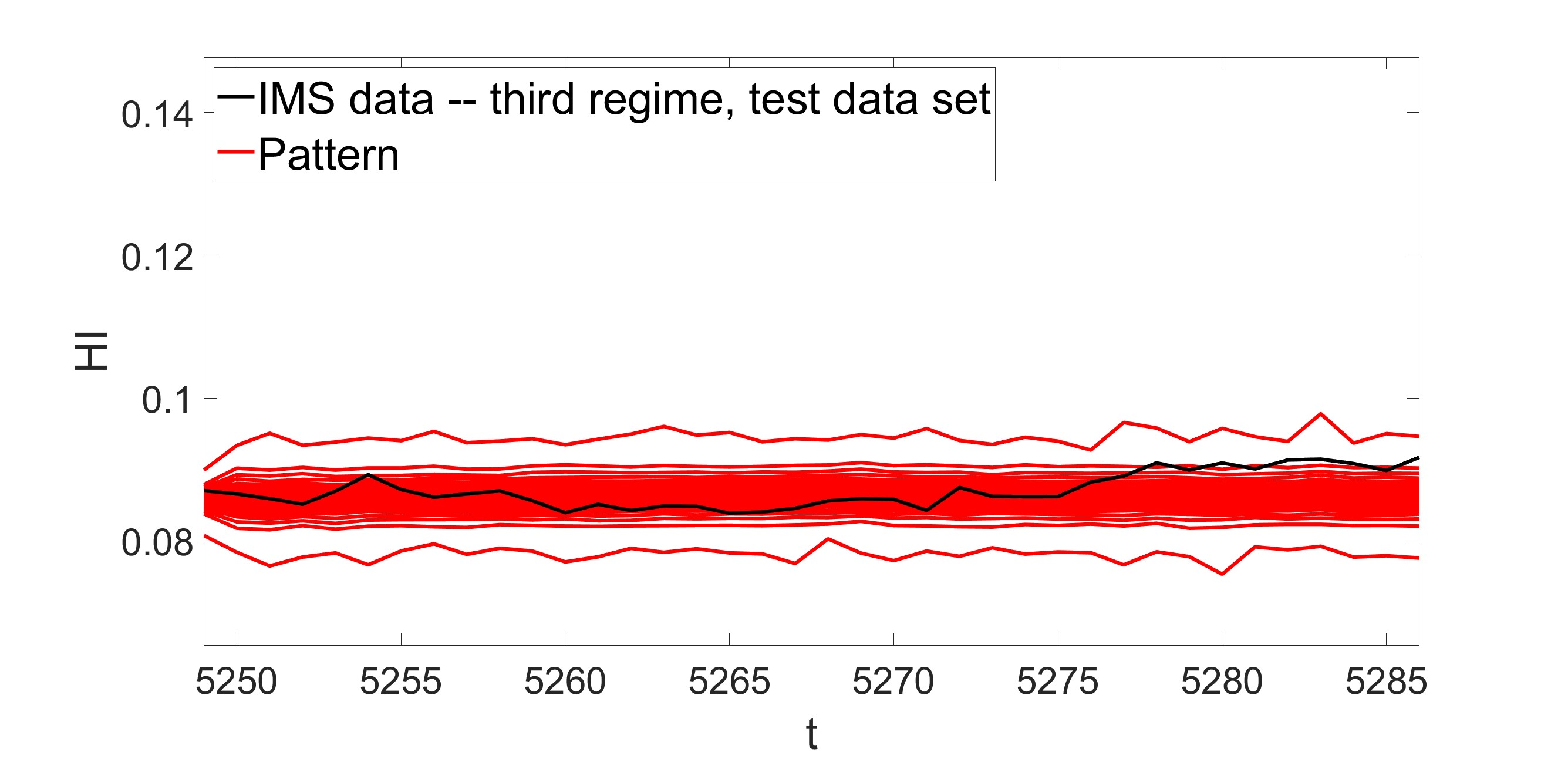}
        \caption{Comparison of the test data from the third regime of the IMS data with the pattern for SQIF}
    \label{IMS_third_regime_space}
    \end{minipage}
\end{figure}

In Fig. \ref{IMS_third_regime_average} we can see that the test part of the third regime of IMS data is very short and includes only 38 observations. The pattern for the MSE and MAPE metrics is in the middle of the IMS data; however, the differences between the pattern and IMS data are larger than for the second regime. Especially if we consider observation numbers larger than 5275, we can see that observations from IMS are significantly higher than the pattern.

This is also clearly visible for the SQIF pattern (see Fig. \ref{IMS_third_regime_space}). Until $t=5278$, every observation of IMS data is between the 5\% and 95\% quantile lines, however, later almost all of the observations are above the 95\% quantile line. There is no such observation in IMS data that exceeds outer quanile lines of orders 0\% and 100\%. 

\begin{figure}[H]
    \centering
    \begin{minipage}[t]{0.48\textwidth}
        \centering
        \includegraphics[width=1\textwidth]{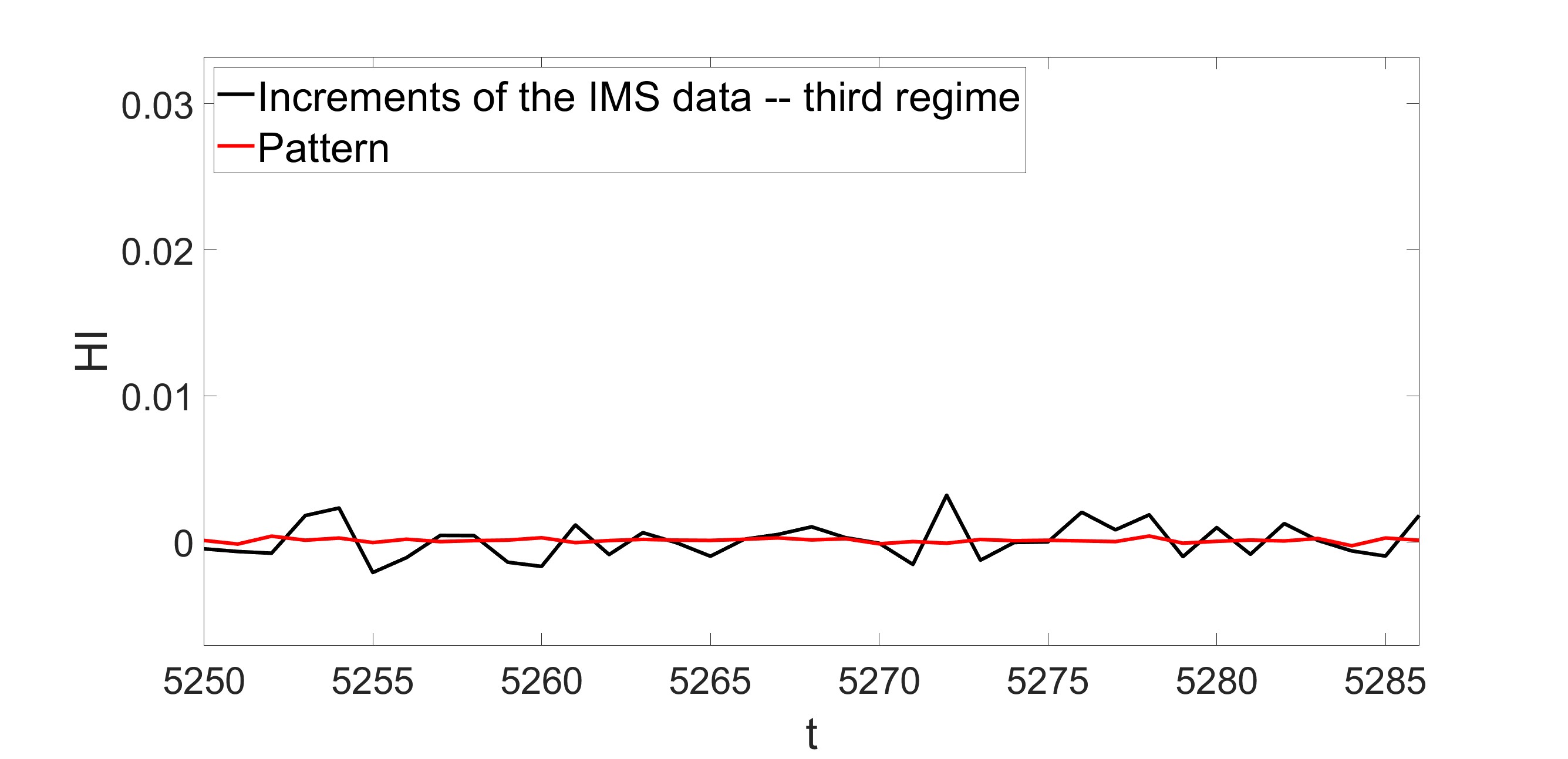}
        \caption{Comparison of the increments of the test data from the third regime of the IMS data with the pattern for Kupiec's POF}
    \label{IMS_third_regime_kupiec_pof}
    \end{minipage}
    \hfill
    \begin{minipage}[t]{0.48\textwidth}
        \centering
        \includegraphics[width=1\textwidth]{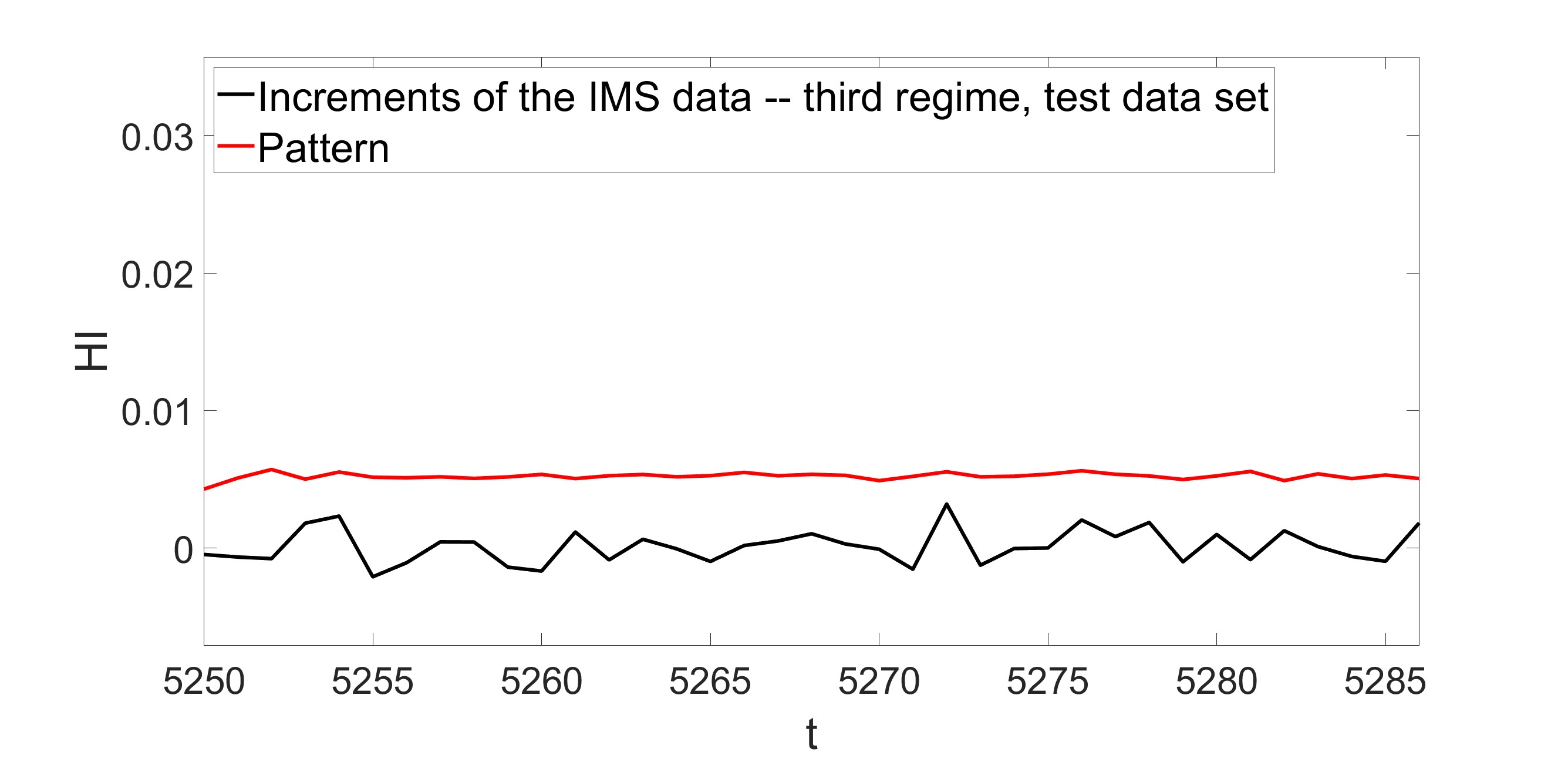}
        \caption{Comparison of the increments of the test data from the third regime of the IMS data with the pattern for Kupiec's TUFF}
    \label{IMS_third_regime_kupiec_tuff}
    \end{minipage}
\end{figure}

In Fig. \ref{IMS_third_regime_kupiec_pof}, the pattern for the Kupiec's POF metric is shown, which is a quantile line of order 51\%. We remind the reader that this pattern is obtained for increments of IMS data. As we can see, observations from IMS data vary around the pattern and there are about 50\% observations below the pattern and about 50\% observations above the pattern.

For Kupiec's TUFF metric, in Fig. \ref{IMS_third_regime_kupiec_tuff} pattern is presented. Here, the order of the quantile line (pattern for the Kupiec's TUFF metric) is obtained {according to the formula (\ref{p_TUFF})}. There are no observations that exceed the presented pattern.

\begin{figure}[H]
    \centering
        \includegraphics[width = 16cm]{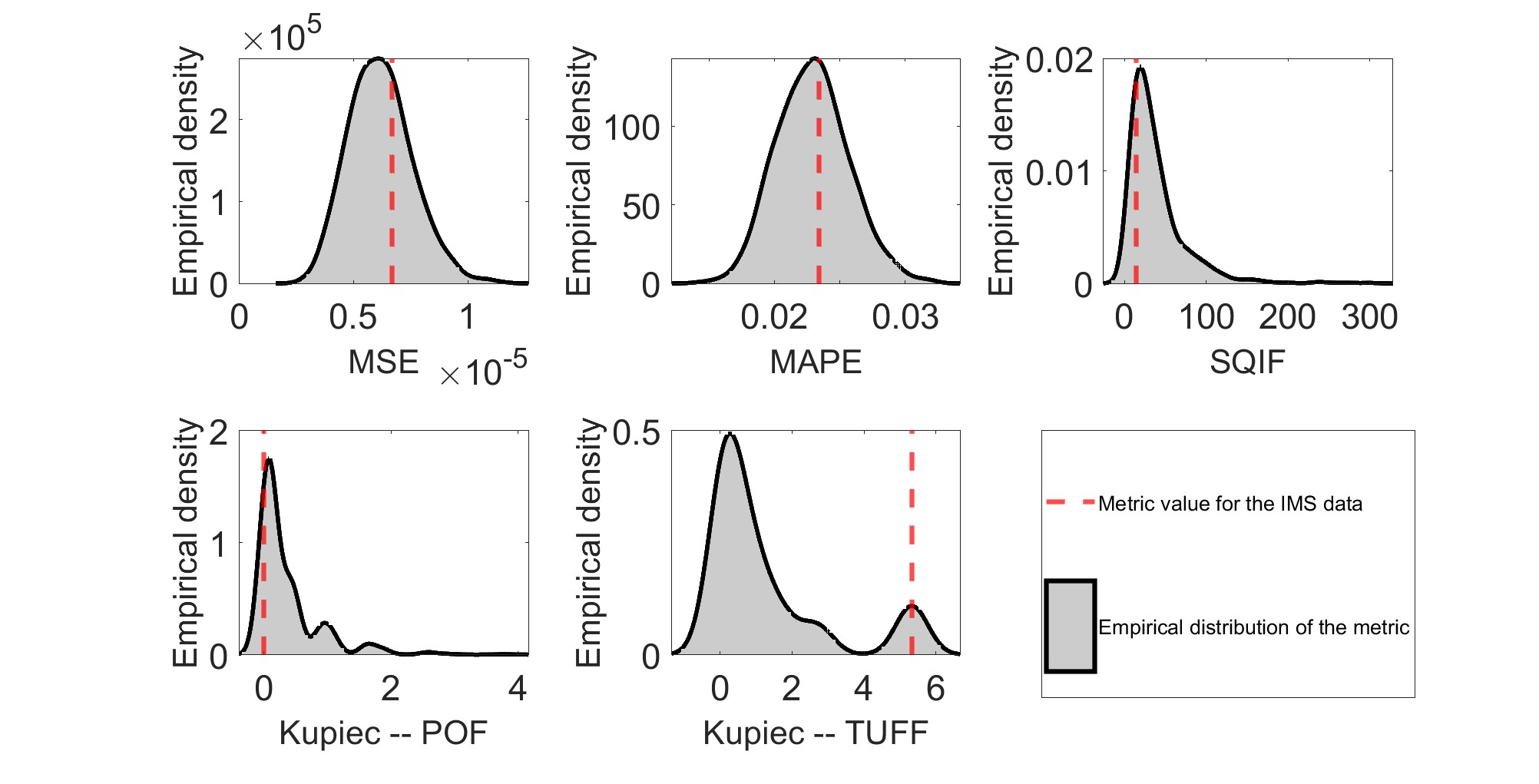}
        \caption{Comparison of metric distribution for 1000 training trajectories and for the test data from the third regime of the IMS data}
    \label{IMS_third_regime_densities}
\end{figure}

In Fig. \ref{IMS_third_regime_densities} we present a final assessment of the prediction for the third regime of IMS data. As we can see, the best results were obtained based on SQIF and Kupiec's POF metrics, where the mass of the distributions for these metrics derived for simulated trajectories of the model is on the right side of the metric for IMS data. This means that for most of the simulated trajectories of the model, larger metric values were obtained. Precise results are presented in Tab. \ref{IMS_third_regime_table}. The prediction evaluation for the SQIF metric is about 70\% and for Kupiec's POF statistic is 80\%.  For every metric, the prediction assessment is at least at 5\%, where the prediction assessment 5\% is achieved for Kupiec's TUFF metric. For MSE and MAPE, similar results are obtained, and these values are 30\% and 40\%, respectively.

\section{Discussion and conclusions}\label{sec_Discussion_Conclusions}


\textcolor{black}{In this paper we propose a novel and versatile approach for evaluating the results of data forecasting. This approach allows for a general assessment (reliable/non-reliable prognosis) or a more detailed evaluation expressed as a percentage. Almost any objective metric that provides a numerical value as a result can be utilized in this method. In this paper, we focus on five criteria: MSE, MAPE, SQIF, Kupiec's POF, and Kupiec's TUFF, and compare the results for both simulated and real data.}

\textcolor{black}{Based on the obtained results, we assess the prediction quality using a binary 0/1 decision at various levels. Our method allows for visual evaluation but also enables measurement according to the employed metrics. We analyze the most commonly observed degradation process regimes, namely degradation and critical stages, characterized by linear trends with linearly changing scales for the degradation regime and exponential trends with exponentially changing scales for the critical regime. For both regimes, the proposed procedure yields reasonable results, as demonstrated in the case of simulated data (see Section \ref{sec_simul_gauss}). Additionally, we apply the procedure to real datasets, namely FEMTO and IMS. Finally, we interpret the obtained values and provide explanations for the observed outcomes.}

\textcolor{black}{Due to the universality of the proposed procedure, we believe that the presented approach can be employed for various prognostic applications. This includes not only problems related to condition monitoring but also for assessing any kind of prognostic processes, such as those in the financial sector.}

\textcolor{black}{One of the primary limitations of the proposed procedure is that the model used to fit the data and perform the forecast must accurately reflect the examined time series.  It is worth to highlight that this paper focuses on the quality of prediction, assuming that the prediction is good and aiming to evaluate how good it is. If the model (or its estimated parameters) used for prediction is inappropriate, it is clear that we cannot discuss the quality of the prognosis.}

\textcolor{black}{Additionally, if we dispose of a data set with a small number of observations, some of the metrics should not be used due to their sensibility to the size of a data set. However, this is a typical constraint for a data modeling problem. \textcolor{black}{The impact of a small data size can be minimized by using resampling techniques like bootstrap, where the prediction assessment can be performed on a new dataset obtained by sampling with replacement from the observations in the prediction period.}
Finally, to obtain a precise percentage assessment, a large number of trajectories of the proposed model are required to be simulated, which can be computationally complex and time-consuming for large data sets from industrial monitoring systems. In most of the available data sets this problem is neglectable}.

\section*{Acknowledgements}
This work is supported by National Center of Science under Sheng2 project No.  UMO-2021/40/Q/ST8/00024 "NonGauMech - New methods of processing non-stationary signals (identification, segmentation, extraction, modeling) with non-Gaussian characteristics for the purpose of monitoring complex mechanical structures".

\bibliography{mybibliography}

\appendix





\section{Visualisations of metric patterns}

\begin{figure}[H]
    \centering
    \begin{minipage}[t]{0.48\textwidth}
        \centering
        \includegraphics[width=1\textwidth]{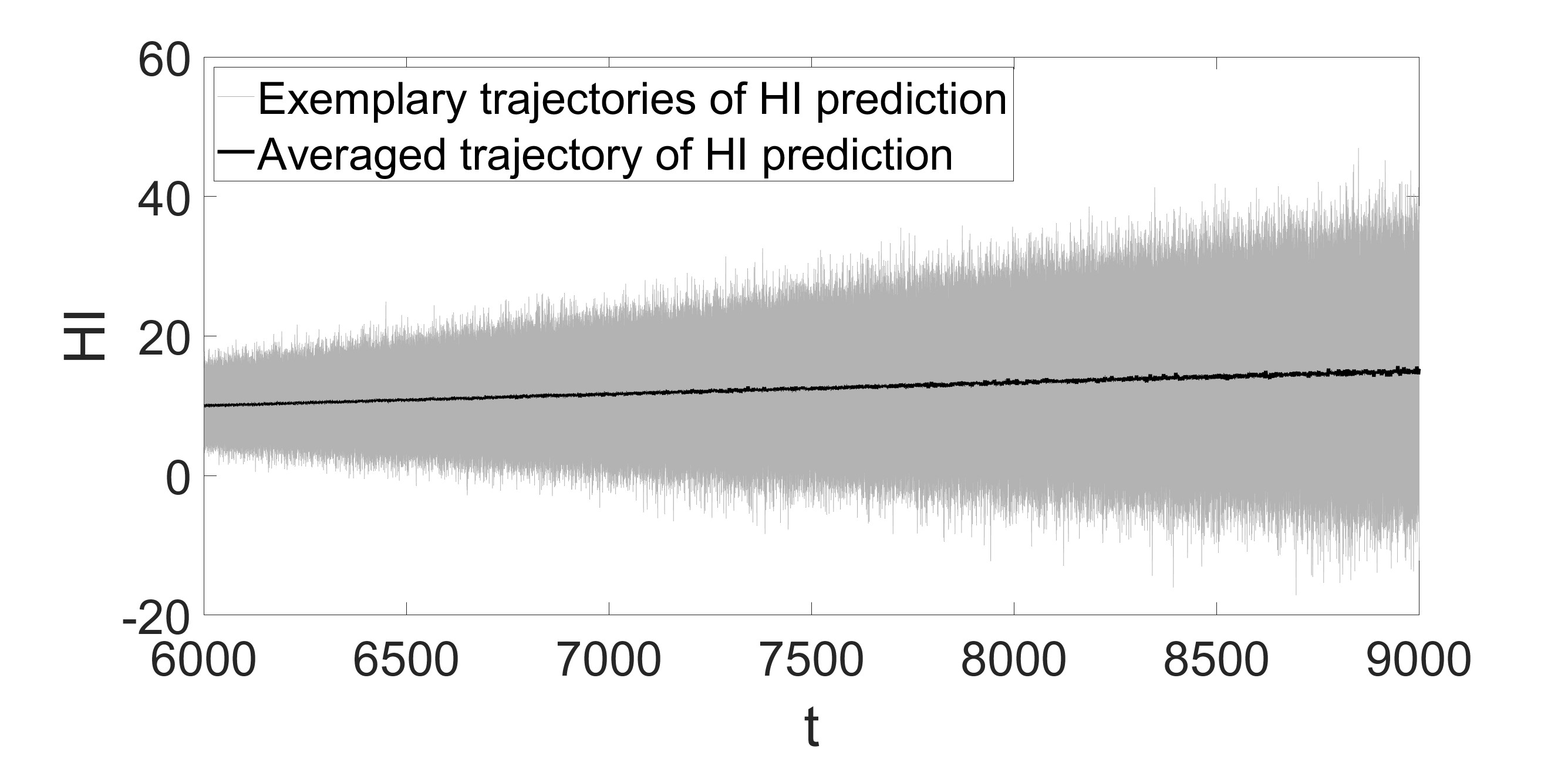}
        \caption{Comparison of exemplary trajectories of HI with averaged trajectory (pattern for MSE and MAPE metrics)}
    \label{block_diagram_avg}
    \end{minipage}
    \hfill
    \begin{minipage}[t]{0.48\textwidth}
        \centering
        \includegraphics[width=1\textwidth]{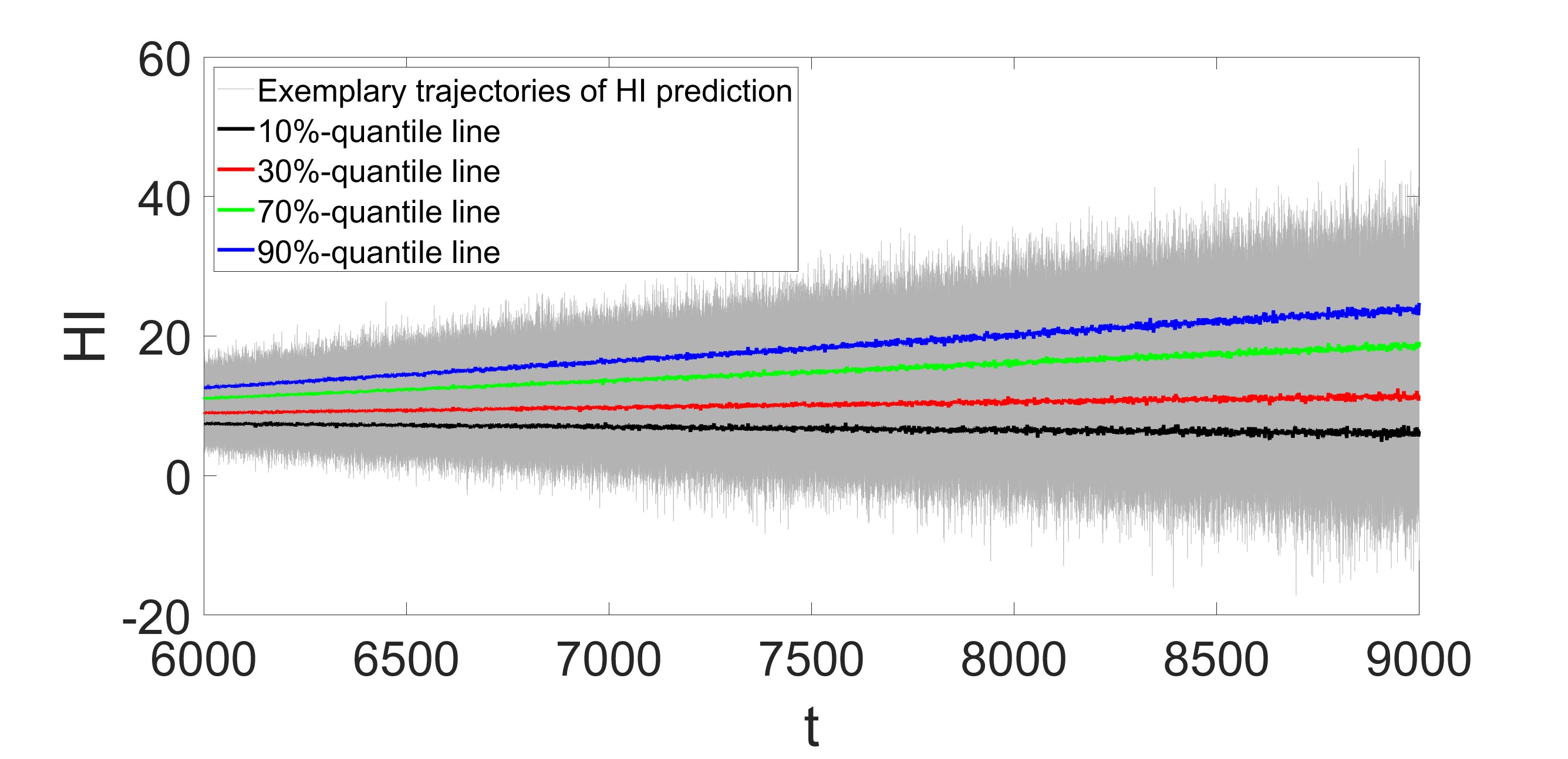}
        \caption{Comparison of exemplary trajectories of HI with selected quantile lines (pattern for SQIF)}
    \label{block_diagram_quantiles}
    \end{minipage}
\end{figure}
\begin{figure}[H]
    \centering
    \begin{minipage}[t]{0.96\textwidth}
        \centering
        \includegraphics[width=0.8\textwidth]{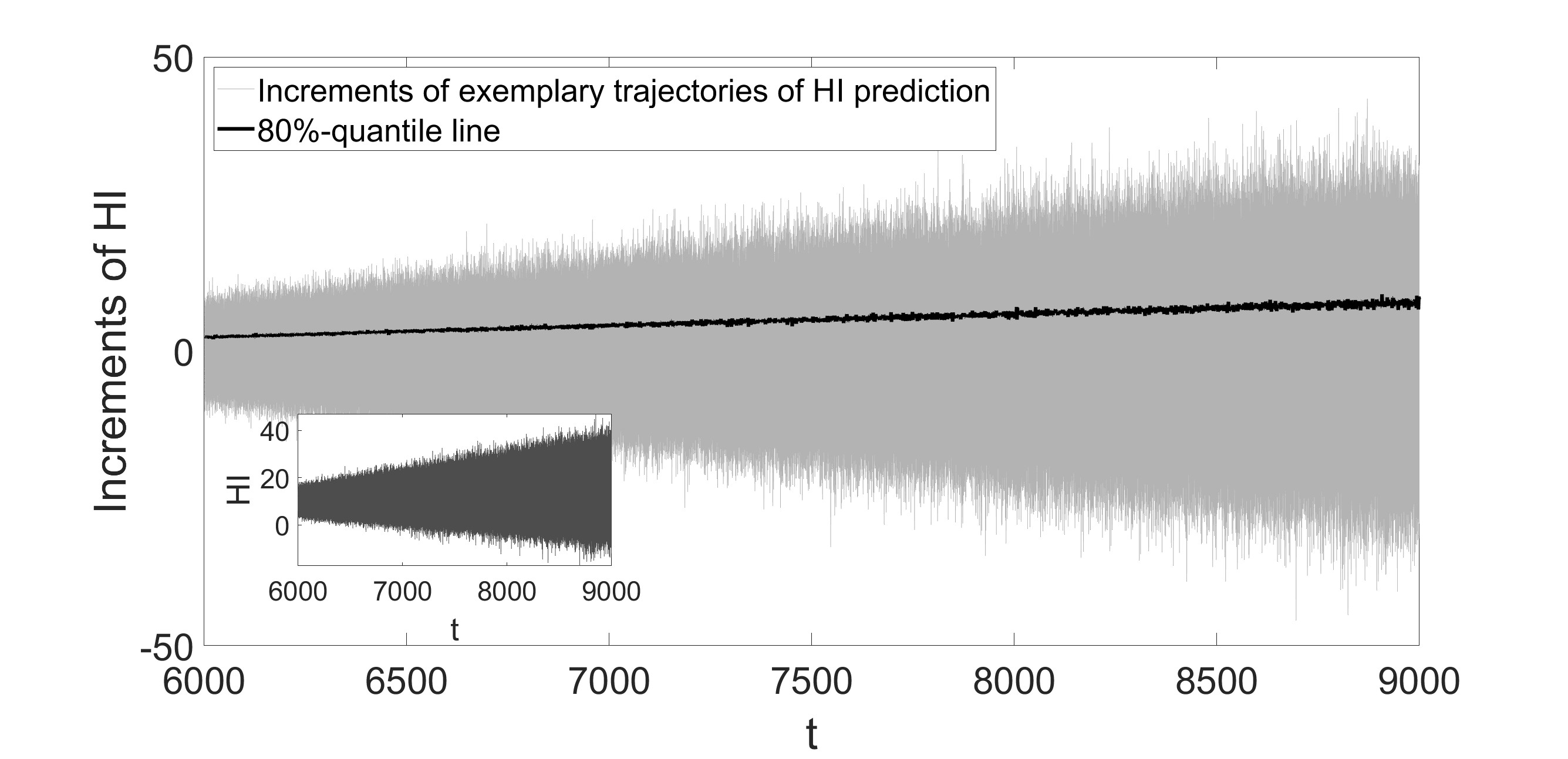}
        \caption{Comparison of increments of exemplary trajectories of HI with 80\% quantile line (exemplary pattern for Kupiec's POF and Kupiec's TUFF); in zoom -- exemplary trajectories of HI.}
    \label{block_diagram_kupiec_quantile}
    \end{minipage}
\end{figure}

\section{FEMTO data}


\begin{table}[H]
    \centering
    \begin{minipage}[t]{0.48\textwidth}
\centering
    \scalebox{0.75}{
    \begin{tabular}{|c|c|c|c|c|c|}
    \hline
    & \multicolumn{5}{c|}{Metric}  \\      \hline
        $\tau^*$ & MSE & MAPE & SQIF & \makecell{Kupiec's \\ POF} & \makecell{Kupiec's \\ TUFF} \\ \hline
        1 & 1 & 1 & 1 & 1 & 1 \\ \hline
        2 & 1 & 1 & 1 & 1 & 1 \\ \hline
        3 & 1 & 1 & 1 & 1 & 1 \\ \hline
        4 & 1 & 1 & 0 & 1 & 1 \\ \hline
        5 & 1 & 1 & 0 & 1 & 1 \\ \hline
        10 & 1 & 1 & 0 & 1 & 1 \\ \hline
        20 & 1 & 1 & 0 & 1 & 1 \\ \hline
        30 & 1 & 1 & 0 & 1 & 0 \\ \hline
        40 & 1 & 1 & 0 & 1 & 0 \\ \hline
        50 & 1 & 1 & 0 & 1 & 0 \\ \hline
        60 & 0 & 1 & 0 & 1 & 0 \\ \hline
        70 & 0 & 1 & 0 & 1 & 0 \\ \hline
        80 & 0 & 1 & 0 & 1 & 0 \\ \hline
        90 & 0 & 1 & 0 & 0 & 0 \\ \hline
    \end{tabular}}
    \caption{Prediction quality assessment for the FEMTO data -- test data set of the second regime, for $\tau^*\in\{1,2,3,4,5,10,20,30,\ldots,90\}$\%. 1 -- good prediction, 0 -- bad prediction.}
    \label{FEMTO_second_regime_table}
    \end{minipage}
    \hfill
    \begin{minipage}[t]{0.48\textwidth}
            \scalebox{0.75}{
    \begin{tabular}{|c|c|c|c|c|c|}
    \hline
    & \multicolumn{5}{c|}{Metric}  \\      \hline
        $\tau^*$ & MSE & MAPE & SQIF & \makecell{Kupiec's \\ POF} & \makecell{Kupiec's \\ TUFF} \\ \hline
        1 & 1 & 1 & 1 & 1 & 1 \\ \hline
        2 & 1 & 1 & 1 & 1 & 1 \\ \hline
        3 & 1 & 1 & 1 & 1 & 1 \\ \hline
        4 & 1 & 1 & 1 & 1 & 1 \\ \hline
        5 & 1 & 1 & 1 & 1 & 1 \\ \hline
        10 & 1 & 1 & 1 & 1 & 1 \\ \hline
        20 & 1 & 1 & 1 & 1 & 1 \\ \hline
        30 & 0 & 1 & 1 & 1 & 1 \\ \hline
        40 & 0 & 1 & 1 & 1 & 1 \\ \hline
        50 & 0 & 1 & 1 & 1 & 1 \\ \hline
        60 & 0 & 0 & 1 & 1 & 1 \\ \hline
        70 & 0 & 0 & 1 & 1 & 1 \\ \hline
        80 & 0 & 0 & 1 & 1 & 1 \\ \hline
        90 & 0 & 0 & 1 & 1 & 1 \\ \hline
    \end{tabular}}
    \caption{Prediction quality assessment for the FEMTO data -- entire third regime, for $\tau^*\in\{1,2,3,4,5,10,20,30,\ldots,90\}$\%. 1 -- good prediction, 0 -- bad prediction.}
    \label{FEMTO_third_regime_table}
    \end{minipage}
\end{table}

\section{IMS data}

\begin{table}[H]
    \centering
    \begin{minipage}[t]{0.48\textwidth}
\centering
    \scalebox{0.75}{
    \begin{tabular}{|c|c|c|c|c|c|}
    \hline
    & \multicolumn{5}{c|}{Metric}  \\      \hline
    $\tau^*$ & MSE & MAPE & SQIF & \makecell{Kupiec's \\ POF} & \makecell{Kupiec's \\ TUFF} \\ \hline
        1 & 1 & 1 & 1 & 1 & 1 \\ \hline
        2 & 1 & 1 & 1 & 1 & 1 \\ \hline
        3 & 1 & 1 & 1 & 1 & 1 \\ \hline
        4 & 1 & 1 & 1 & 1 & 1 \\ \hline
        5 & 1 & 1 & 1 & 1 & 1 \\ \hline
        10 & 1 & 1 & 1 & 1 & 1 \\ \hline
        20 & 1 & 1 & 1 & 1 & 1 \\ \hline
        30 & 1 & 1 & 0 & 1 & 1 \\ \hline
        40 & 1 & 1 & 0 & 1 & 1 \\ \hline
        50 & 1 & 1 & 0 & 1 & 1 \\ \hline
        60 & 0 & 1 & 0 & 1 & 1 \\ \hline
        70 & 0 & 1 & 0 & 1 & 1 \\ \hline
        80 & 0 & 0 & 0 & 1 & 1 \\ \hline
        90 & 0 & 0 & 0 & 0 & 1 \\ \hline
    \end{tabular}}
    \caption{Prediction quality assessment for the IMS data -- test data set of the second regime, for $\tau^*\in\{1,2,3,4,5,10,20,30,\ldots,90\}$\%. 1 -- good prediction, 0 -- bad prediction.}
    \label{IMS_second_regime_table}
    \end{minipage}
    \hfill
    \begin{minipage}[t]{0.48\textwidth}
    \scalebox{0.75}{
    \begin{tabular}{|c|c|c|c|c|c|}
    \hline
    & \multicolumn{5}{c|}{Metric}  \\   \hline
    $\tau^*$ & MSE & MAPE & SQIF & \makecell{Kupiec's \\ POF} & \makecell{Kupiec's \\ TUFF} \\ \hline
        1 & 1 & 1 & 1 & 1 & 1 \\ \hline
        2 & 1 & 1 & 1 & 1 & 1 \\ \hline
        3 & 1 & 1 & 1 & 1 & 1 \\ \hline
        4 & 1 & 1 & 1 & 1 & 1 \\ \hline
        5 & 1 & 1 & 1 & 1 & 1 \\ \hline
        10 & 1 & 1 & 1 & 1 & 0 \\ \hline
        20 & 1 & 1 & 1 & 1 & 0 \\ \hline
        30 & 1 & 1 & 1 & 1 & 0 \\ \hline
        40 & 0 & 1 & 1 & 1 & 0 \\ \hline
        50 & 0 & 0 & 1 & 1 & 0 \\ \hline
        60 & 0 & 0 & 1 & 1 & 0 \\ \hline
        70 & 0 & 0 & 1 & 1 & 0 \\ \hline
        80 & 0 & 0 & 0 & 1 & 0 \\ \hline
        90 & 0 & 0 & 0 & 0 & 0 \\ \hline
    \end{tabular}}
    \caption{Prediction quality assessment for the IMS data -- test data set of the third regime, for $\tau^*\in\{1,2,3,4,5,10,20,30,\ldots,90\}$\%. 1 -- good prediction, 0 -- bad prediction.}
    \label{IMS_third_regime_table}
    \end{minipage}
\end{table}

\end{document}